\documentclass[article]{aa}         % for the letters

\usepackage{graphicx}
\usepackage{txfonts}

\usepackage{natbib}
\usepackage{upgreek}
\usepackage{subcaption}
\usepackage{url}
\usepackage[rightcaption]{sidecap}
\usepackage{multirow}

\bibpunct{(}{)}{;}{a}{}{,}

\newcommand{\Teff}{\ensuremath{T_{\rm eff}}}                      % Effective temperature symbol
\newcommand{\logg}{\ensuremath{\log g}}                           % log(g) symbol
\newcommand{\vsini}{\ensuremath{v\sin i}}                         % v sin i simbol
\newcommand{\Msun}{\ensuremath{\,{\rm M}_\odot}}                  % Solar mass symbol
\newcommand{\Rsun}{\ensuremath{\,{\rm R}_\odot}}                  % Solar radius symbol
\newcommand{\Msunnom}{\hbox{$\mathcal{M}^{\rm N}_\odot$}}         % Solar mass (IAU units) symbol
\newcommand{\Rsunnom}{\hbox{$\mathcal{R}^{\rm N}_\odot$}}         % Solar radius (IAU units) symbol
\newcommand{\Lsunnom}{\hbox{$\mathcal{L}^{\rm N}_\odot$}}         % Solar luminosity (IAU units) symbol
\newcommand{\kms}{\,km\,s$^{-1}$}                                 % km/s symbol
\newcommand{\spd}{SPD}                                            % Spectral disentangling
\newcommand{\mc}[1]{\multicolumn{2}{c}{#1}}

\newcommand{\er}[3]{\ensuremath{#1^{+#2}_{-#3}}}

\begin{document}

\title{High-mass eclipsing binaries: a testbed for models of interior structure and evolution\thanks{Based on observations made with the ESO 3.6 m Telescope and the HARPS spectrograph, operated on La Silla, Chile by the European Southern Observatory.}
\thanks{Table\,A.1 is only available in electronic form at the CDS via anonymous ftp to cdsarc.cds.unistra.fr (130.79.128.5) or via        https://cdsarc.cds.unistra.fr/viz-bin/cat/J/A+A/}}
\subtitle{Accurate fundamental properties and surface chemical composition for V1034\,Sco, GL\,Car, V573\,Car and V346\,Cen}

\author{K.~Pavlovski\inst{1}
\and
J.~Southworth\inst{2}
\and
A.~Tkachenko\inst{3}
\and
T.~Van~Reeth\inst{3}
\and
E.~Tamajo\inst{4}
}

\institute{Department of Physics, Faculty of Science, University of Zagreb,
10\,000 Zagreb, Croatia \\
\email{pavlovski@phy.hr}
\and
Astrophysics Group, Keele University, Staffordshire, ST5 5BG, UK
\and
Institute of Astronomy, KU Leuven, Celestijnenlaan 200D, 3001 Leuven, Belgium
\and
University of Applied Sciences, 10\,410 Velika Gorica, Croatia
}

%  \date{Received April 1, 2019; accepted April 30, 2019}

% \abstract{}{}{}{}{}
% 5 {} token are mandatory
% {} leave it empty if necessary

\abstract
% context heading (optional)
{}
% aims heading (mandatory)
{The surface chemical compositions of stars are affected by physical processes which bring the products of thermonuclear burning to the surface. Despite their potential in understanding the structure and evolution of stars, elemental abundances are available for only a few high-mass binary stars. We aim to enlarge this sample by determining the physical properties and photospheric abundances for four eclipsing binary systems containing high-mass stars: V1034\,Sco, GL\,Car, V573\,Car and V346\,Cen. The components have masses 8--17\Msun\ and effective temperatures from 22\,500 to 32\,200 K, and are all on the main sequence.}
% methods heading (mandatory)
{We present new high-resolution and high signal-to-noise spectroscopy from HARPS, and analyse them using spectral disentangling and NLTE spectral synthesis. We model existing light curves and new photometry from the TESS satellite.}
% results heading (mandatory)
{We measure the stellar masses to 0.6--2.0\% precision, radii to 0.8--1.7\% precision, effective temperatures to 1.1--1.6\% precision, and abundances of C, N, O, Mg and Si. The abundances are similar to those found in our previous studies of high-mass eclipsing binaries; our sample now comprises 25 high-mass stars in 13 binary systems. We also find tidally-excited pulsations in V346\,Cen.}
% conclusions heading (mandatory)
{We reinforce our previous conclusions: interior chemical element transport is not as efficient in binary star components as in their single-star counterparts in the same mass regime and evolutionary stage, possibly due to the effects of tidal forces. Our ultimate goal is to provide a larger sample of OB-type stars in binaries which would enable a thorough comparison to stellar evolutionary models, as well as to single high-mass stars.}


\titlerunning{High-mass eclipsing binaries: a testbed for models of interior structure and evolution}
\authorrunning{Pavlovski et al.}

\maketitle

%%%%%%%%%%%%%%%%%%%%%%%%%%%%%%%%%%%%%%%%%%%%%%%%%%%%%%%%%%%%%%%%%%%%%%%%%%%%%%%%%%%%%%%%%%%%%%%%%%%%%%%%%%%%%%%%%%%%%%%%%%%%%%%%%%%%%%%%%%%%%%%%%%%%%%%%%%%%%%%%%%%%%%%%%%%%%%%%%%%%%%%%%%%%%%%%%%%%%%%%

\section{Introduction}

The interior structure and evolution of a star are largely determined by its mass and chemical composition at formation. Precise and accurate observational constraints on these fundamental physical quantities are required for the validation, calibration and improvement of theoretical models of the interior structure and evolution of stars. Despite being much more complex than single stars, binary star systems are a treasure trove for testing stellar structure and evolution models and understanding how these might be improved. In the case of eclipsing binaries (EBs) where both components are detected spectroscopically, it is possible to measure their masses and radii with high precision and accuracy using only orbital mechanics and geometry. Detached systems are particularly valuable as they are expected to evolve as single stars without alteration of their evolution by mass transfer episodes.

\begin{table*} \centering
\caption{Basic characteristics of binary systems studied in this work.
}
\begin{tabular}{lccccccc} \hline
Binary     & Other       & Orbital    & $V_{\rm max}$ & Spectral            & Age   & Cluster    & Apsidal         \\
system     & designation & period (d) & (mag)         & types               & (Myr) & membership & period (yr)     \\
\hline
V1034\,Sco  & CPD$-$41$^{\rm o}$7742   & 2.44  & 8.80  & O9.5\,V + B1-1.5\,V & 3--8 & NGC\,6231    & $23.4\pm0.8$    \\
GL\,Car     & HD~306168                & 2.42  & 9.74  &   B0.5\,V + B1\,V   & 2.0  & NGC\,3572    & $25.20\pm0.02$  \\
V573\,Car   & CPD$-$59$^{\rm o}$2628   & 1.47  & 9.47  & O9.5\,V + B0.3\,V   & 2    & Trumpler\,16 & -               \\
V346\,Cen   & HD~101837                & 6.32  & 9.57  &    B1/3II/III       & 10   & Stock~14    & $306 \pm 4$     \\
\hline
\end{tabular}
\label{tab:sample}
\tablefoot{References to the quantities are given in Section~\ref{sec:sample}.  }
\end{table*}

The role of precise empirical mass measurements is difficult to overstate for validating and calibrating modern and sophisticated stellar models. \citet{Herrero1992} presented a study of 25 luminous galactic OB-type stars and reported a discrepancy between the masses inferred from their spectra (via wind theory) and those predicted by evolutionary models. The authors termed the effect the ``mass discrepancy'' and emphasised the difficulty in attributing it to either of the two theories (wind and stellar evolution) involved. Since then, many attempts have been made to diagnose the cause of the mass discrepancy in intermediate- to high-mass stars.

Given the high precision and accuracy that SB2 detached eclipsing binaries (dEBs) allow us to achieve in measurements of mass and surface gravity \citep[e.g.,][]{Torres_2010}, these objects are important in studying the mass discrepancy. \citet{Burkholder1997} studied seven early-type spectroscopic binaries with masses below 15\Msun\ and reported a good agreement between masses inferred from binary dynamics and those estimated with evolutionary models in all cases where the stars are non-interacting. \citet{Guinan2000} and \citet{Pavlovski_2009} presented independent studies of the high-mass SB2 dEB V380\,Cyg and reported a substantial mass discrepancy for the evolved primary component, in the sense that its dynamical mass is too low compared to the predictions of standard stellar models. Whereas \citet{Guinan2000} showed that the discrepancy could be resolved by introducing extra near-core mixing in the form of convective core overshooting, \citet{Pavlovski_2009} found that rotationally induced mixing in models was insufficient to explain the mass discrepancy. Indeed, \citet{Tkachenko_2014a} demonstrated that only the combined effects of rotation and convective core overshooting can account for the mass discrepancy observed in V380\,Cyg.

Recently, \citet{Massey2012}, \citet{Morrell2014}, \citet{Mahy2015}, and \citet{Pavlovski_2018} reported systematic discrepancies between the Keplerian and evolutionary masses of stars less massive than 30\Msun. \citet{Mahy_2020a} found a good agreement between the spectroscopic and dynamical masses for 26 early-type binary components whereas their evolutionary masses appear to be systematically overestimated. These results hint towards models of interior structure and evolution being the primary cause of the mass discrepancy. \citet{Tkachenko_2020} and \citet{Johnston_2021} demonstrated that the problem cannot be attributed to differences in observation and analysis methods between research groups, and instead showed that the mass discrepancy progressively increases with the evolutionary stage of the star. In particular, the authors found that higher convective core masses were required in models of stellar structure and evolution for stars that are born with a convective core. The effects of excess core mass can be efficiently mimicked with an enhanced mixing in the near-core regions, irrespective of the true cause(s) of the mixing.

Connecting the treatment of interior mixing in stellar evolution models and the mass discrepancy requires extra observational constraints. Surface chemical composition measurements are ideal because chemical abundance patterns are expected to be substantially altered by various mechanisms of interior mixing and chemical element transport in stars. For example, \citet{Heap2006} found surface nitrogen enrichment in 80\% of their sample stars and speculated on the role of rotation in causing this phenomenon. Overall, these observational findings are in good agreement with predictions from rotating stellar evolution models for high-mass stars \citep{Meynet_Maeder_2000, Maeder_Meynet_2000, Heger_2000, Heger_Langer_2000,Langer_2012} with the caveat that the observed nitrogen enrichments \citep{Heap2006} are larger than those predicted by the models.

\citet{Hunter_2008,Hunter_2009} studied a large sample of intermediate- to high-mass stars in the Magellanic Clouds. Some of their findings corroborate the theory of rotationally-induced mixing while others contradict it (e.g., rapidly and slowly rotating stars without and with substantial surface nitrogen enrichment, respectively). Slowly-rotating, nitrogen-enriched stars were also found by \citet{Markova2018} and the authors suggested that inadequacies of the models in these particular cases might be related to the efficiency of rotational mixing. At the same time, \citet{Pavlovski_2018} presented a detailed study of the surface chemical compositions in several high-mass SB2 dEBs and found no dependence of the abundances of carbon (C), nitrogen (N) and oxygen (O) on either the projected rotational velocity (\vsini) or surface gravity (\logg) of the star.

Whilst changes in the photospheric CNO abundances of high-mass B-type single stars have been found \citep{Przybilla_2010, Nieva_Przybilla_2012, Maeder_2014, Cazorla_2017a, Cazorla_2017b, Markova2018}, the role of rotationally-induced mixing in the formation of these chemical abundance patterns remains poorly quantified. At the same time, \citet{Aerts_2014} demonstrated that neither \vsini\ nor the rotational frequency of a star has significant predictive power for the surface N abundance. Instead, the latter correlates strongly with the effective temperature (\Teff) of the star and the frequency of its dominant acoustic oscillation mode. Furthermore, \citet{Rogers2013} showed that internal gravity waves (IGWs) excited at the convective-radiative boundary near the core in high-mass stars are efficient in transporting angular momentum and chemicals on short timescales and over large distances. \citet{Pedersen_2018} demonstrated that the IGW-driven functional form of the interior mixing profile is a good candidate to simultaneously explain the observed properties of gravity-mode oscillations and surface abundances in B-type stars.

SB2 dEBs are at the forefront of efforts to resolve deficiencies in theoretical stellar models. The masses and radii of the component stars can be measured precisely and independently of models, and the requirement for the stars to have the same age and initial chemical composition at formation provides an additional stringent constraint on theoretical models. Moreover, measured masses and radii give a precise surface gravity which can be used to break the degeneracy between \Teff\ and \logg\ in spectral analysis, boosting the accuracy of measurements of the surface chemical compositions of the stars. The DEBCat\footnote{\url{http://www.astro.keele.ac.uk/jkt/debcat/}} catalogue of dEBs \citep{Southworth_2015} currently lists approximately 300 examples with precisely-measured masses and radii, but only a small fraction of high-mass systems have useful constraints on their photospheric chemical abundances \citep{Serenelli_2021}. In this study, we aim to enlarge the sample of high-mass SB2 dEBs with accurately determined surface chemical abundance patterns. In Sections~\ref{sec:sample} and \ref{sec:data}, we present the sample and high-quality spectroscopic data used in this study. Section~\ref{sec:orbits} covers the determination of the spectroscopic orbits of the stars, Sections~\ref{sec:atmos} and \ref{sec:abund} the inference of their atmospheric parameters and chemical abundances, and Section~\ref{sec:lc} the light curve analysis. The chemical compositions, ages and distances to the binary stars analysed are compared in Section~\ref{sec:parents} to the properties of their parent clusters. We finish with a discussion (Section~\ref{sec:discussion}) and conclusions (Section~\ref{sec:conclusions}). It is important to state that the various analyses presented in this work were performed iteratively to ensure internal consistency in the derived results.

%%%%%%%%%%%%%%%%%%%%%%%%%%%%%%%%%%%%%%%%%%%%%%%%%%%%%%%%%%%%%%%%%%%%%%%%%%%%%%%%%%%%%%%%%%%%%%%%%%%%%%%%%%%%%%%%%%%%%%%%%%%%%%%%%%%%%%%%%%%%%%%%%%%%%%%%%%%%%%%%%%%%%%%%%%%%%%%%%%%%%%%%%%%%%%%%%%%%%%%%

\section{Sample} \label{sec:sample}

We selected four main-sequence (MS) dEBs for study, based on the masses of the components, their membership of open clusters or associations, and on their visibility during the telescope time we were allocated. Basic information on these targets is given in Table~\ref{tab:sample}. All eight stars have a spectral type of late-O or early-B and a surface gravity between 3.7 or 4.2 dex, so are in a relatively early evolutionary phase. All systems except V573\,Car reside in eccentric orbits. All are confirmed members of open clusters, although we did not impose any additional constraints on their ages and/or chemical compositions from the cluster membership. All of our targets except GL\,Car were included in the homogeneous sample of \citet{Tkachenko_2020}.

V1034\,Sco is located in the core of the open cluster NGC\,6231, which in turn is near the centre of the Sco~OB1 association. A detailed spectroscopic and X-ray study was presented by \citet{Sana_2003}. Light curves have been presented and analysed by \citet{Sana_2005} and \citet{Bouzid_2005}. The most recent analysis is that published by \citet{Rosu_2022b}, who determined physical properties of the components and measured apsidal motion from our spectra (retrieved from the ESO archive) and the light curves from \citet{Bouzid_2005}.

\begin{figure*} \centering
\includegraphics[width=\textwidth]{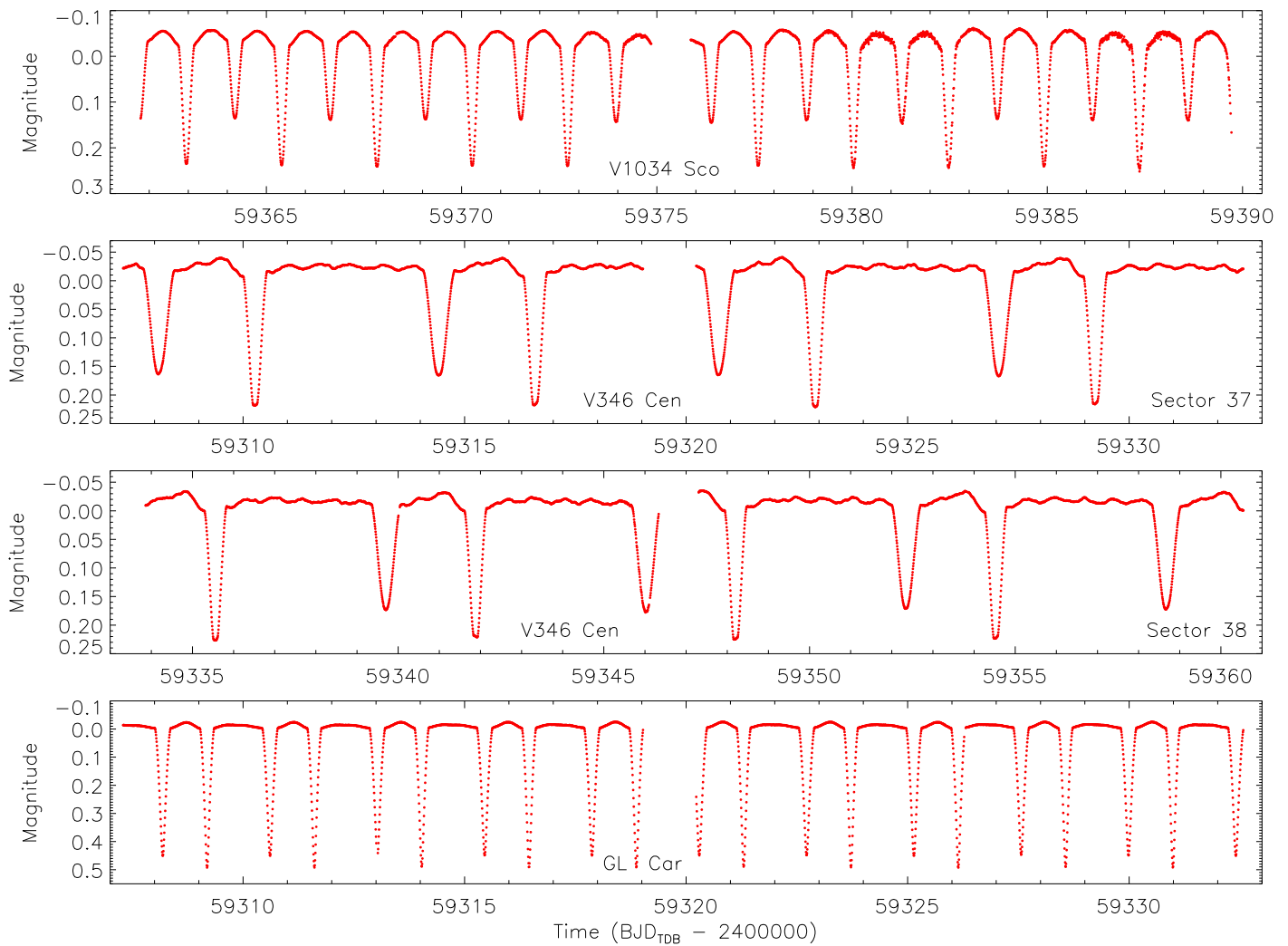}
\caption{\label{fig:tess} Light curves used in the current study from our reduction of data from the TESS satellite.  They have been normalised to zero magnitude for display purposes.}
\end{figure*}

GL\,Car is a dEB studied by the Copenhagen group \citep{Gimenez_Clausen_1986} using $uvby$ photometry. The system shows a significant orbital eccentricity ($e=0.157$) and fast apsidal motion with $U = 25.20 \pm 0.02$~yr \citep{Gimenez_Garcia-Pelayo_1983, Gimenez_Clausen_1986, Wolf_2008}. We know of no previous time-series spectroscopy of the system, so the current work provides the first measurements of its physical properties.

V573\,Car is one of the brightest stars in very young open cluster Trumpler\,16, although its membership is disputed \citep{Kaltcheva_Georgiev_1993}. Its spectroscopic binary nature was found by \citet{Walborn_1982}, and the discovery of eclipses was made by \citet{Freyhammer_2001} during a study of the nearby massive binary system $\eta$~Carinae. \citet{Freyhammer_2001} obtained extensive $uvby$ photometry and, combined with radial velocities (RVs) from \citet{Levato_1991}, determined the physical properties of the system.

V346\,Cen contains early B-type components \citep{Houk_Cowley_1975} in an orbit with a significant eccentricity. Apsidal motion is present with a period of $U = 306 \pm 4$~yr \citep{Gimenez_1986a, Drobek_2013}. High-quality light curves in the Str\"omgren $uvby$ system were obtained and analysed by the Copenhagen group \citep{Gimenez_1986a, Gimenez_1986b}. The only full spectroscopic dataset for this system is our own HARPS data, available through the ESO archive, and which were already analysed by \citet{Mayer_2016}.

\section{Observations} \label{sec:data}

\subsection{Spectroscopy}

The spectra presented in this work were all taken in one observing run\footnote{ESO proposal 083.D-0040(A), PI J.\ Southworth} over the nights 2--7 April 2009 using the High-Accuracy Radial-velocity Planet Searcher (HARPS) cross-dispersed \'echelle spectrograph \citep{Mayor_2003} at the 3.6-m telescope at ESO La Silla.  HARPS achieves extreme RV precision due to a high mechanical stability, being fed by two optical fibres, sited in a vacuum chamber, and calibrated by a Th-Ar emission lamp. We operated HARPS in the high-efficiency EGGS mode, which has a resolving power of $R = 80\,000$, and used the second fibre to obtain the sky background during each observation. Each spectrum consists of 72 orders incident on two CCDs, covering 3780--6900\,\AA\ with a gap at 5304--5337\,\AA\ between the CCDs.

We reduced the spectra using semi-automatic IRAF\footnote{IRAF is distributed by the National Optical Astronomy Observatory, which are operated by the Association of the Universities for Research in Astronomy, Inc., under cooperative agreement with the NSF.} scripts. Reduction of the spectra included the standard steps: bias subtraction, flat-field correction, spectral order localisation, extraction, and wavelength calibration. Normalisation of extracted spectral orders was performed by fitting ninth-order polynomial functions to selected continuum points in the blaze function. Since the Balmer lines cover up to three consecutive spectral orders, these were normalised by interpolating the blaze functions from adjacent orders as described by \citet{Kolbas_2015}. The HARPS blaze functions are very stable so the normalisation and merging of even these difficult orders produced very satisfactory results.

\subsection{Photometry}

% V1034:
% Sector 12 (2019-May-21 to 2019-Jun-19, in cycle 1): observed in camera 1.
% Sector 39 (2021-May-26 to 2021-Jun-24, in cycle 3): observed in camera 1.
% Sector 66 (2023-Jun-02 to 2023-Jul-01, in cycle 5): observed in camera 1.
%
% V346:
% Sector 10 (2019-Mar-26 to 2019-Apr-22, in cycle 1): observed in camera 3.
% Sector 11 (2019-Apr-22 to 2019-May-21, in cycle 1): observed in camera 3.
% Sector 37 (2021-Apr-02 to 2021-Apr-28, in cycle 3): observed in camera 3.
% Sector 38 (2021-Apr-28 to 2021-May-26, in cycle 3): observed in camera 3.
% Sector 64 (2023-Apr-06 to 2023-May-04, in cycle 5): observed in camera 3.
%
% V573:
% Sector 10 (2019-Mar-26 to 2019-Apr-22, in cycle 1): observed in camera 3.
% Sector 36 (2021-Mar-07 to 2021-Apr-02, in cycle 3): observed in camera 3.
% Sector 37 (2021-Apr-02 to 2021-Apr-28, in cycle 3): observed in camera 3.
% Sector 63 (2023-Mar-10 to 2023-Apr-06, in cycle 5): observed in camera 3.
% Sector 64 (2023-Apr-06 to 2023-May-04, in cycle 5): observed in camera 3.
%
% GL:
% Sector 10 (2019-Mar-26 to 2019-Apr-22, in cycle 1): observed in camera 3.
% Sector 11 (2019-Apr-22 to 2019-May-21, in cycle 1): observed in camera 3.
% Sector 37 (2021-Apr-02 to 2021-Apr-28, in cycle 3): observed in camera 3.
% Sector 64 (2023-Apr-06 to 2023-May-04, in cycle 5): observed in camera 3.

Our analysis below originally relied on published ground-based light curves, as will be discussed in Section~\ref{sec:lc}. In the course of this work, additional data became available from the NASA Transiting Exoplanet Survey Satellite (TESS), a space-based mission that has observed most of the celestial sphere in sectors of 27.4~d duration \citep{Ricker+15jatis}. The TESS datasets used in the current study are shown in Fig.~\ref{fig:tess}. 
Additional data are shown in Fig.~\ref{fig:tessnotused}, and may be useful in future for period or apsidal motion studies.

\begin{figure*}
\begin{tabular}{cc}
\includegraphics[width=85mm]{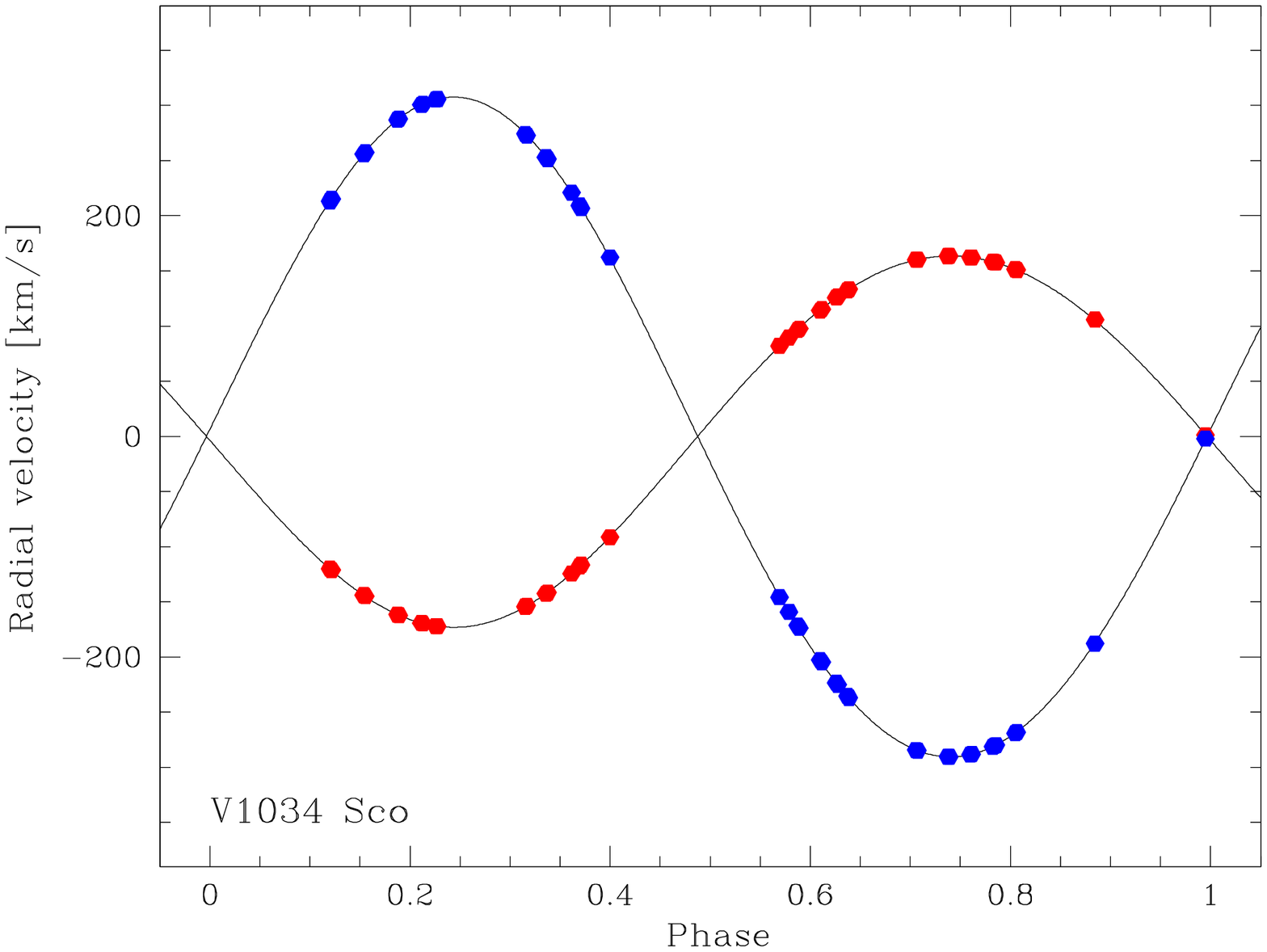} &
\includegraphics[width=85mm]{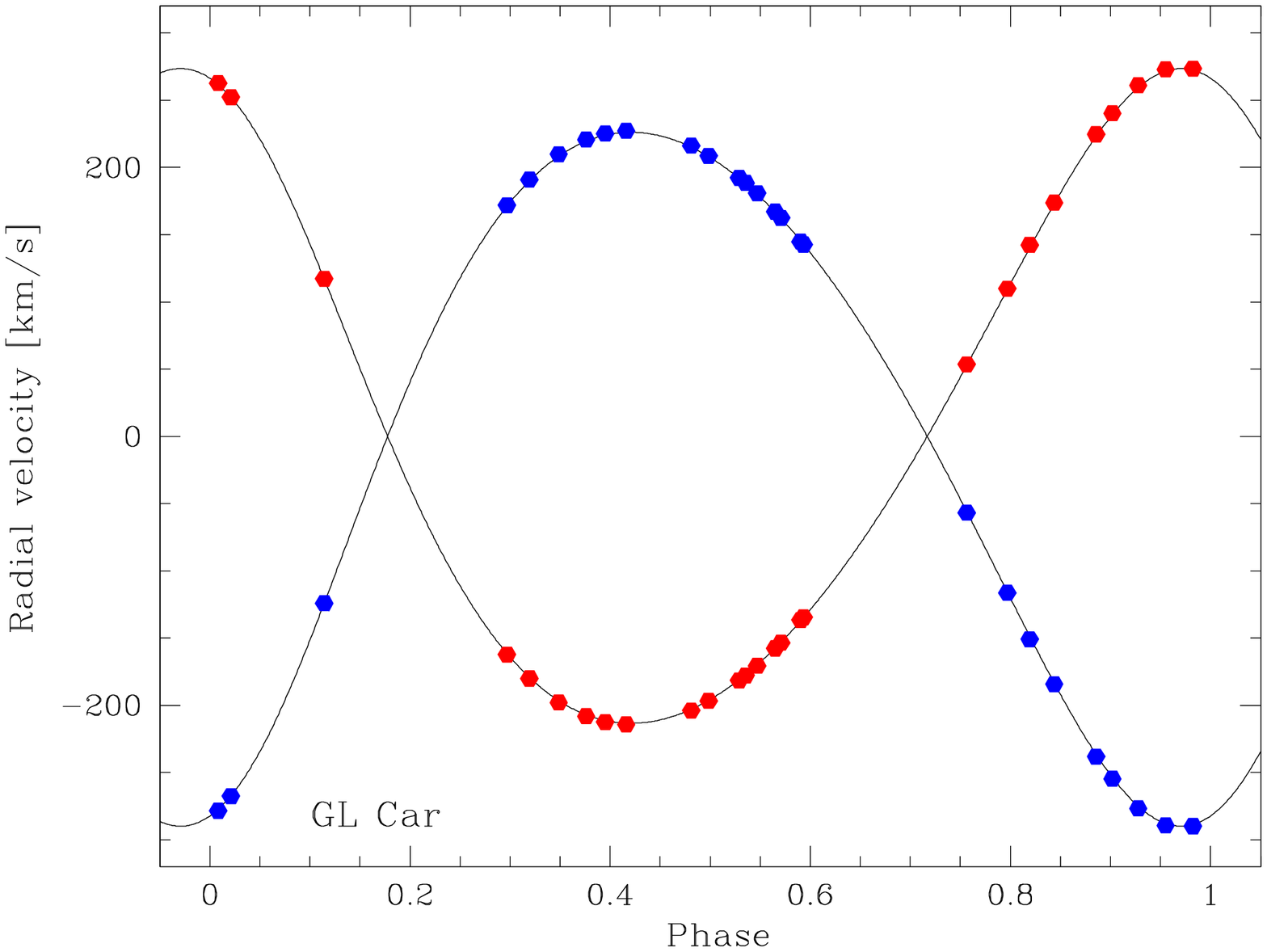} \\
\includegraphics[width=85mm]{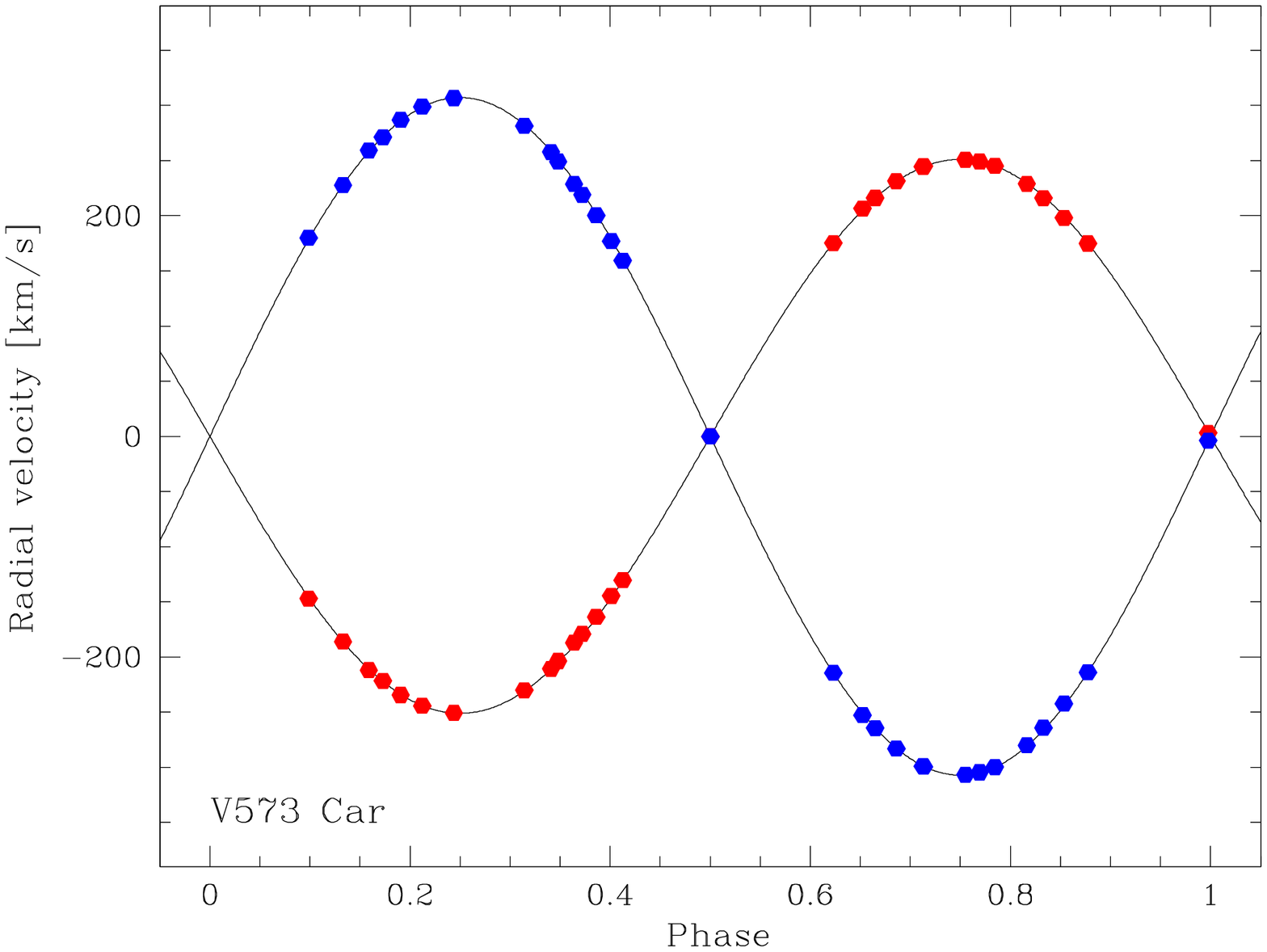}   &
\includegraphics[width=85mm]{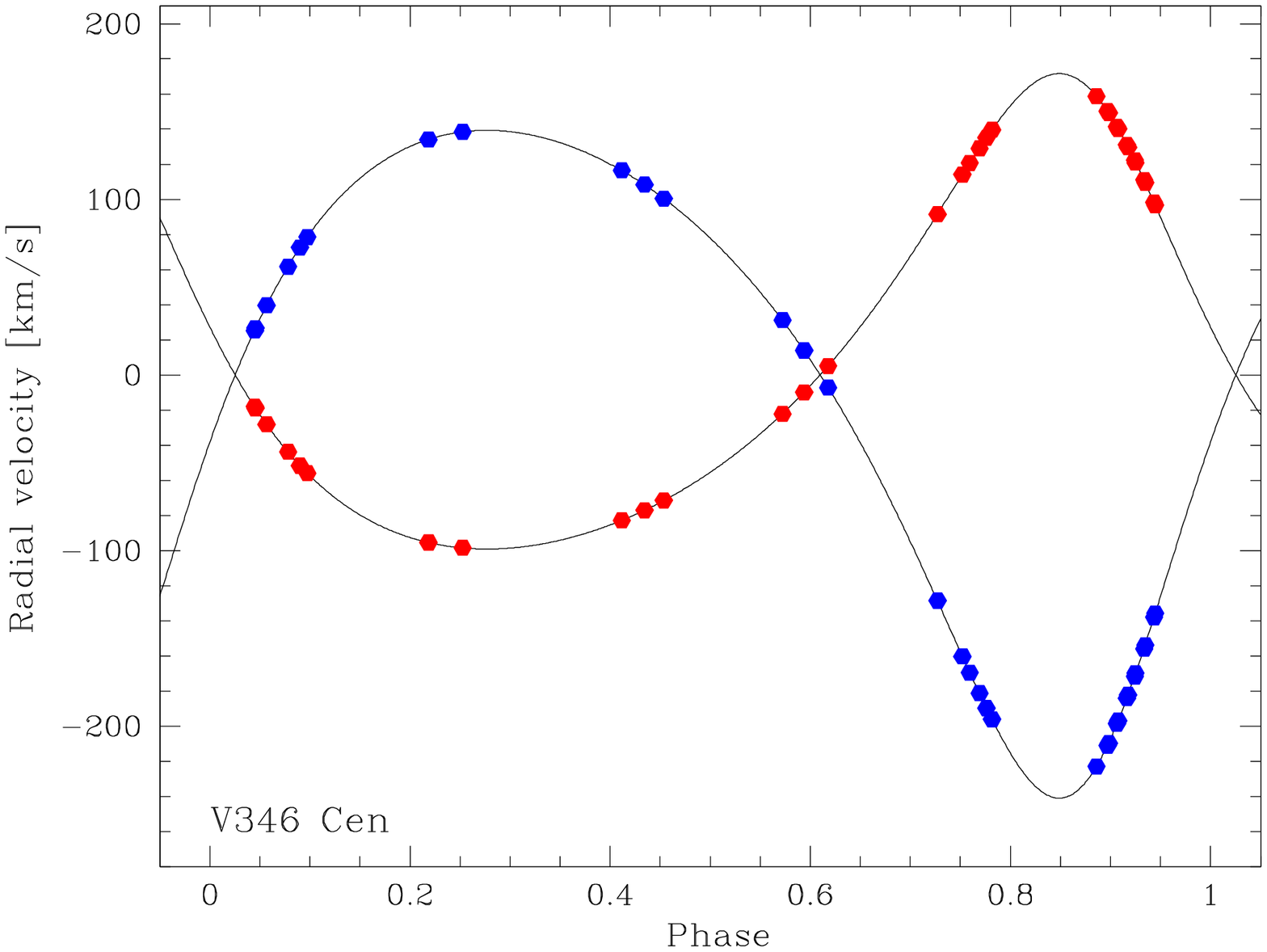} \\
\end{tabular}
\caption{\label{fig:RVs} Visualisation of the spectroscopic orbits of our targets. The best-fitting orbits are shown with black lines and the RVs of the stars at the times of observation with red symbols for the primary component, and blue symbols for the secondary component. Note that these are not measured RVs, hence the uncertainties in RVs are not assigned to individual symbols, because we calculate orbital parameters directly from all observed spectra for each system  (see Section \ref{sec:orbits:spd}).}
\end{figure*}

TESS observed V1034\,Sco in sectors 12 (1800~s cadence) and 39 (600~s cadence). We extracted the light curves using custom aperture masks. V1034\,Sco is in a crowded field and the TESS pixels subtend a large angle (21$^{\prime\prime}$) so the light curves contain a significant amount of third light. Our analysis was based on sector 39 due to the better temporal sampling.

V346\,Cen was observed using TESS in sectors 10 and 11 (1800~s cadence), and 37 and 38 (600~s cadence). The light curves available for download from MAST\footnote{https://mast.stsci.edu/portal/Mashup/Clients/Mast/Portal.html} are very affected by the field crowding but are nevertheless much better than the ground-based data for this object. We based our analysis on the data from sectors 37 and 38.

TESS observed V573\,Car in sectors 10, 36 and 37. Because this object is very close to the extremely bright $\eta$~Car binary system, the standard data products from TESS \citep{Jenkins+16spie} are unreliable. We therefore extracted photometry from the halo of V573\,Car by making a careful customised pixel selection for the aperture mask, with the aim to maximise the collected flux of V573\,Car compared to the flux of $\eta$\,Car. The resulting light curves are of relatively low quality and suffer from a large and varying amount of third light, so we did not use these in our analysis. We note that V573\,Car was just outside the field of view of TESS during sector 11 but we were still able to extract a light curve using halo photometry.

GL\,Car was observed by TESS in sectors 10 and 11 (1800~s cadence), and 37 (600~s cadence). The light curves available on MAST have eclipses that are too deep so we again extracted our own photometry from the TESS full-frame images using custom aperture masks.

\section{Spectroscopic orbits} \label{sec:orbits} \label{sec:orbits:spd}

\begin{table*} \centering
\caption{\label{tab:specorb} Parameters of the spectroscopic orbits for the four targets determined by \spd.} 
\begin{tabular}{lccccccc} \hline\hline
Binary    & $P$   & $T_{\rm peri}$ & $e$  & $\omega$       &    $K_{\rm A}$    & $K_{\rm B}$     & $q$       \\
system    & (d)      & (BJD)          &      & (deg)          &    (km\,s$^1$)    & (km\,s$^1$)     &                  \\
\hline
V1034\,Sco & 2.440656 & 51934.356$\pm$0.032 & 0.029 $\pm$ 0.003 & 191  $\pm$ 12  & 168.3  $\pm$ 0.3  & 299.0 $\pm$ 1.1 & 0.563 $\pm$ 0.002 \\
GL\,Car    & 2.422238 & 54901.182$\pm$0.015 & 0.146 (fixed)       & 32.2 $\pm$ 2.7 & 244.6  $\pm$ 1.8  & 259.3 $\pm$ 1.6 & 0.943 $\pm$ 0.009 \\
V573\,Car  & 1.469332 & - & 0.0               & 90             & 250.61 $\pm$ 0.71 & 306.3 $\pm$ 1.1 & 0.818 $\pm$ 0.011 \\
V346\,Cen  & 6.321835 & 50452.543$\pm$0.016 & 0.289 $\pm$ 0.006 & 22.2 $\pm$ 1.3 & 135.3  $\pm$ 0.6  & 190.1 $\pm$ 0.7 & 0.712 $\pm$ 0.038 \\
\hline
\end{tabular}
\end{table*}

The spectra of binary systems containing high-mass stars are difficult to analyse for several reasons. First, the \vsini\ values are typically large, smearing out the spectral lines and causing the lines from the two components to blend together even around the phases of maximum RV difference. Second, there are relatively few spectral lines that are strong enough to provide useful RV information. We therefore determined the spectroscopic orbits of the stars using the method of spectral disentangling (\spd). This method was introduced by \citet{Simon_Sturm_1994} in wavelength space and by \citet{Hadrava_1995} in Fourier space. It represents the observed composite spectra of a binary system as a sum of the individual spectra of the two stars shifted in RV according to their orbital motion. \spd\ makes it possible to quantitatively analyse time-series spectra of SB2 systems even when line blending is strong \citep{Hensberge_2000,Pavlovski_Hensberge_2005}. Note that no  template spectra are needed for \spd, thus avoiding any biases due to template mismatch \citep{Hensberge_Pavlovski_2007}.

We used the {\sc FDBinary}\footnote{\url{http://sail.zpf.fer.hr/fd3}} code \citep{Ilijic_2004} to perform \spd\ in Fourier space using Fast Fourier Transform (FFT). For each object we analysed all spectra simultaneously to determine the disentangled spectra of the two stars and their spectroscopic orbital parameters. We fitted directly for the orbital parameters, without the intermediate step of calculating RVs.  The orbital parameters were the orbital period, $P$, time of periastron pasage, $T_{\rm peri}$, eccentricity, $e$, argument of periastron, $\omega$, and velocity semiamplitudes, $K_{\rm A}$ and $K_{\rm B}$. The orbital periods were held fixed as they are well determined from previous analyses. The orbital solutions are given in Table~\ref{tab:specorb} in which the mass ratio ($q = K_{\rm A}/K_{\rm B}$) is also given.

We also disentangled individual short segments of spectra in order to concentrate on spectral lines of interest, avoid interstellar lines, and achieve reasonable computation times. The Balmer lines were not used in the determination of the spectroscopic orbits because they are much wider than the changes in RV of the stars over an orbital cycle. The best fits were obtained using the downhill simplex algorithm \citep{Press_1992}. We found 100 runs with 1000 iterations each to be sufficient to ensure the global minimum was found whilst keeping the required computation time manageable. Convergence was achieved quickly because of the high quality of the HARPS spectra and the availability of preliminary orbital parameters from the literature. Uncertainties in the results were obtained using 10\,000 bootstrapping simulations \citep{Pavlovski_2018}. Fig.~\ref{fig:RVs} is a visualisation of the spectroscopic orbits of the four targets and the phase distribution of our spectra.

\subsection{V1034\,Sco}

V1034\,Sco has been found to have a small eccentricity \citep{Hill_1974, Levato_Morrell_1983}. We have been able to measure precise velocity amplitudes for the components (Table~\ref{tab:specorb}) which highlight the low mass ratio of the system.

Our results are in good agreement with those from \citet{Sana_2003,Sana_2005} and agree within the errorbars with those from \citet{Rosu_2022b}. We conclude that the RV semi-amplitudes of the components of V1034\,Sco are now well-determined since the the accuracy achieved is about 0.2\% for the primary and 0.4\% for the secondary. The argument of periastron is quite uncertain due to the small eccentricity, and is much better determined from the photometric analysis in Section~\ref{sec:lc}.

\subsection{GL\,Car}

To the best of our knowledge, our spectroscopic orbit for GL\,Car is the first one published. The mass ratio is in fairly good agreement with the photometric value of $q = 0.943\pm0.009$ found by \citet{Gimenez_Clausen_1986}. We fixed the eccentricity to a value of 0.146, which is precisely known from analyses of its apsidal motion \citep{Wolf_2008}. We fitted for the argument of periastron, which is well-determined when the eccentricity is fixed.

\subsection{V537~Car}

This is the only binary system with a circular orbit in our sample. The HARPS spectra densely cover both quadratures. We also obtained spectra during the primary and secondary minimum, but did not use these in our analysis because the eclipses are not total. 

A spectroscopic orbit for V573\,Car has previously been published by \citet{Freyhammer_2001} but based on only two newly obtained spectra and eight spectra taken from \citet{Levato_Malaroda_1982}. The velocity amplitudes measured by these authors are quite uncertain but agree with ours to within the errorbars.

\subsection{V346\,Cen}

V346\,Cen has a significant eccentricity of $e = 0.289\pm0.006$. Our HARPS spectra have good phase coverage and \spd\ quickly converged to a stable solution (Table~\ref{tab:specorb}). Our results are in reasonable agreement with the only previous spectroscopic analysis of this system \citep{Mayer_2016}, as expected because they used the same spectra.

However, the velocity amplitudes we measured are both 1.5\kms\ lower than those of \citet{Mayer_2016}. We attribute this to differences in the methods employed in the two analyses. In particular, \citet{Mayer_2016} employed cross-correlation \citep{Zucker_Mazeh_1994} to determine RVs, using as templates the disentangled spectra themselves. This approach is mathematically incorrect and extensive numerical experiments have shown that it is not reliable \citep{Ilijic_2001}; \spd\ has instead been shown to be the best approach to determining spectroscopic orbits \citep{Southworth_Clausen_2007}. Our approach yields masses that are smaller by 0.26\Msun\ and 0.11\Msun\ than those found by \citet{Mayer_2016} for the primary and secondary component of the system, respectively, which is larger than the quoted uncertainties.

\begin{figure*}
{ \centering
\includegraphics[width=40mm]{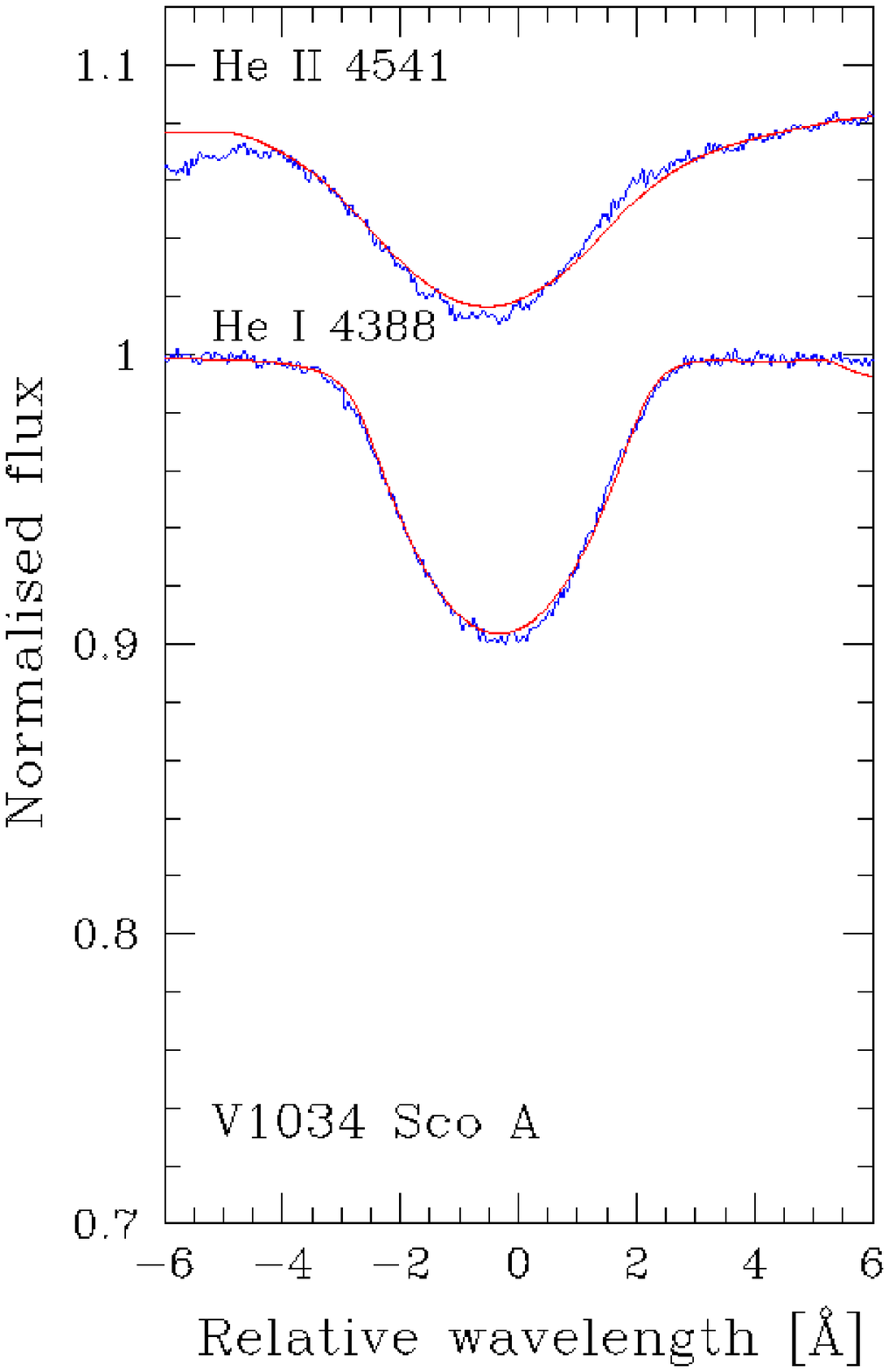} ~~~~
\includegraphics[width=40mm]{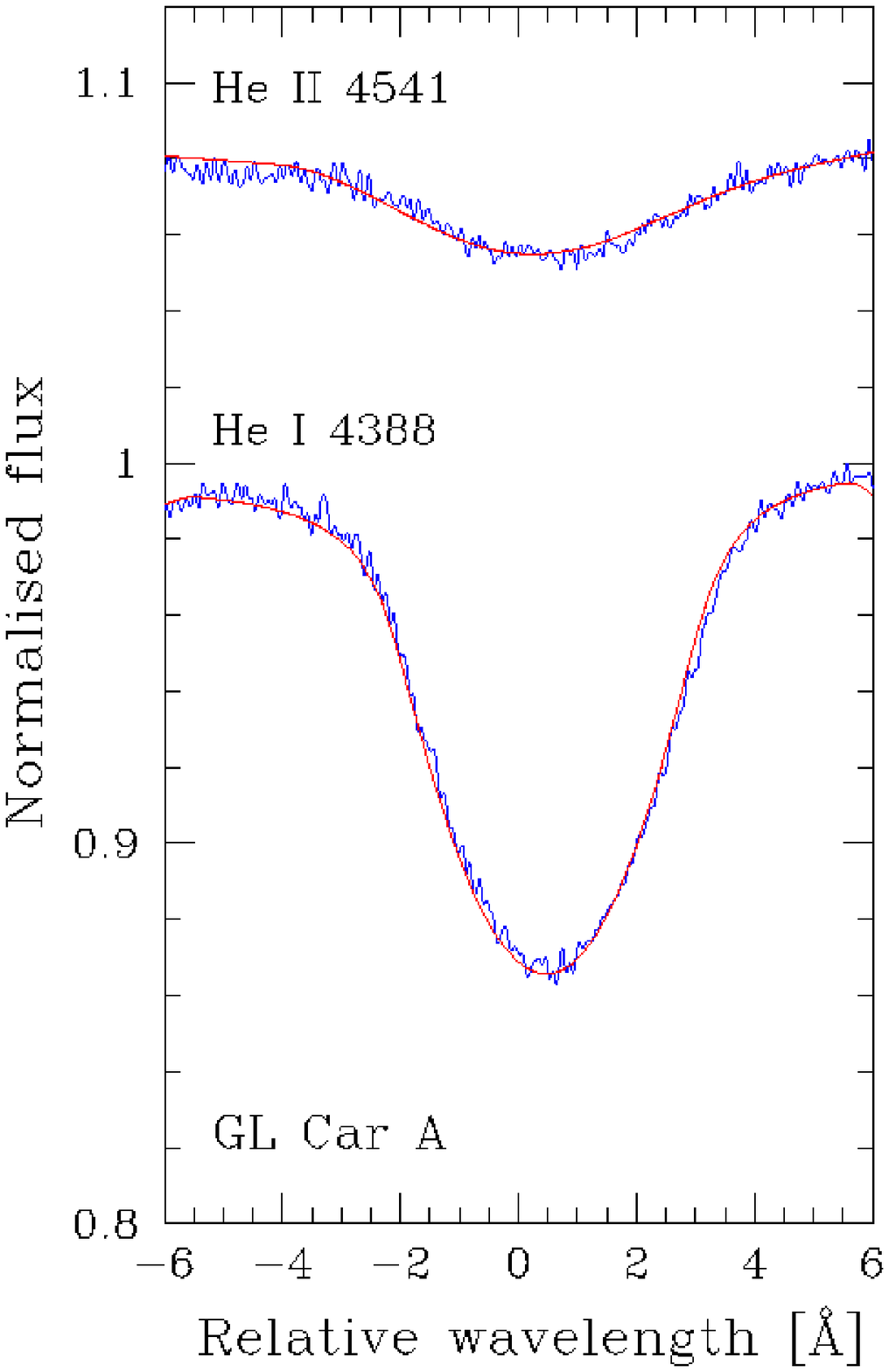} ~~~~
\includegraphics[width=40mm]{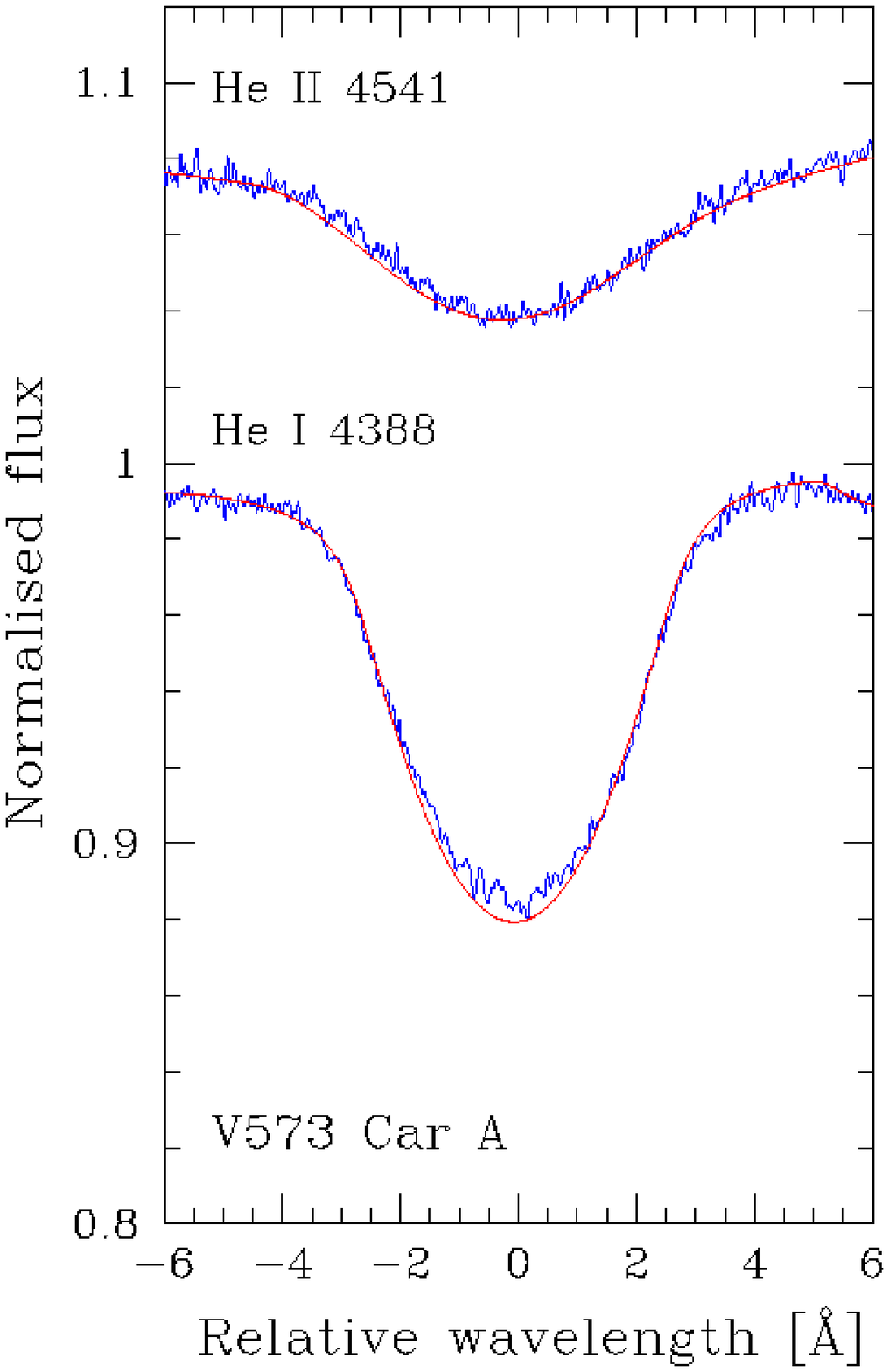} ~~~~
\includegraphics[width=40mm]{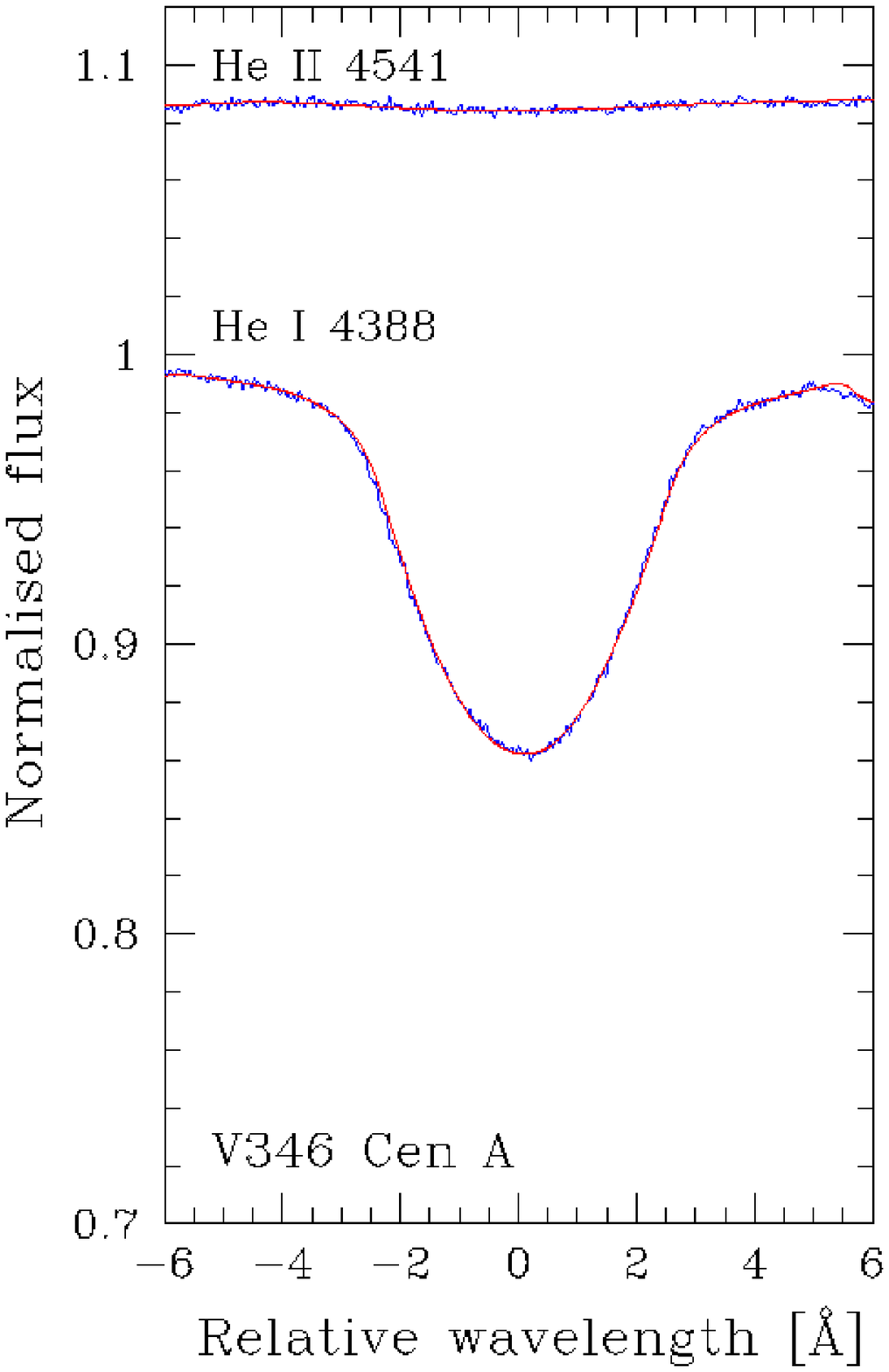}\\ \vspace{0.5cm}
\includegraphics[width=40mm]{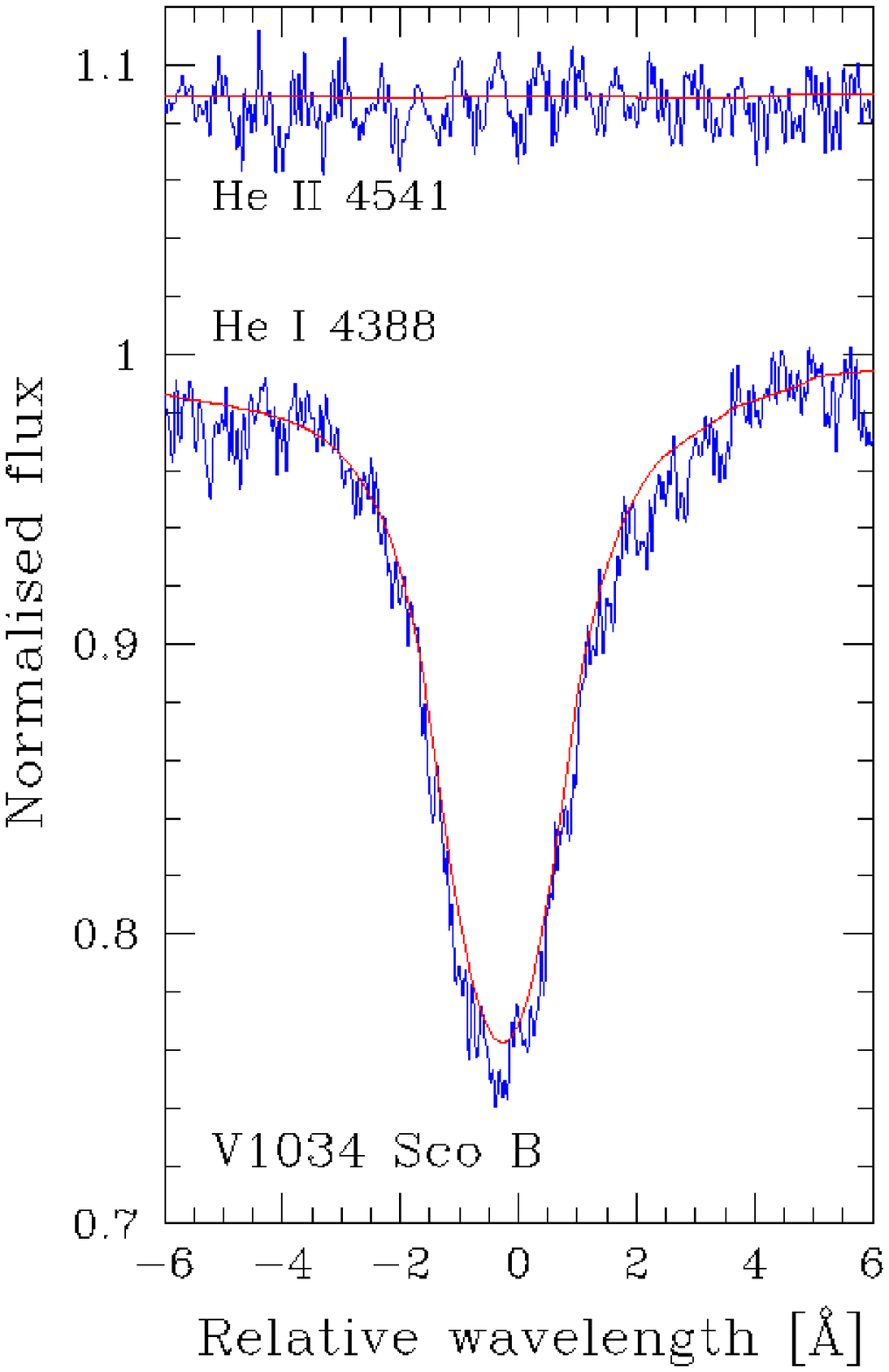} ~~~~
\includegraphics[width=40mm]{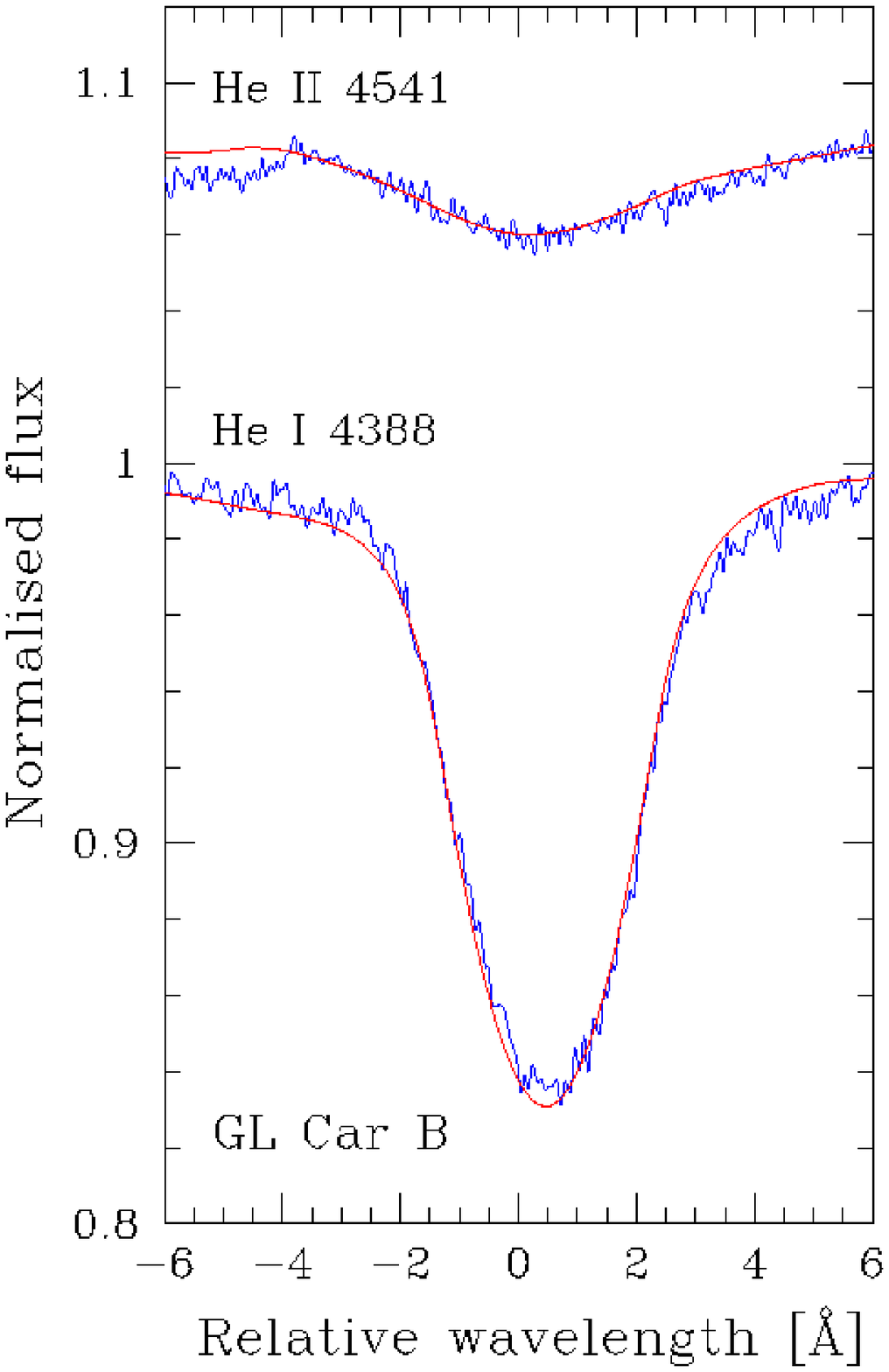} ~~~~
\includegraphics[width=40mm]{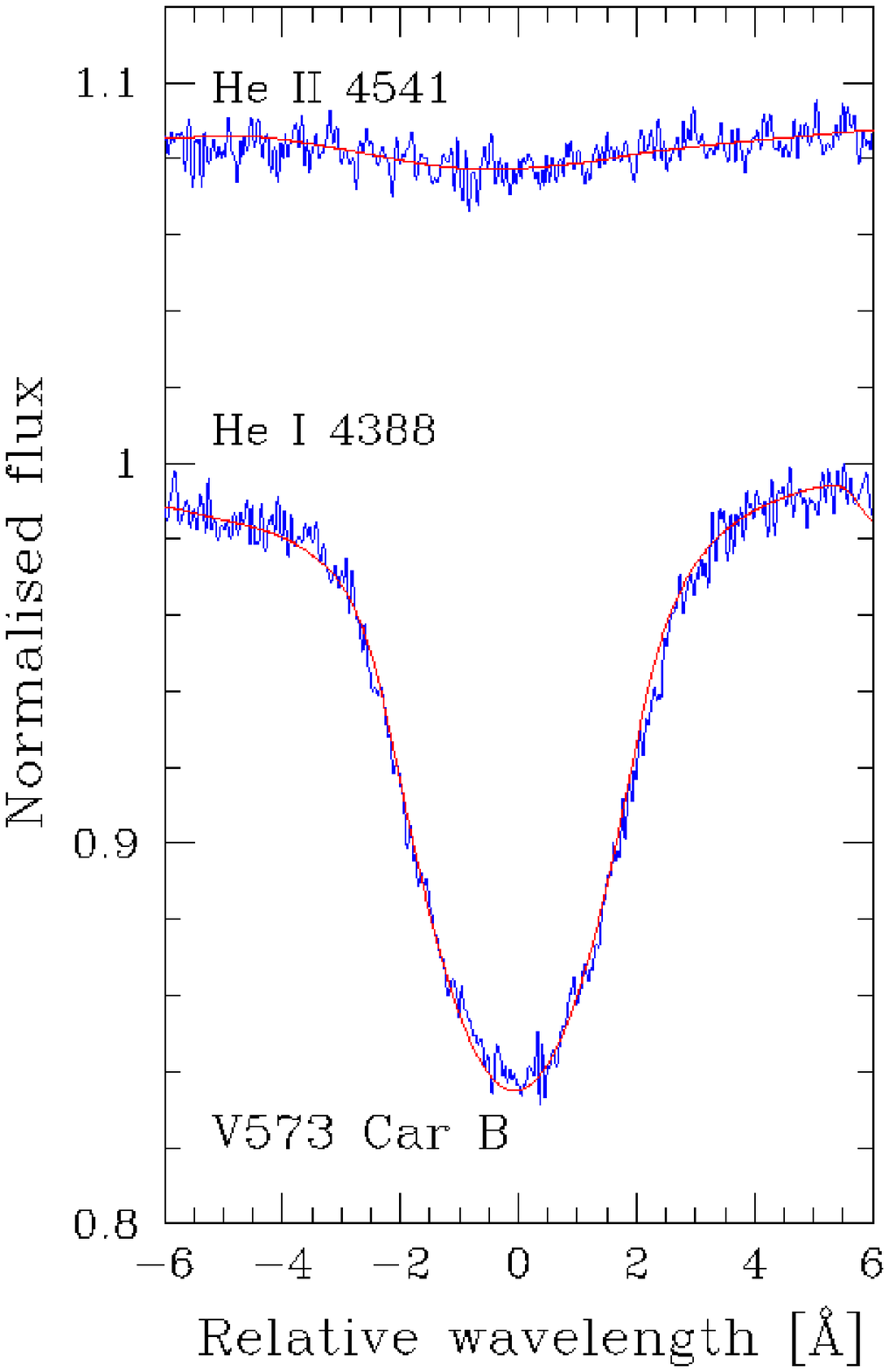} ~~~~
\includegraphics[width=40mm]{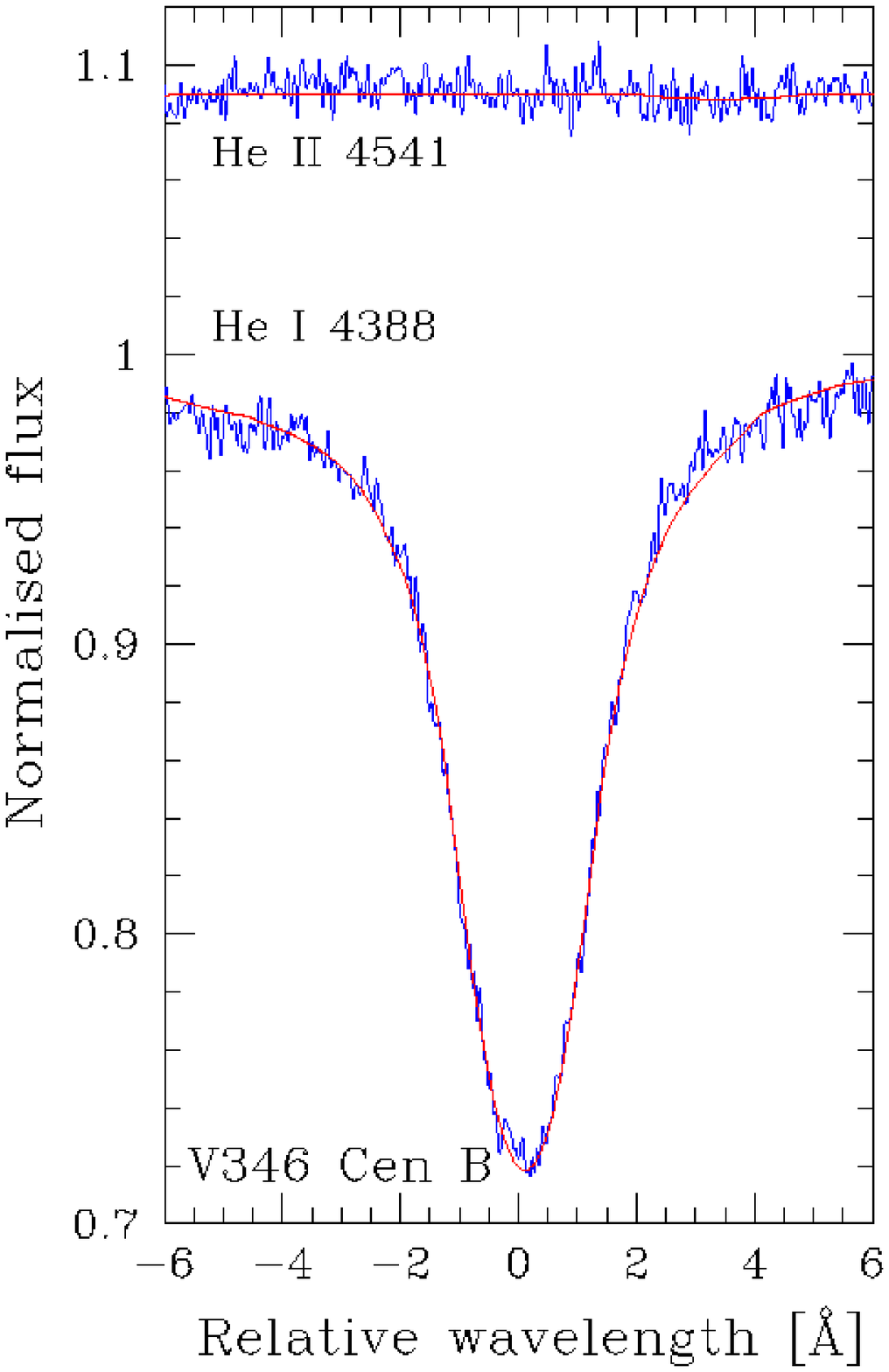} \\
}
\caption{\label{fig:helfit} Fits to the \ion{He}{i} 4388\,\AA\ and \ion{He}{ii} 4541\,\AA\ lines. The ionisation balance of \ion{He}{i} and \ion{He}{ii} was used in the determination of \Teff\ for the stars.
The blue data are the disentangled spectra and the red lines the best fits. The upper row is for the primary stars and the lower row is for the secondary stars. The lower S/N for the secondary components arises because they are fainter than the primary components. The absence of \ion{He}{ii} 4541\,\AA\ absorption in V1034\,Sco~B and V346\,Cen~B is obvious and indicates that $\Teff < 23\,000$\,K.}
\end{figure*}

\section{Atmospheric parameters} \label{sec:atmos}

For determination of the atmospheric parameters and individual abundances of C, N, O, Mg and Si, we employed a hybrid NLTE approach as described in detail in \citet{Nieva_Przybilla_2007,Nieva_Przybilla_2012}. A hybrid NLTE approach means that the modelling combines hydrostatic, plane-parallel, and line-blanketed model atmospheres in local thermodynamic equilibrium (LTE) with line formation calculated in NLTE. We used the {\sc Atlas9} code \citep{Kurucz_1979, Castelli_Kurucz_2003} for the calculations of model atmospheres. Then emergent fluxes and line profiles were calculated with the codes {\sc Detail} and {\sc Surface} \citep{Giddings_1980, Butler_Giddings_1985}. In {\sc Detail} the coupled radiative transfer and statistical equilibrium equations are solved, while {\sc Surface} was used for the calculations of NLTE synthetic spectra. The following model atoms were used in these calculations: \ion{H}{i} \citep{Przybilla_Butler_2004}, \ion{He}{i/ii} \citep{Przybilla_2005}, \ion{C}{ii/iii} \citep{Nieva_Przybilla_2006}, \ion{N}{ii} \citep{Przybilla_Butler_2001}, \ion{O}{i/ii} \citep{Becker_Butler_1988, Przybilla_2000}, \ion{Mg}{ii} \citep{Przybilla_2001}, and \ion{Si}{ii/iii/iv} \citep{Becker_Butler_1990}.

\begin{table} \centering
\caption{\label{tab:atmospar} The atmospheric parameters derived from optimal fitting of disentangled spectra of the components to a grid of NLTE spectra.}
\begin{tabular}{lccc} \hline\hline
Star         & \Teff         & \vsini\   & $\xi_{\rm t}$  \\
             & (K)           & (km\,s$^{-1}$) & (km\,s$^{-1}$) \\
\hline
V1034\,Sco A  & 32\,200 $\pm$ 500 & 169.8 $\pm$ 2.6 & 5 $\pm$ 1 \\
V1034\,Sco B  & 25\,800 $\pm$ 300 & ~94.5 $\pm$ 3.3 & 5 $\pm$ 1 \\
GL\,Car A     & 30\,950 $\pm$ 500 & 180.1 $\pm$ 2.2 & 4 $\pm$ 1 \\
GL\,Car B     & 30\,400 $\pm$ 500 & 134.6 $\pm$ 3.5 & 2 $\pm$ 1 \\
V573\,Car A   & 31\,900 $\pm$ 400 & 184.6 $\pm$ 2.7 & 5 $\pm$ 1 \\
V573\,Car B   & 28\,700 $\pm$ 350 & 155.4 $\pm$ 3.1 & 3 $\pm$ 1 \\
V346\,Cen A   & 26\,100 $\pm$ 300 & 165.2 $\pm$ 2.8 & 5 $\pm$ 1 \\
V346\,Cen B   & 22\,500 $\pm$ 300 & ~89.1 $\pm$ 2.3 & 5 $\pm$ 1 \\
\hline \\
\end{tabular}
\end{table}

\begin{table*} \centering
\caption{\label{tab:cno} Abundances determined for the stars in our sample of binary systems.}
\begin{tabular}{lccccccc} \hline\hline
Star  &   $\log \epsilon({\rm C})$  & $\log \epsilon({\rm N})$  & $\log \epsilon({\rm O})$ & [N/C] & [N/O] & $\log \epsilon({\rm Mg})$ & $\log \epsilon({\rm Si})$      \\
\hline
V1034\,Sco A  & 8.39 $\pm$ 0.12 & 7.71 $\pm$ 0.12 & 8.76 $\pm$ 0.07 & $-0.68 \pm 0.17$ & $-1.05 \pm 0.14$ & 7.67 $\pm$ 0.14 & 7.56 $\pm$ 0.01 \\
V1034\,Sco B  & 8.27 $\pm$ 0.05 & 7.67 $\pm$ 0.08 & 8.69 $\pm$ 0.12 & $-0.57 \pm 0.09$ & $-1.02 \pm 0.14$ & 7.45 $\pm$ 0.07 & 7.46 $\pm$ 0.14 \\
GL\,Car A     & 8.18 $\pm$ 0.08 & 7.69 $\pm$ 0.14 & 8.74 $\pm$ 0.12 & $-0.67 \pm 0.16$ & $-1.04 \pm 0.18$ & 7.52 $\pm$ 0.12 & 7.50 $\pm$ 0.12 \\
GL\,Car B     & 8.21 $\pm$ 0.12 & 7.72 $\pm$ 0.11 & 8.76 $\pm$ 0.12 & $-0.52 \pm 0.16$ & $-0.85 \pm 0.16$ & 7.50 $\pm$ 0.13 & 7.44 $\pm$ 0.14 \\
V573\,Car A  & 8.30 $\pm$ 0.08 & 7.63 $\pm$ 0.10 & 8.67 $\pm$ 0.05 & $-0.57 \pm 0.12$ & $-0.52 \pm 0.11$ & 7.58 $\pm$ 0.08 & 7.57 $\pm$ 0.12 \\
V573\,Car B  & 8.28 $\pm$ 0.05 & 7.76 $\pm$ 0.07 & 8.61 $\pm$ 0.04 & $-0.55 \pm 0.09$ & $-0.94 \pm 0.08$ & 7.45 $\pm$ 0.05 & 7.54 $\pm$ 0.13 \\
V346\,Cen A   & 8.13 $\pm$ 0.05 & 7.68 $\pm$ 0.05 & 8.70 $\pm$ 0.04 & $-0.45 \pm 0.07$ & $-1.02 \pm 0.06$ & 7.70 $\pm$ 0.13 & 7.45 $\pm$ 0.16 \\
V346\,Cen B   & 8.33 $\pm$ 0.06 & 7.72 $\pm$ 0.09 & 8.80 $\pm$ 0.07 & $-0.61 \pm 0.11$ & $-1.08 \pm 0.11$ & 7.40 $\pm$ 0.14 & 7.35 $\pm$ 0.17 \\
\hline
This work    & 8.27 $\pm$ 0.08 & 7.69 $\pm$ 0.05 & 8.70 $\pm$ 0.06 & $-0.58 \pm 0.07$ & $-1.01 \pm 0.07$& 7.59 $\pm$ 0.14 & 7.49 $\pm$ 0.08 \\
OB binaries\tablefootmark{a}  & 8.25 $\pm$ 0.07 & 7.69 $\pm$ 0.06 & 8.71 $\pm$ 0.05 & $-0.56 \pm 0.08$ & $-1.02 \pm 0.07$& 7.56 $\pm$ 0.12 & 7.45 $\pm$ 0.09 \\
B stars\tablefootmark{b} & 8.33 $\pm$ 0.04 & 7.79 $\pm$ 0.04 & 8.76 $\pm$ 0.05 & $-0.54 \pm 0.06$ & $-0.97 \pm 0.06$ & 7.56 $\pm$ 0.05 & 7.50 $\pm$ 0.05 \\
\hline
\end{tabular}
\tablefoot{
 The \Teff\ and \logg\ values used for the construction of the model atmospheres are given in Tables \ref{tab:atmospar} and \ref{tab:absdim}, respectively.
\\
\tablefoottext{a}{The abundances found for OB binaries in our previous work \citep{Pavlovski_2018}}
\tablefoottext{b}{The `present-day cosmic abundances' for B stars \citep{Nieva_Przybilla_2012}}
}
\end{table*}

We used the disentangled spectra generated in the previous section to determine the \Teff, \vsini, and microturbulent velocity ($\xi_{\rm t}$) for each of the eight stars in our sample. This process was greatly helped by the availability of \logg\ values from the measured masses and radii (see Section~\ref{sec:lc}) so our analysis was performed iteratively. The disentangled spectra were still in the common continuum of the binary system so needed to be renormalised to the continuum of the individual component stars. This was done iteratively alongside the light curve analysis, to arrive at light ratios that were consistent between the two types of the analysis \citep{Ilijic_2004,Pavlovski_Hensberge_2005}.

The exception to the process above was GL\,Car, for which the light curve solutions suffered from a degeneracy which caused the light ratio to be highly uncertain. We therefore fitted the disentangled spectra to obtain the \Teff, \vsini\, and \logg\ values and the light ratio directly, using the approach of \citet{Tamajo_2011} and \citet{Kolbas_2015}. After iteration with the light curve solution, \logg\ was fixed for the final measurements of the remaining parameters. We have found  that such spectroscopically determined light ratios can be competitive with those from light curve analysis \citep{Pavlovski_2009, Pavlovski_2018, Pavlovski_2022}.

Since we are dealing with late-O, and early-B type stars, the helium ionisation balance (\ion{He}{i}/\ion{He}{ii}) is a sensitive indicator of \Teff. Our spectra cover a broad spectral range and thus allowed us to use a large number of lines:
4009, 4026, 4388, 4437, 4471, 4713, 4921, 5015, 5047, 5875 and 6678\,\AA\ for \ion{He}{i} and 4200, 4541, 4686, and 5411\,\AA\ for \ion{He}{ii}. Once a first set of parameters was obtained, we made the light ratio a free parameter to check its reliability. As a further check we also fitted the H$\delta$, H$\gamma$ and H$\beta$ lines, during which we excluded wavelengths affected by interstellar absorption (specifically the red wing of H$\beta$). We did not base our \Teff\ measurements on the Balmer lines because their large width makes them susceptible to errors due to continuum normalisation.

The He line strengths also depend on $\xi_{\rm t}$, which for hot stars can be obtained by minimising the scatter in the O abundances. We started with the assumptions of solar He abundance and $\xi_{\rm t} = 2$\kms, and subsequently relaxed each of them before refitting. Convergence was fast, taking either one or two iterations for all eight stars. Once this was achieved, we repeated our optimal fitting of the disentangled spectra described above. The results of this process are given in Table~\ref{tab:atmospar}. Below we compare our results to published determinations for each system, except for GL\,Car for which there is no other analysis based on modern spectroscopy.

\subsection{V1034\,Sco}

In the most recent study, \citet{Rosu_2022b} analysed disentangled spectra of the components obtained from the HARPS spectra obtained in our observing run, and available at the ESO archive. The \Teff s they derived are within 1$\sigma$ uncertainty of our results. This is encouraging, especially as \citet{Rosu_2022b} used a different NLTE spectrum synthesis code to us.

\citet{Rosu_2022b} also fitted for surface gravity, using the wings of the Balmer and some \ion{He}{i} lines, whereas we prefer the surface gravities determined with a high precision from the masses and radii. The two analyses agree to within 2$\sigma$, but the uncertainties of the values from \citet{Rosu_2022b} are much larger ($\pm$0.10 dex) than our own ($\pm$0.01 dex).

\subsection{V573\,Car}

\citet{Freyhammer_2001} studied V573\,Car using two high-resolution spectra from the FEROS spectrograph, taken near opposite quadratures. They fitted the spectra with NLTE synthetic spectra for the \ion{He}{i}, \ion{He}{ii}, H$\delta$ and H$\gamma$ lines. The helium lines of the components are not completely resolved at quadrature due to the high \vsini, and the Balmer lines are not resolved.

The agreement between their and our \Teff\ measurements is well within the 1$\sigma$ errorbars for component A, but only within 2$\sigma$ for component B. We attribute this to the very small number of spectra available to \citet{Freyhammer_2001} compared to our own extensive dataset. We also find that fitting disentangled spectra is superior to fitting individual observed spectra because it avoids problems with blending of lines from the two stars. Moreover, disentangled spectra have a higher S/N than individual observed spectra.

\subsection{V346\,Cen}

The atmospheric parameters for both components of V346\,Cen were determined by \citet{Mayer_2016} from the same set of the HARPS spectra as we used. In their analysis, \citet{Mayer_2016} used optimal fitting of disentangled spectra in similar manner as we did, for fixed surface gravities and microturbulent velocities (fixed to $\xi_{\rm t} = 2$\kms) and using a grid of synthetic spectra from \citet{Lanz_Hubeny_2007}. They found a large discrepancy for the secondary: their spectroscopic analysis gave $20\,991\pm190$~K and their light curve analysis gave $25\,376\pm18$~K. The former value is in much better agreement with our result (Table~\ref{tab:atmospar}), and the complete absence of the \ion{He}{ii} lines demands $T_{\rm eff} < 23\,000$~K (based on a detailed examination of theoretical spectra for the \ion{He}{ii} 4686\,\AA\ line). \citet{Mayer_2016} did not discuss the \ion{He}{ii} lines in the secondary's spectrum at all. They also gave unrealistically small uncertainties for \Teff: $\pm$25~K for the primary, $\pm$190~K for the secondary from spectroscopy, and $\pm$18~K for the secondary from the light curve analysis.  Such low uncertainties are typical for formal errors of the fitting algorithms, but are unrealistic.

\section{Abundance analysis} \label{sec:abund}

\begin{figure}
%\centering
%\begin{tabular}{cc}
\includegraphics[width=43mm]{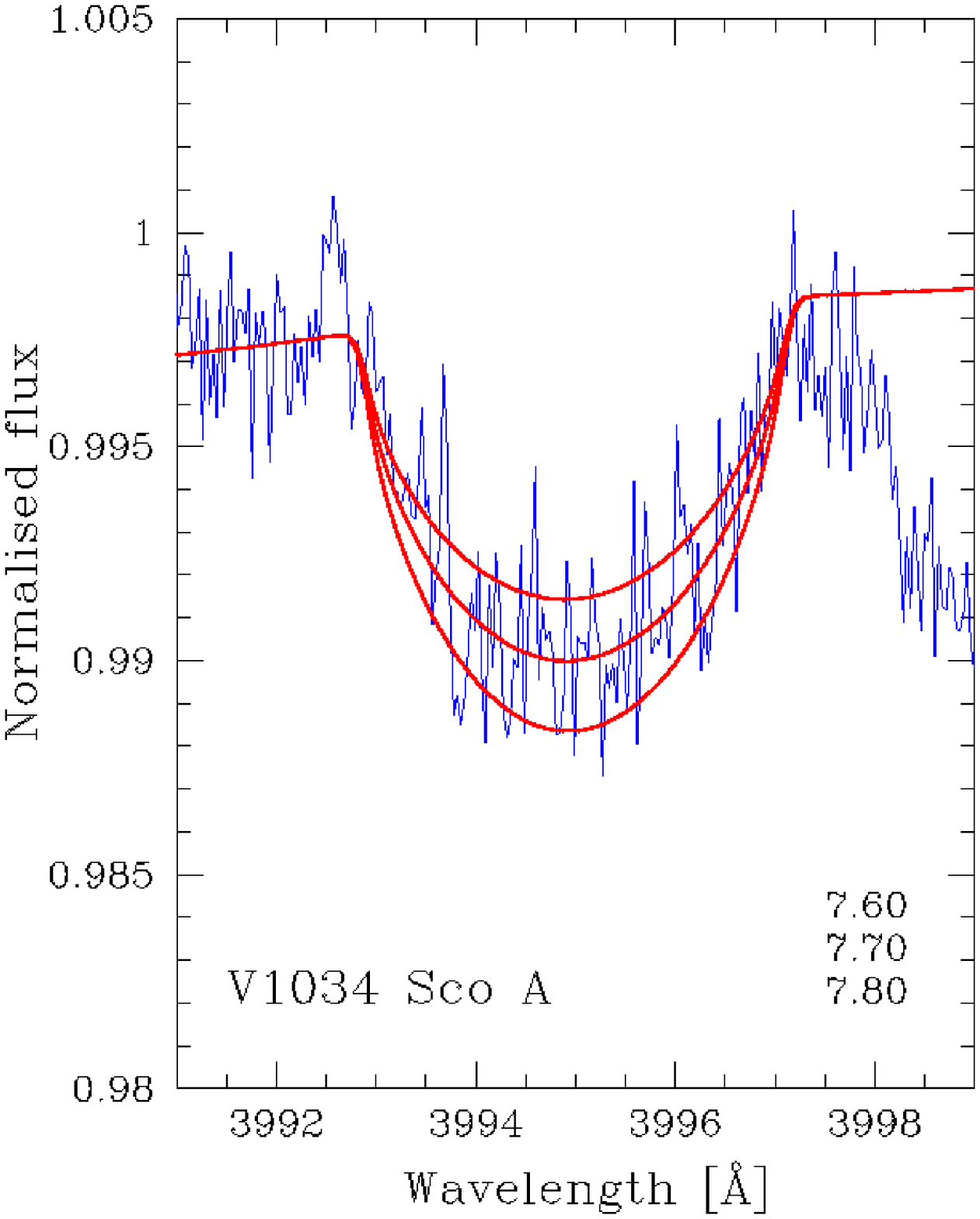}
\includegraphics[width=43mm]{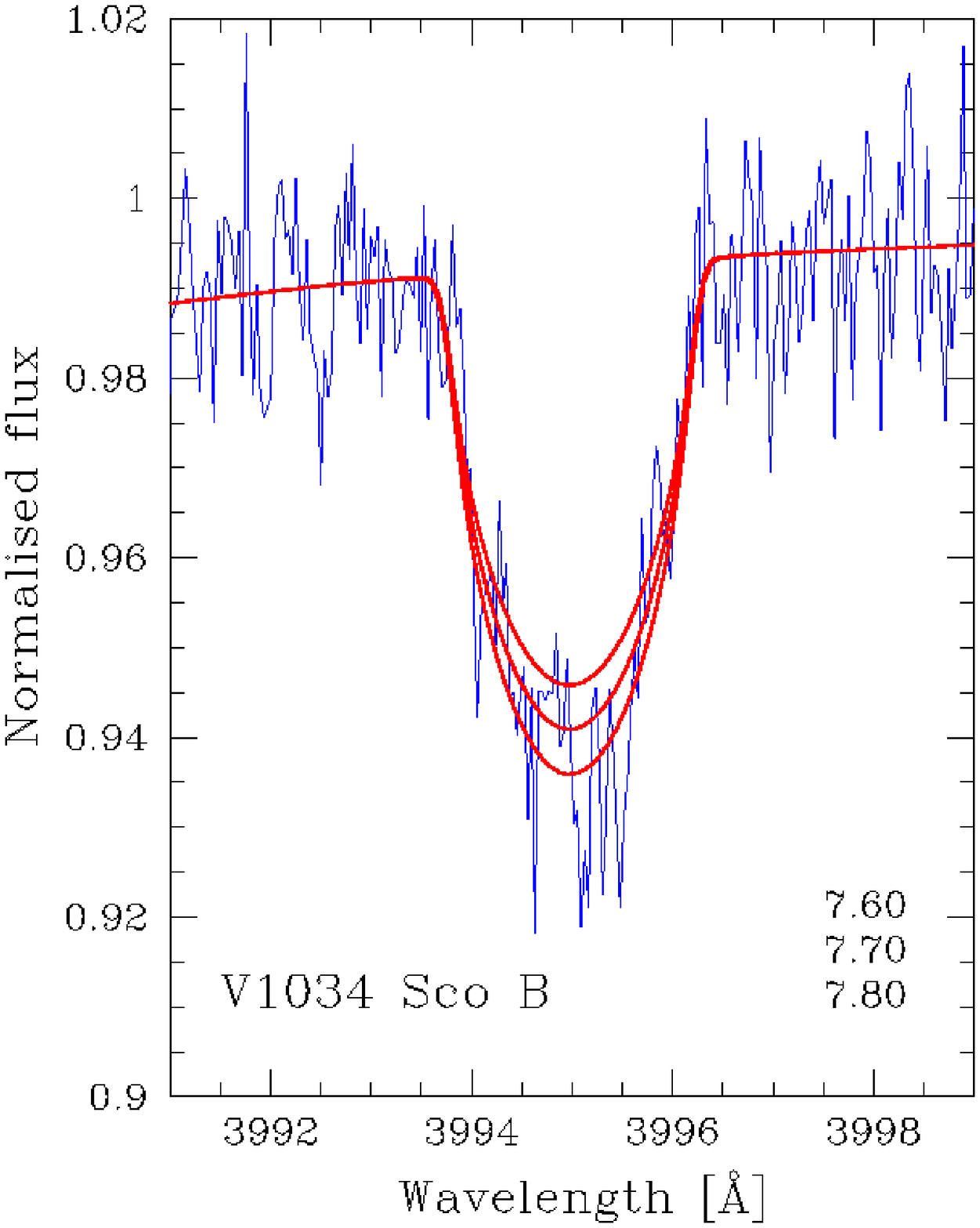} \\
\caption{\label{fig:nitro1034} Example of fits to N lines for our target stars. In this case the
\ion{N}{ii} 3995\,\AA\ line is shown for V1034\,Sco (primary star on the left, secondary star on the right).
The blue lines show the renormalised disentangled spectra of the stars. The red lines show synthetic spectra
from our precalculated grid for three different abundances (labelled on the bottom right corner in each panel).}
%\end{tabular}
\end{figure}

\begin{figure}[t]
%\centering
%\begin{tabular}{cc}
\includegraphics[width=88mm]{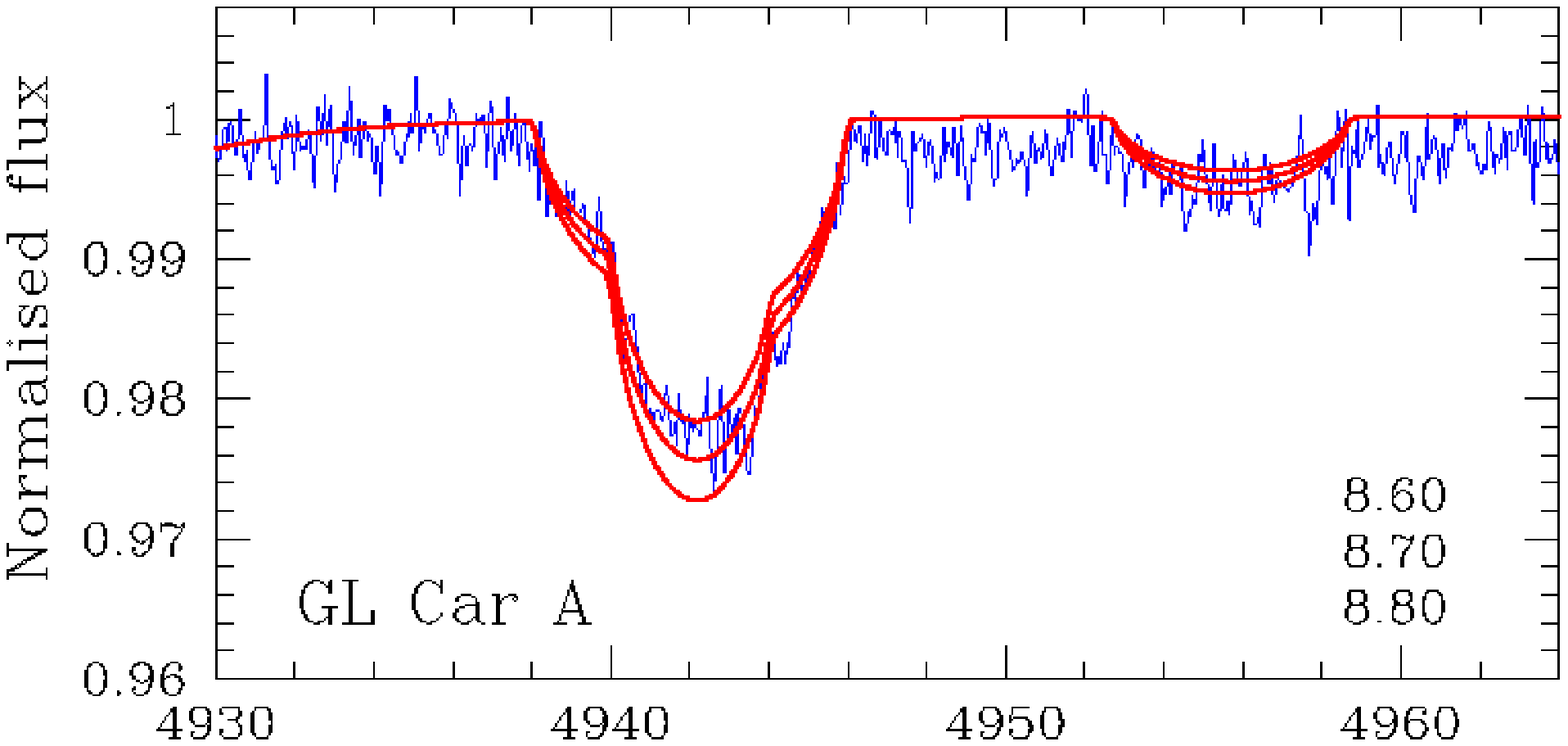} \\ \\
\includegraphics[width=88mm]{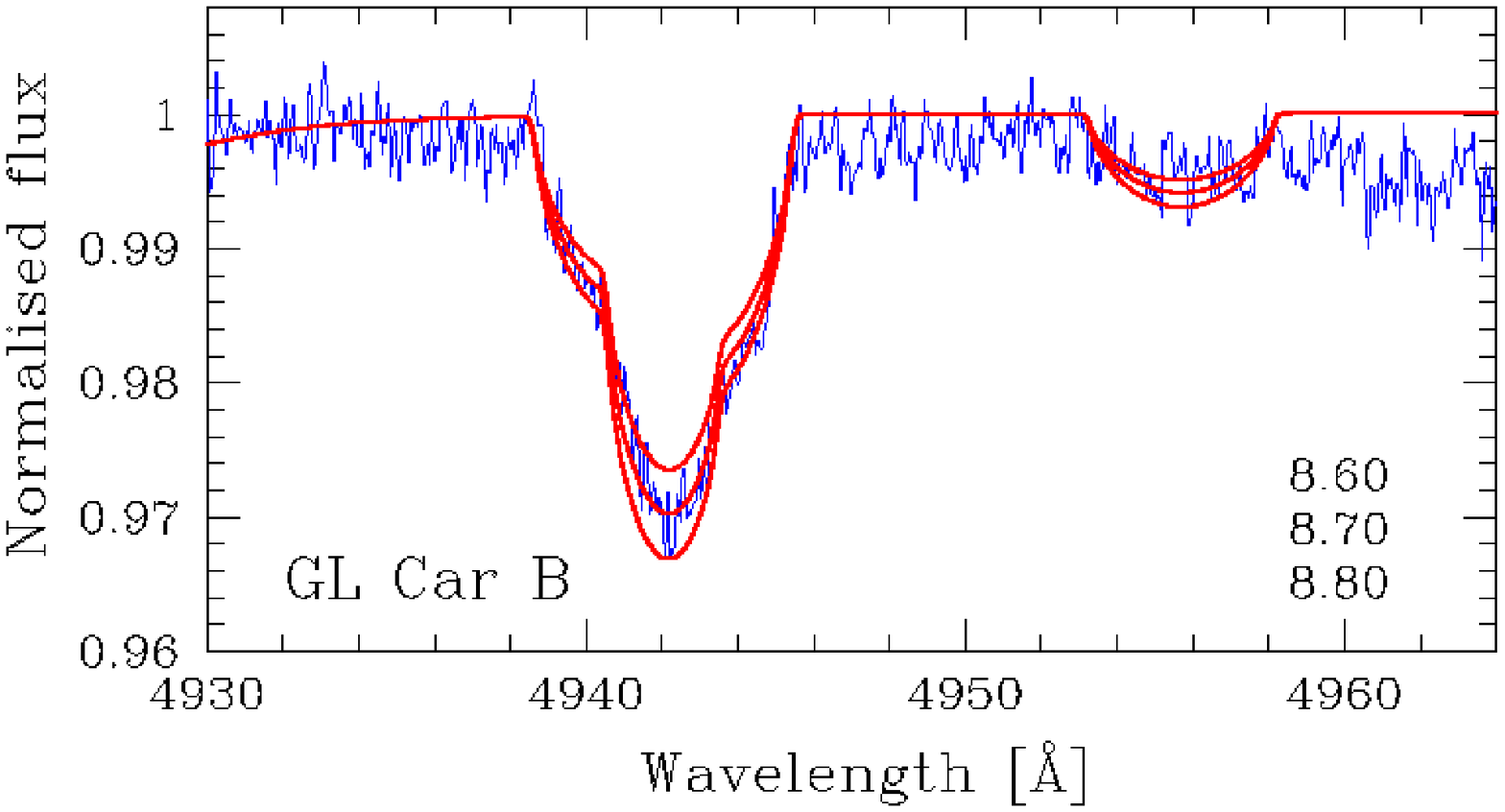} \\
\caption{\label{fig:oxyglcar} Same as Fig.~\ref{fig:nitro1034} but for O lines in the components
of GL\,Car.  The complex blend of \ion{O}{ii} lines at 4941 and 4943 and the \ion{O}{ii} line at 4955\,\AA\ are shown.     }
%\end{tabular}
\end{figure}

\begin{figure}[t]
%\centering
%\begin{tabular}{cc}
\includegraphics[width=88mm]{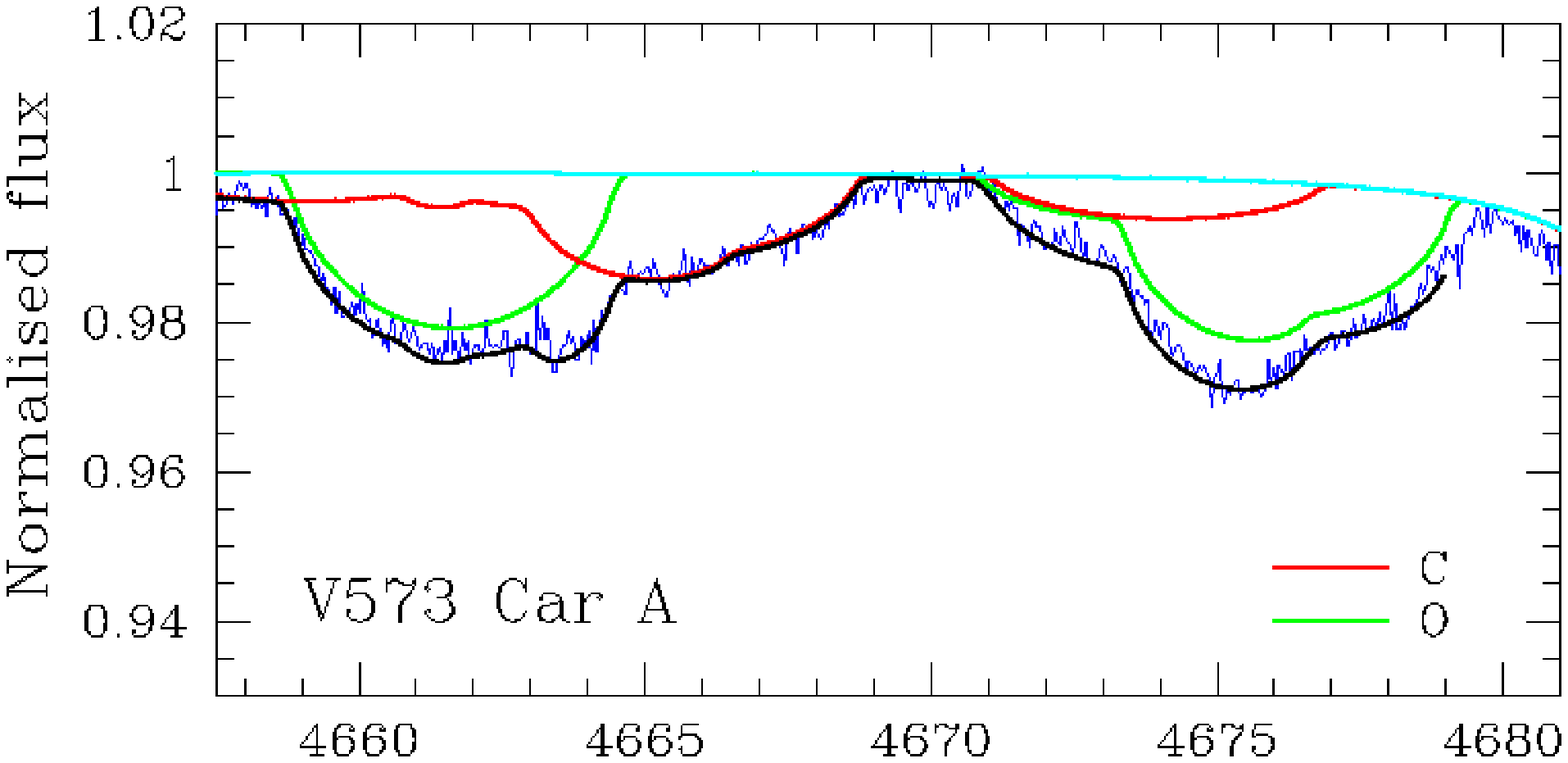} \\ \\
\includegraphics[width=88mm]{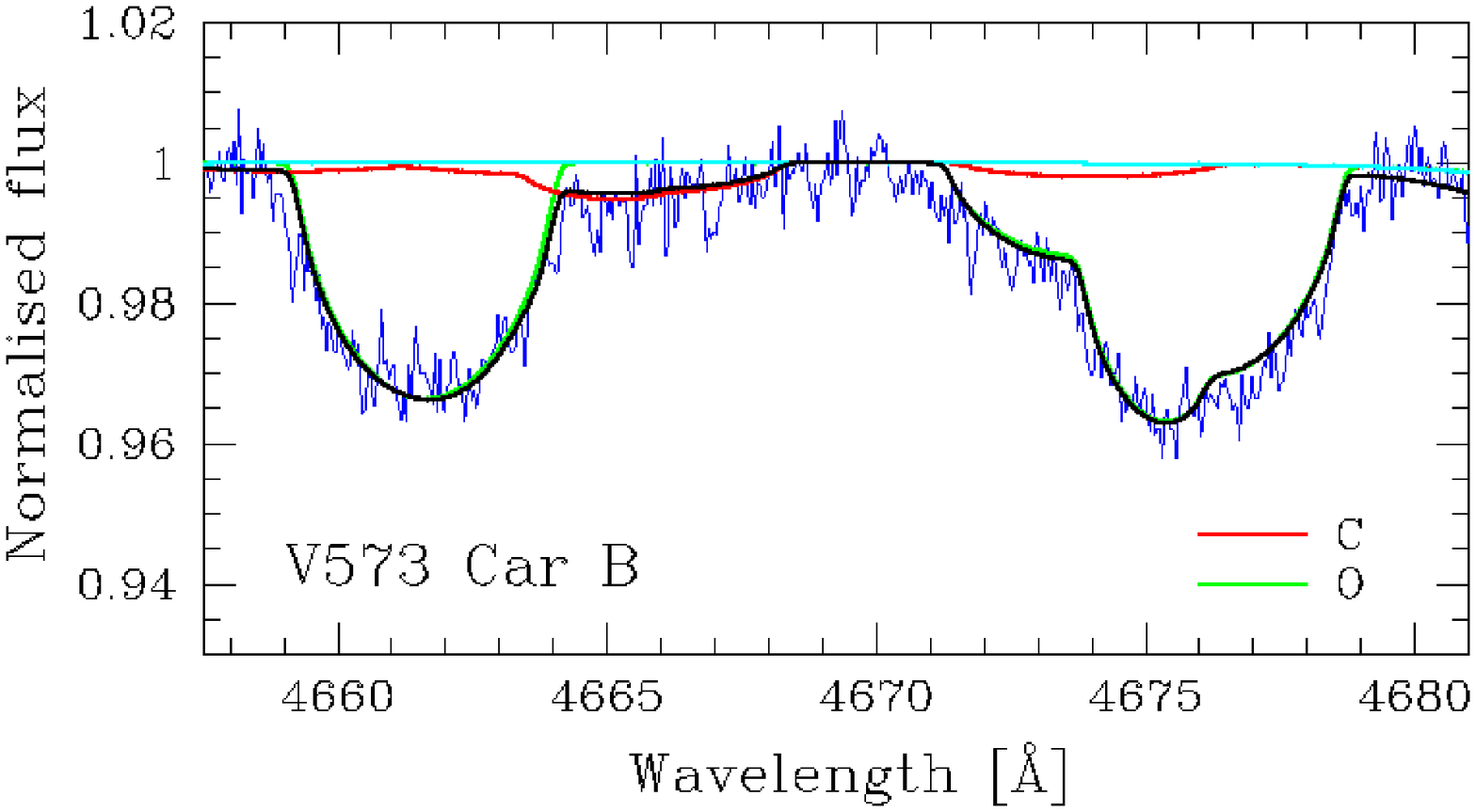} \\ \\
\includegraphics[width=43mm]{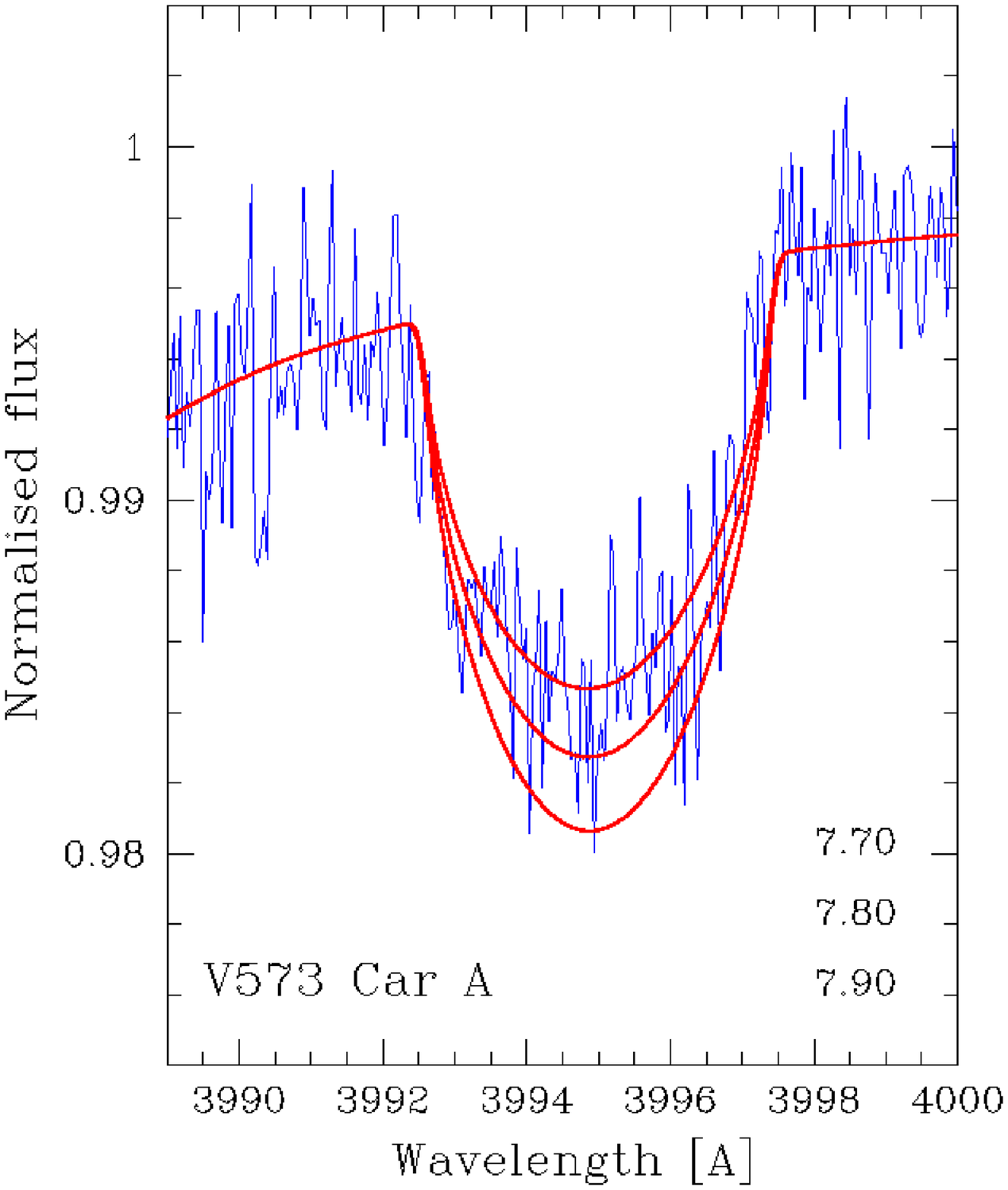}
\includegraphics[width=43mm]{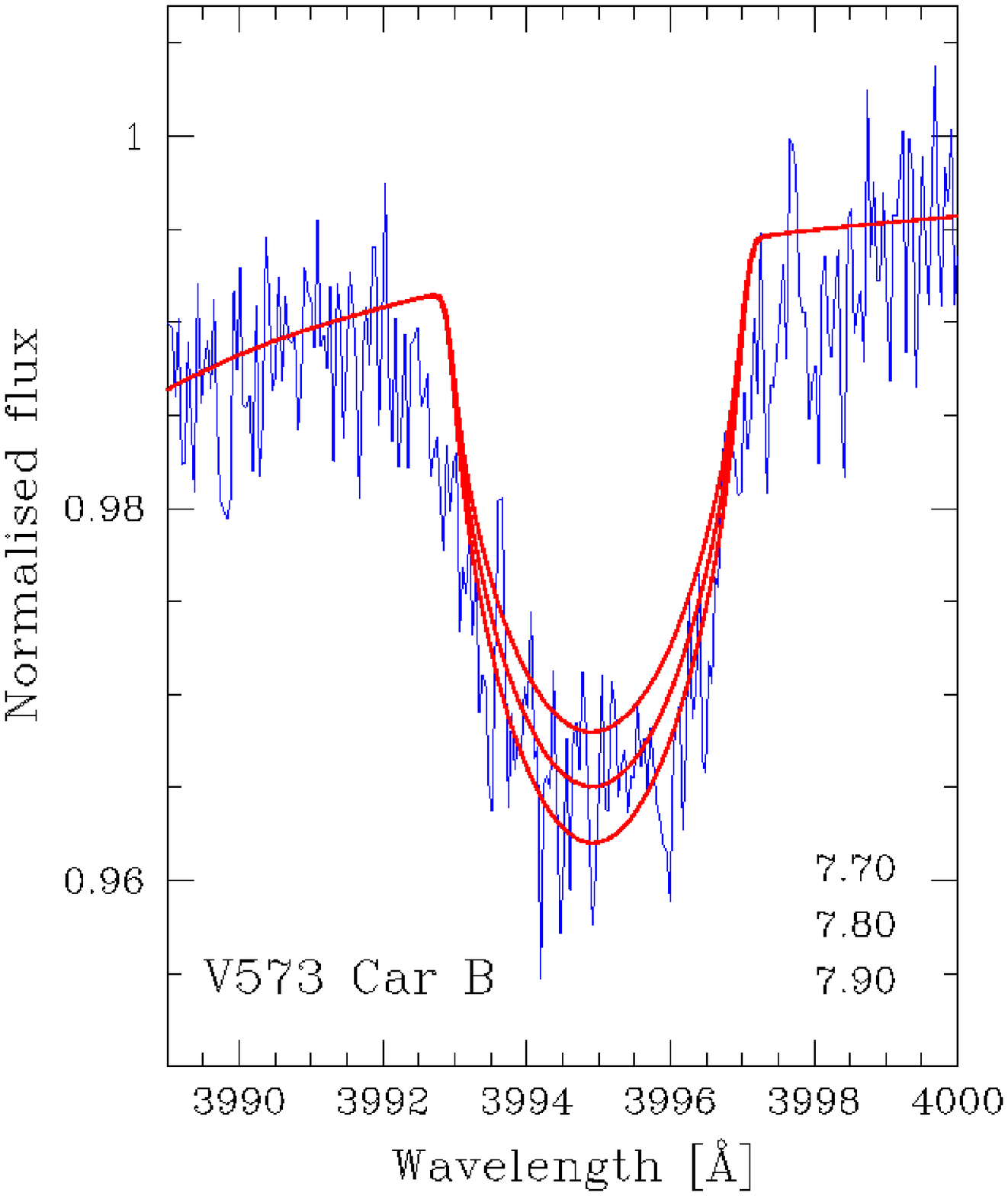} \\
\caption{\label{fig:nitro573} Same as Fig.~\ref{fig:nitro1034} but for lines in the components of V573\,Car.
The upper two panels show \ion{O}{ii} lines at 4661, 4673 and 4676\,\AA\ (shown with green lines), which are blended with \ion{C}{ii} lines at
4659, 4663, 4665 and 4673\,\AA\ (shown with red lines). The calculated synthetic spectrum is shown using a black line. The lower two panels show the \ion{N}{ii} 3995\,\AA\ lines.}
%\end{tabular}
\end{figure}

\begin{figure}[t]
%\centering
%\begin{tabular}{cc}
 \includegraphics[width=88mm]{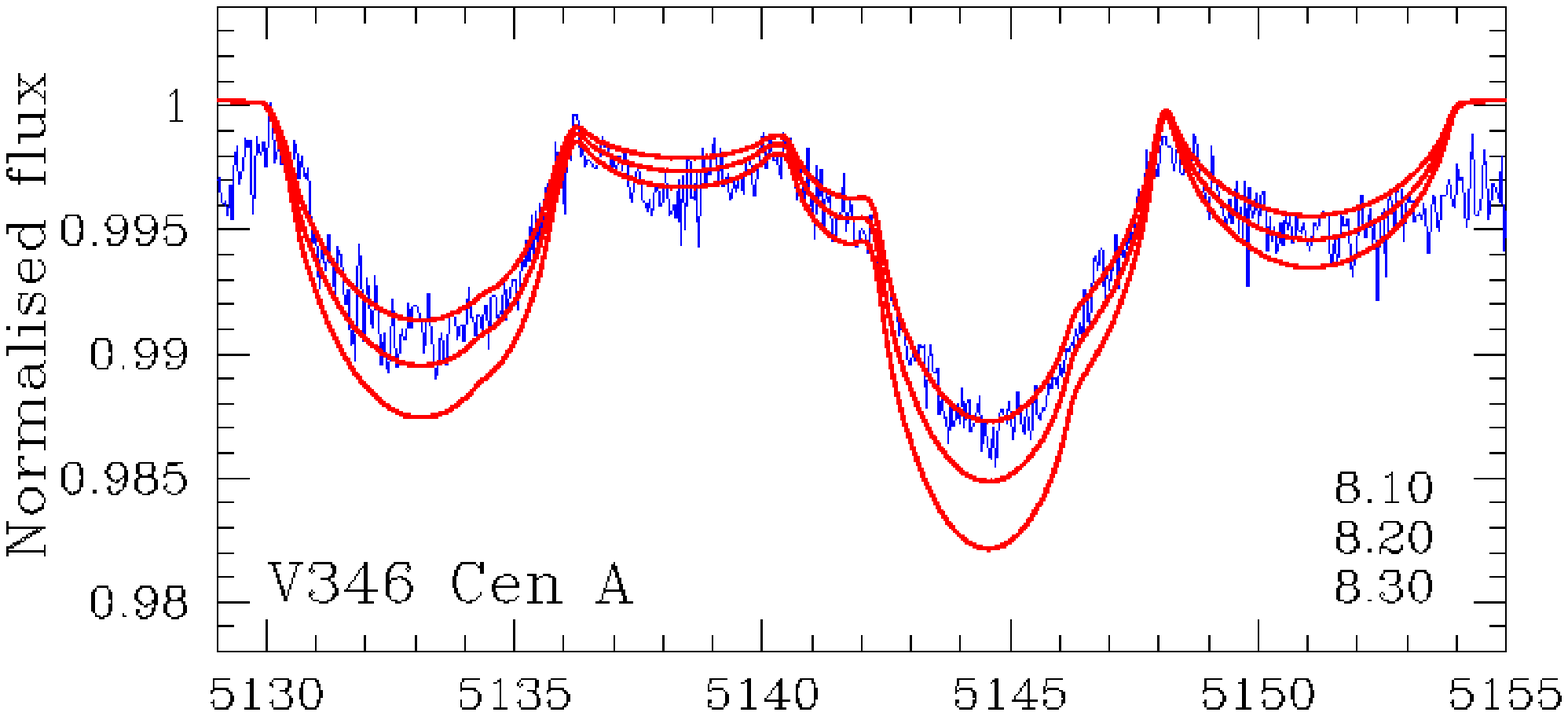} \\ \\
 \includegraphics[width=88mm]{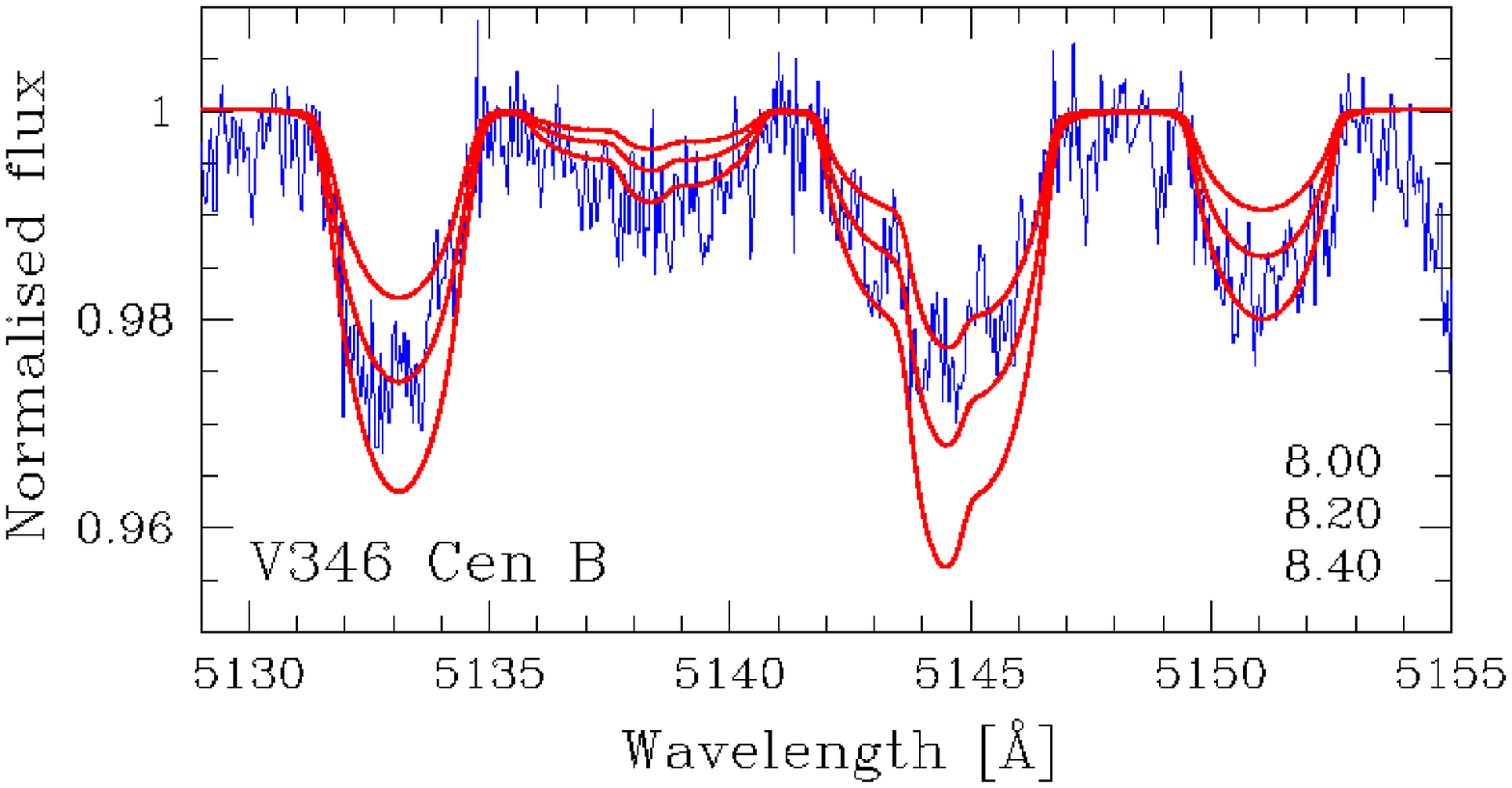} \\  \\
\includegraphics[width=43mm]{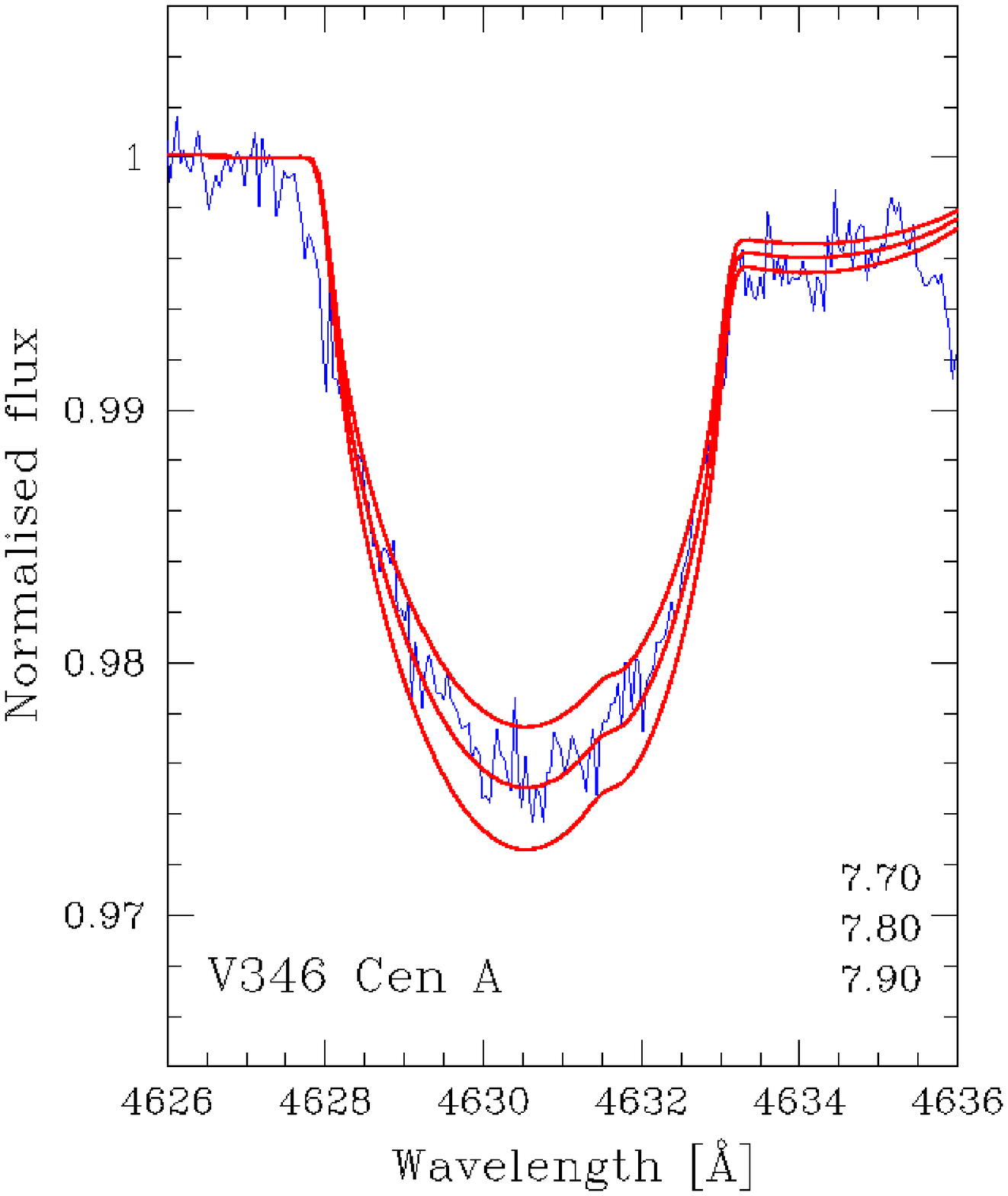}
\includegraphics[width=43mm]{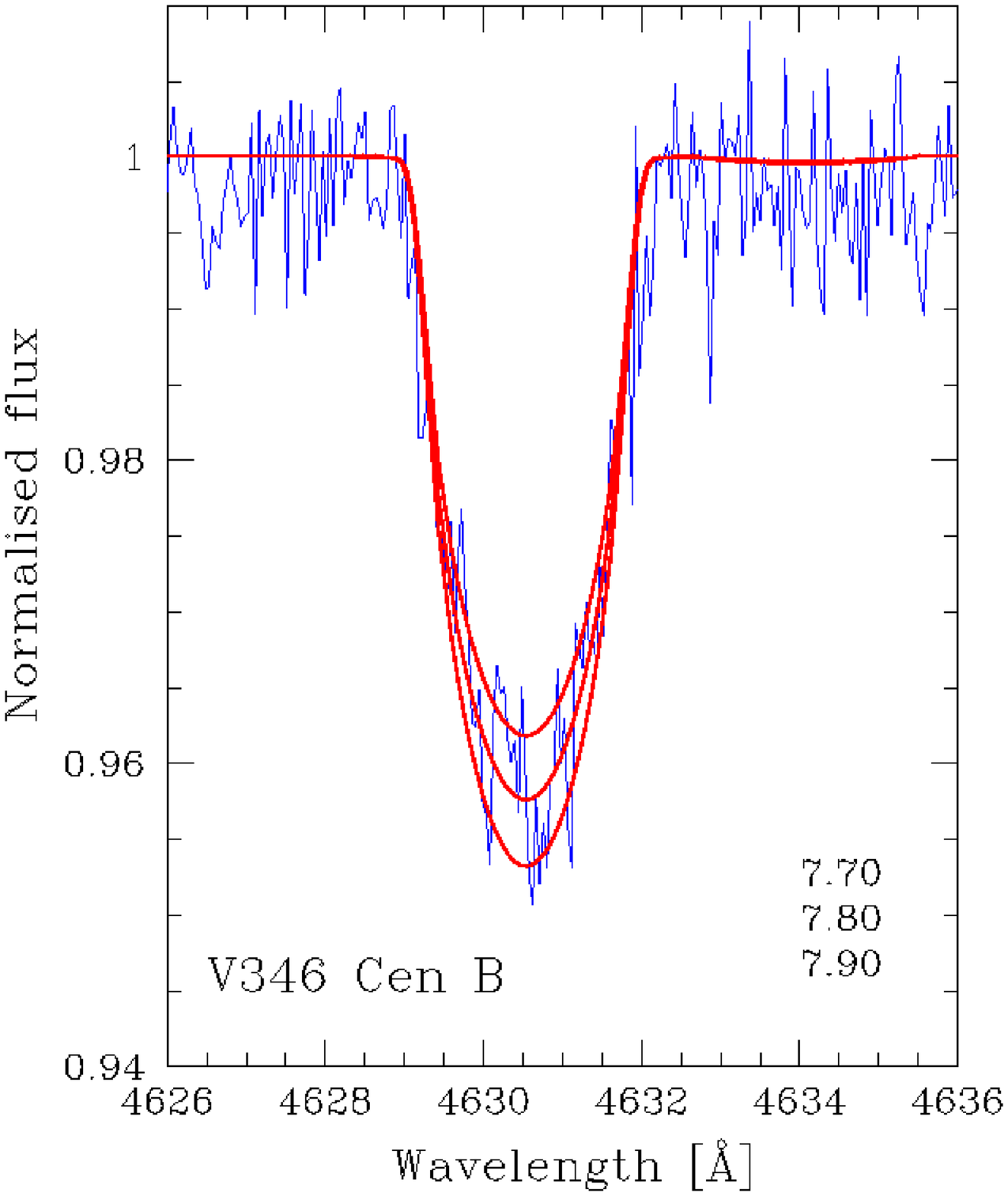}
\caption{\label{fig:carb346} Same as Fig.~\ref{fig:nitro1034} but for V346\,Cen. The upper two panels show
the \ion{C}{ii} lines at 5133--5151\,\AA, and the bottom two panels show the \ion{N}{ii} 4630.5\,\AA\ line for the two components.}
%\end{tabular}
\end{figure}

With the \Teff, $\xi_{\rm t}$ and \vsini\ from Section~\ref{sec:atmos}, and \logg\ from the masses and radii of the stars (Section~\ref{sec:lc}), we have all quantities needed for determining surface abundances. We calculated model atmospheres for the \Teff\ and \logg\ values of the components with the {\sc atlas9} code. Then a grid of synthetic spectra was calculated in NLTE with {\sc detail}, and {\sc surface}. The following species were considered: C, N, O, Mg and Si.  Spectra for a broad range of elemental abundances were calculated, spanning $\pm0.05$~dex in steps of 0.05~dex, around the `present-day cosmic abundances' determined in \citet{Nieva_Przybilla_2012}  ($\log \epsilon({\rm C})=8.25$, $\log \epsilon({\rm N})=7.69$, $\log \epsilon({\rm O})=8.71$, $\log \epsilon({\rm Mg})=7.56$, and $\log \epsilon({\rm Si})=8.45$). These were broadened by the instrumental broadening, and a rotational kernel. The microturbulent velocity was taken into account in the line profile calculations with {\sc surface} as determined from minimising the scatter in the O abundances and given in Table~\ref{tab:atmospar}.
Abundances were determined by minimising the residuals ($\chi^2$ criterion) between the renormalised disentangled spectrum and the synthetic spectrum. In the renormalisation of the disentangled spectra to their individual continuum, the light ratio obtained in the light curve analysis (Table~\ref{tab:wd}) was used, except for GL\,Car where the spectroscopically determined light ratio was employed.

The number of lines available for the abundance determination of a particular element varies due primarily to \Teff. For the \Teff\ range covered by our target stars, the spectral lines of CNO are quite varied. The most numerous spectral lines are for O, which is why we used them to determine $\xi_{\rm t}$. The spectral lines of C are the least numerous, and the broad wavelength coverage of HARPS spectra is of vital importance. We show examples of the disentangled and synthetic spectra in Figs.\ \ref{fig:nitro1034} to \ref{fig:carb346} for selected CNO lines.

The results for all five elements are given in Table~\ref{tab:cno}, as well as the indices [N/C] and [N/O]. Uncertainties were calculated including the standard deviation of the mean for available spectral lines, and the uncertainty due to uncertainties in \Teff\ and $\xi_{\rm t}$. Uncertainty in the abundances due to uncertainties in the surface gravity are negligible since $\log g$ is determined to high precision from the masses and radii.

\begin{figure*}
    
\centering
\includegraphics[width=6.cm]{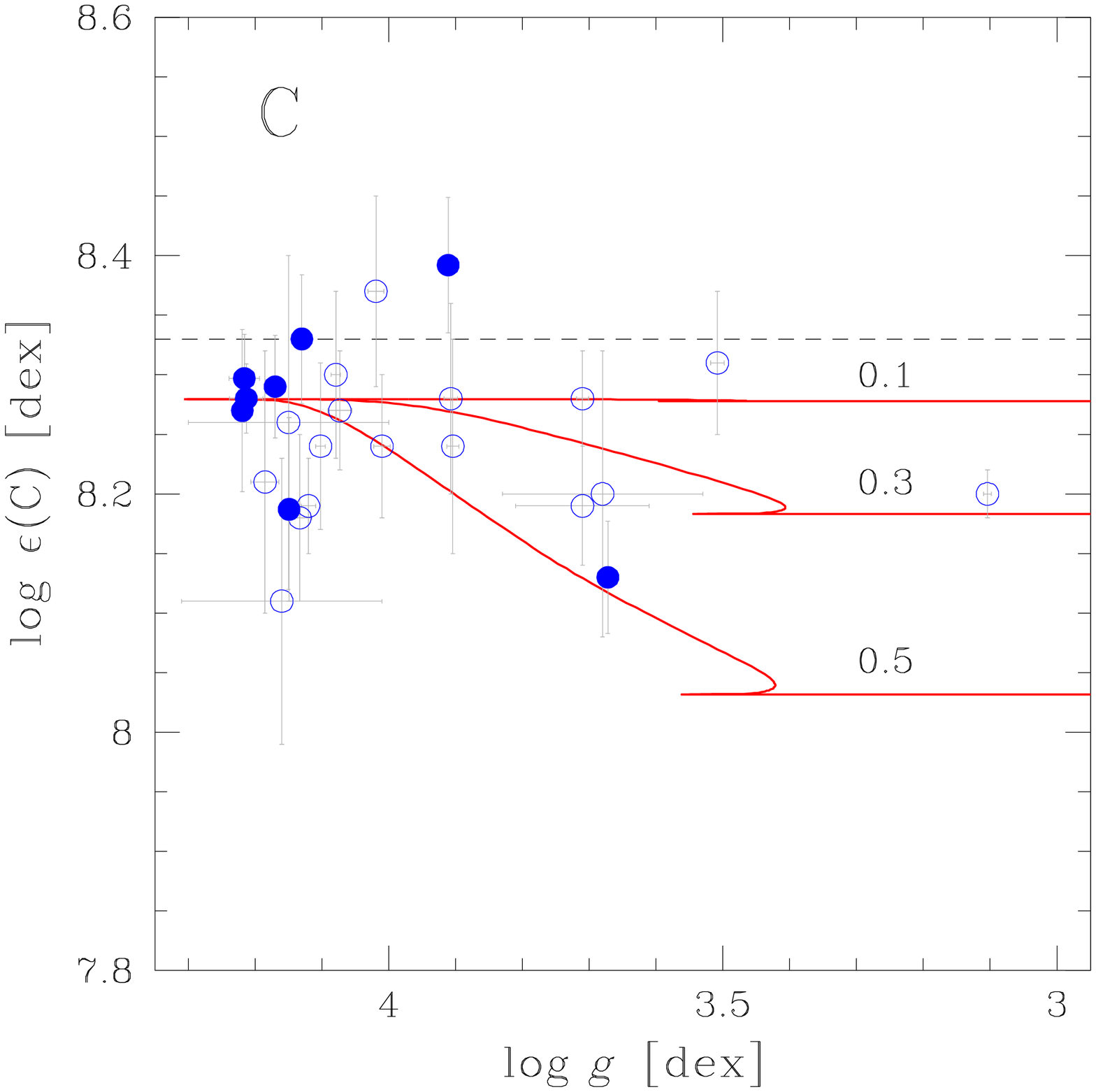}
\includegraphics[width=6cm]{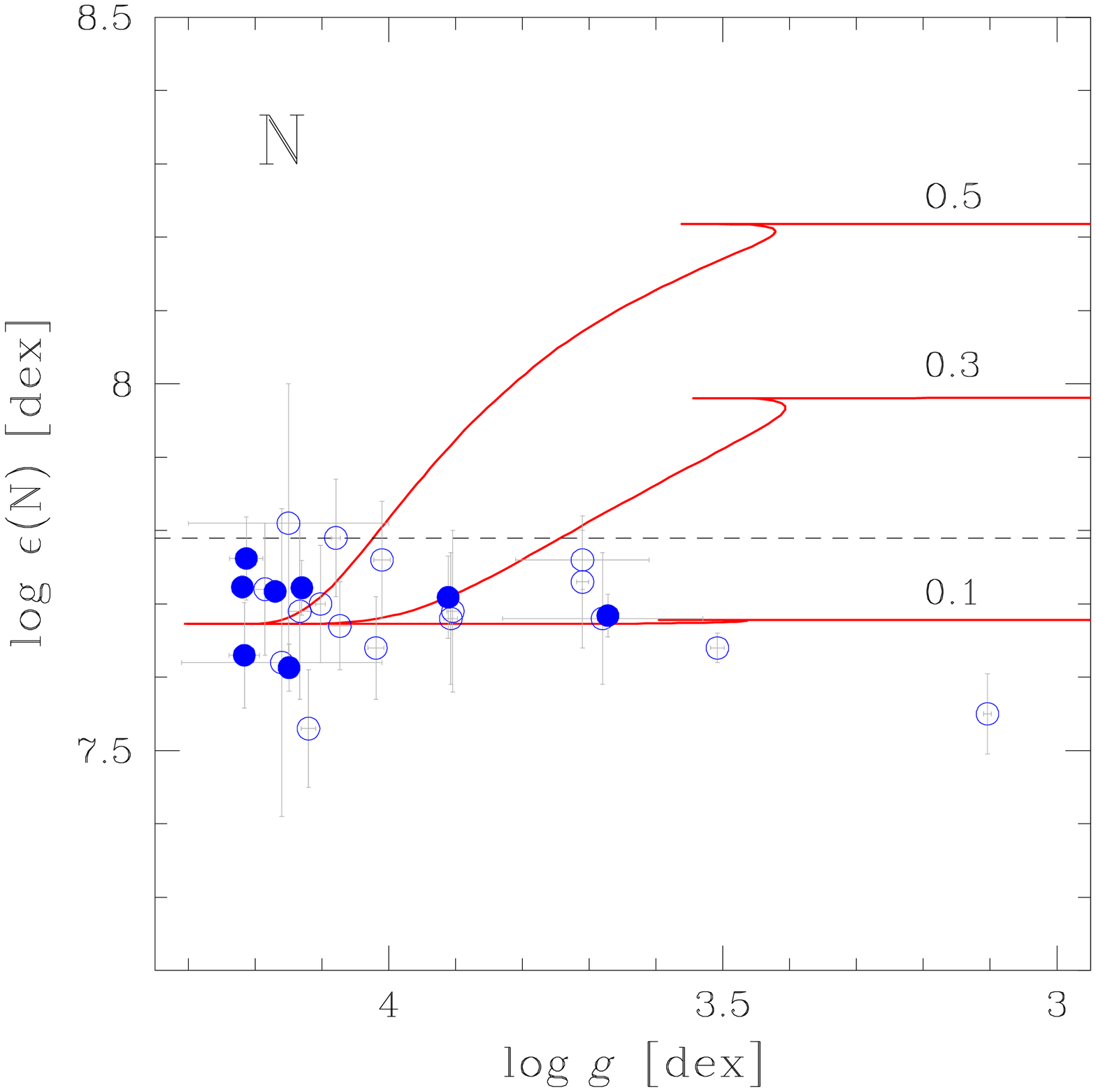}
\includegraphics[width=6cm]{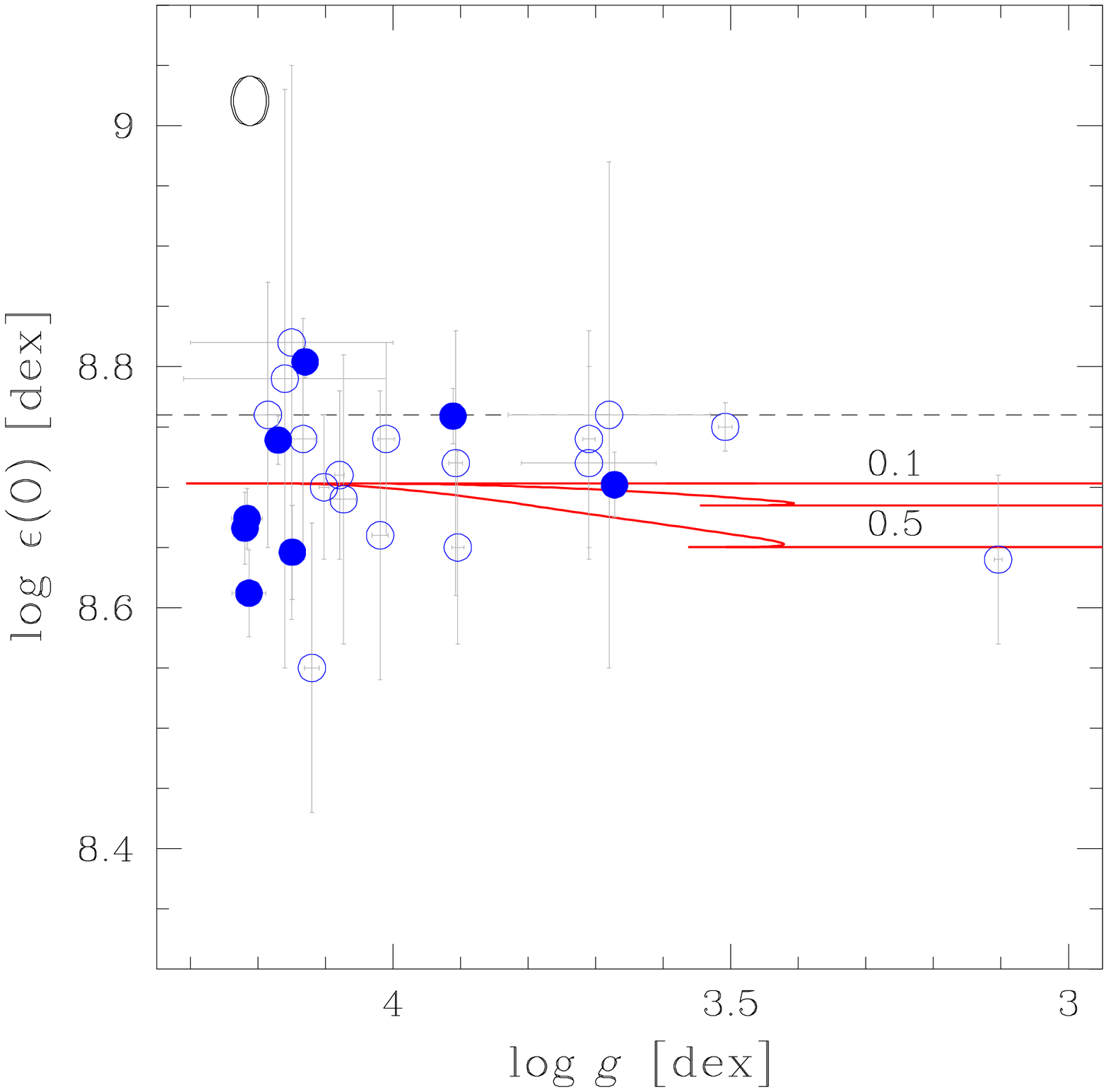}
\caption{\label{fig:abund:all}  Individual abundances of carbon (left), nitrogen (middle), and oxygen (right) as a function of surface gravity. The stars in the present sample are indicated with filled blue circles while stars taken from our previous studies are shown with open blue circles. The surface gravity (obtained from the binary solution) is used as a proxy for stellar evolution.  Solid red lines show theoretical evolutionary tracks for a 15\Msun\ star and three values of the initial rotational velocity $\Omega/\Omega_{\rm crit}$ = 0.1, 0.3, and 0.5 \citep{Georgy_2013}. The cosmic standard abundance values of \citet{Nieva_Przybilla_2012} are indicated with horizontal dashed lines.}
\end{figure*}

\begin{figure}
\centering
\includegraphics[width=8.4cm]{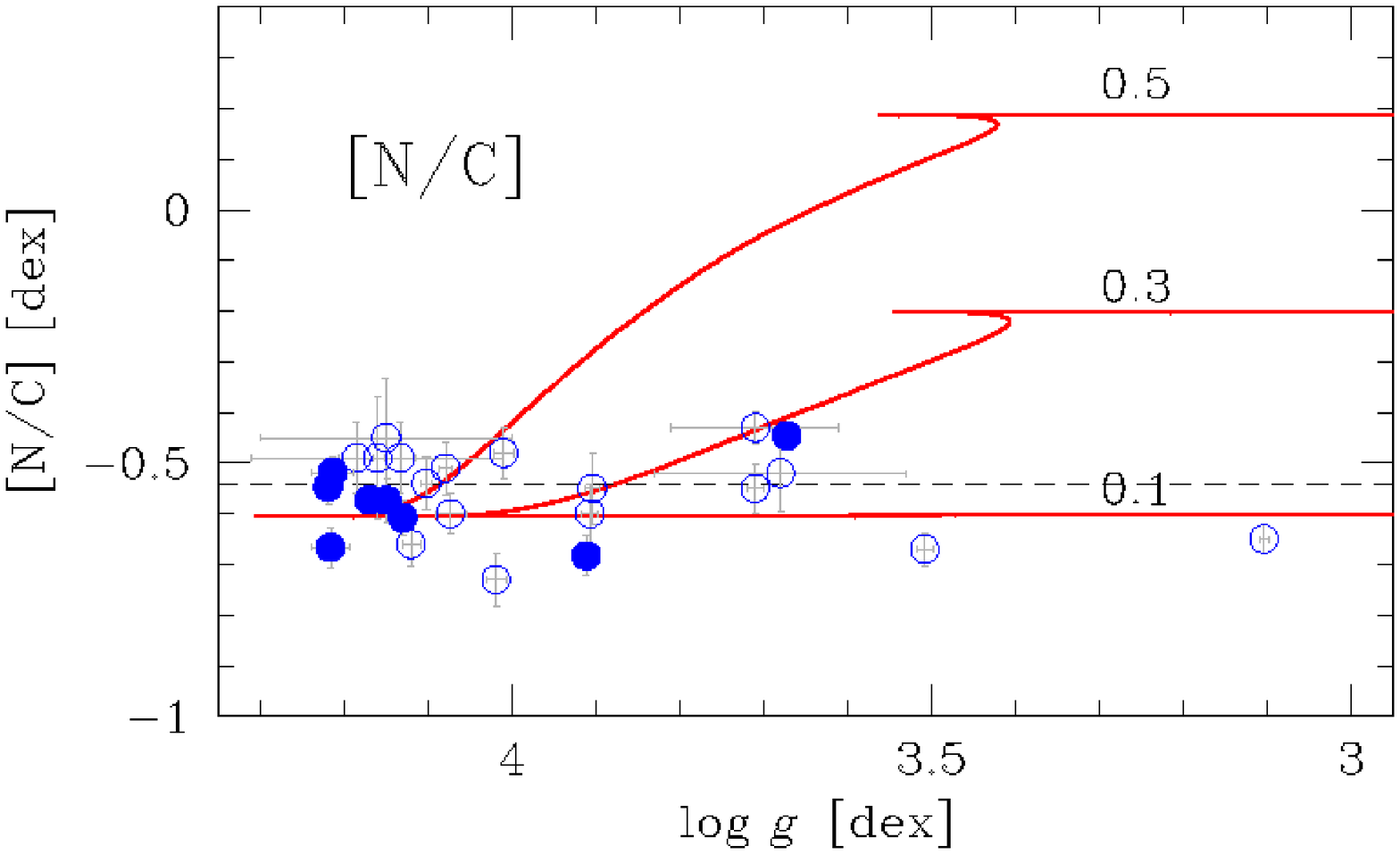} 
\vspace{0.3cm}

\includegraphics[width=8.4cm]{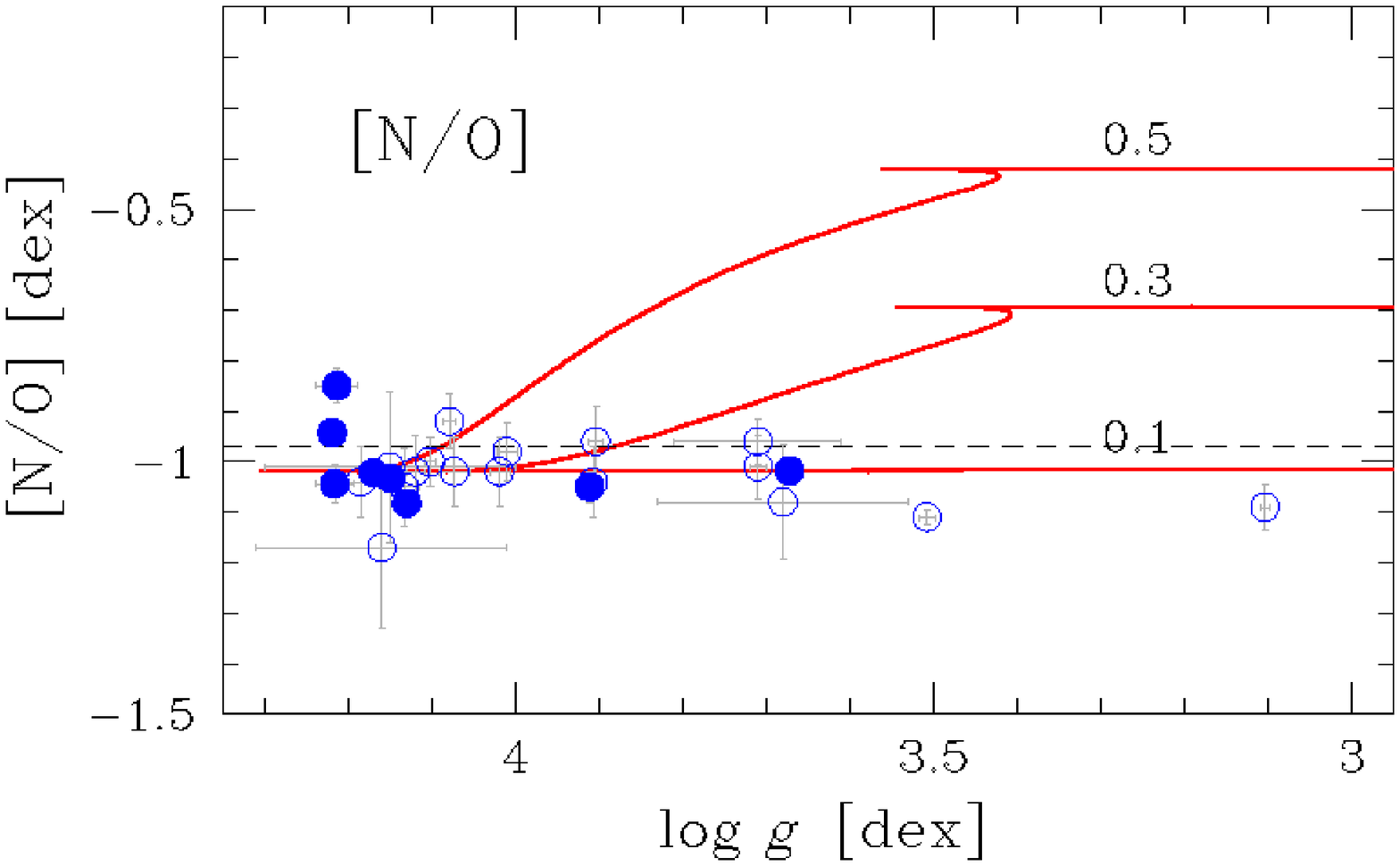}
\caption{\label{fig:cnomodels:all} Same as in Fig.~\ref{fig:abund:all} but for the [N/C] and [N/O] abundance indices.}
\end{figure}

\begin{table}  \centering
\caption{\label{tab:abucomp} Comparison of abundances determined for targets in present work to their parent clusters.
}
\begin{tabular}{lccc} \hline \hline
NGC\,6231 & & & \\ \hline
Element & Kilian et al.  & Mathys et al.  & This work  \\
           & (1994) & (2002) & V1034\,Sco \\
           \hline
 $\log \epsilon({\rm C})$   & 8.37$\pm$0.05  & 8.29$\pm$0.17  &  8.33$\pm$0.08 \\
$\log \epsilon{\rm (N)}$    & 7.85$\pm$0.05  & 7.85$\pm$0.10  &  7.69$\pm$0.08  \\
$\log \epsilon({\rm O})$    & 8.61$\pm$0.05  & 8.30$\pm$0.42  &  8.73$\pm$0.12  \\
$\log \epsilon ({\rm Mg})$  & 7.39$\pm$0.04  & --             &  7.56$\pm$0.11 \\
$\log \epsilon ({\rm Si})$  &  7.15$\pm$0.06 & --             &  7.51$\pm$0.10  \\
$[{\rm N/C}]$  &  $-$0.52$\pm$0.07  & $-$0.44$\pm$0.20  & $-$0.64$\pm$0.11  \\
$[{\rm N/O}]$  &  $-$0.76$\pm$0.07  & $-$0.45$\pm$0.40  & $-$1.04$\pm$0.14  \\
\hline
\hline
NGC\,3293  & & & \\ \hline
Element & Hunter et al.  & Morel et al.  & V573\,Car  \\
           & (2009) & (2022) & GL\,Car \\
           \hline
$\log \epsilon({\rm C})$   & 7.97$\pm$0.19  & 8.13$\pm$0.16  & 8.24$\pm$0.17  \\
$\log \epsilon{\rm (N)}$   & 7.60$\pm$0.15  & 7.72$\pm$0.14  & 7.70$\pm$0.27  \\
$\log \epsilon({\rm O})$   & 8.65$\pm$0.17  &  --            & 8.70$\pm$0.18  \\
$\log \epsilon ({\rm Mg})$ & 7.22$\pm$0.16  & 7.45$\pm$0.18  & 7.50$\pm$0.20  \\
$\log \epsilon ({\rm Si})$ & 7.42$\pm$0.09  & 7.56$\pm$0.25  & 7.51$\pm$0.26  \\
$[{\rm N/C}]$  &  $-$0.37$\pm$0.21  &  $-$0.40$\pm$0.21  &  $-$0.54$\pm$0.27  \\
$[{\rm N/O}]$  &  $-$1.05$\pm$0.26  &  --                &  $-$1.00$\pm$0.11  \\
\hline
\end{tabular}
\tablefoot{
V1034\,Sco is member of the open cluster NGC\,6231 for which abundance analyses were published by \citet{Kilian_1994}
and \citet{Mathys_2002}. For the open clusters NGC\,3572 and Trumpler\,16, parent clusters of GL\,Car and V573\,Car,
respectively, no abundance studies are available. We used abundance studies of the open cluster NGC\,3293, since it
is part of the large Car\,OB1 complex, as are NGC\,3572 and Trumpler\,16.
}
\end{table}

 A fairly good agreement between the abundances in both components of the same binary system is seen from examination of Table~\ref{tab:cno}. The most notable difference is for the abundance of magnesium (Mg) which in three cases (V1034\,Sco, V573\,Car and V346\,Cen) is modestly larger than the uncertainties. The Mg abundances are inferred from a single line, \ion{Mg}{ii} 4481\,\AA, so their uncertainties are larger than for other species. For other elements, the observed abundance differences are mostly below 0.1 dex, well within the 1$\sigma$ uncertainty interval. Apart from Mg, the largest deviations are for the C abundance in V346\,Cen ($\log \epsilon({\rm C})_{\rm A} - \log \epsilon({\rm C})_{\rm B} = -0.20 \pm 0.08$\,dex), the N abundance in V573\,Car ($-0.13 \pm 0.12$), and the O abundance in V346\,Cen ($-0.10 \pm 0.08$). V346\,Cen\,A has the lowest C abundance among the eight stars, with $\log \epsilon(C) = 8.13\pm0.05$, almost 0.20 dex less than the mean C abundance. Contrary to this, the N abundance for the same star is normal. It is also worth noting that V346\,Cen\,A is the most evolved in our sample of eight OB stars.

 Fig.~\ref{fig:abund:all} compares individual CNO abundances determined for the eight stars in this work to our previous abundance measurements in high-mass binaries. The new determinations are shown in solid blue circles, whilst our previous results are represented with open blue circles. Our previous determinations are for 17 high-mass stars in nine binary systems: V578\,Mon \citep{Pavlovski_Hensberge_2005, Garcia_2014, Pavlovski_2018}, V453\,Cyg \citep{Pavlovski_Southworth_2009, Pavlovski_2018}, V380\,Cyg \citep{Pavlovski_2009, Tkachenko_2014a}, $\sigma$\,Sco \citep{Tkachenko_2014b}, $\alpha$\,Vir \citep{Tkachenko_2016}, CW\,Cep \citep{Johnston_2019}, AH\,Cep \citep{Pavlovski_2018}, V478\,Cyg \citep{Pavlovski_2018} and the primary component in V621\,Per (Southworth et al.\ in prep.). There are no discernable systematics between the new and previous sample. This is expected because we are consistently using the same reduction and analysis tools, so the 25 stars (in 13 binary systems) represent an homogeneous sample. In the last three rows of Table~\ref{tab:cno} the mean values for abundances in the present sample, in OB binaries studied previously by us, and the `present cosmic abundance standard' -- an abundance pattern evaluated for B-type stars by \citet{Nieva_Przybilla_2012} -- are given for comparison. As already mentioned, both our samples are in perfect agreement and there are no outliers. However, it can be seen that the CNO abundances for OB stars in binary systems are below the cosmic abundance standard, with very few exceptions. \citet{Nieva_Przybilla_2012} determined elemental abundances for sample of sharp-lined early B-type stars, enabling a very high accuracy. This was not an option for our work because our sample stars are all in short-period binary systems so are either moderate or fast rotators. Thus the spectral lines are usually broad and overlapping, making the choice of suitable spectral lines for abundance determination more limited, and thus affecting the accuracy of the results.

 Fig.~\ref{fig:abund:all} presents a comparison of the inferred CNO abundances with theoretical evolutionary tracks of a 15\Msun\ star computed for three values of the initial rotational velocity $\Omega/\Omega_{\rm crit} = 0.1$, 0.3 and 0.5 \citep{Georgy_2013}. In their model calculations \citet{Georgy_2013} used the following initial abundances: $\log \epsilon({\rm C})=8.28$, $\log \epsilon({\rm N})=8.67$, $\log \epsilon({\rm O})=8.55$. The values for the abundances of C and N are in fair agreement with our present and preious findings (c.f.~Table~\ref{tab:abucomp}, but differ by 0.15\,dex for the O abundance. Therefore, we empirically `corrected' the initial O abundance in the theoretical models, and shift the O abundance upwards in  Fig.~\ref{fig:abund:all}, and accordingly for [N/O] in Fig.~\ref{fig:cnomodels:all}.

One can see that the models predict significant depletion and enhancement of C and N, respectively, as the rotation rate of the star increases, while only a marginal depletion is predicted for O. Moreover, these abundance trends are substantial at the start of the main-sequence evolution already. However, the individual abundances of CNO elements measured by us do not follow the relations predicted by the models: instead we observe a scatter of values around or slightly below the cosmic standard abundance values of \citet{Nieva_Przybilla_2012}. 

Furthermore, the N to C abundance ratio index [N/C] is a sensitive probe of the stellar evolution model predictions, as can be seen in Fig.~\ref{fig:cnomodels:all}. The models suggest a noticeable increase of the surface N abundance with respect to the abundance of C (top panel) and O (bottom panel) as the rotation rate of the star $\Omega/\Omega_{\rm crit}$ increases. Similar to the individual elemental abundances discussed above, we do not observe the increase in the [N/C] and [N/O] indices as the surface gravity of the star decreases. Moreover, the bulk of our abundance measurements cluster around the mean [N/C] and [N/O] values found by \citet{Nieva_Przybilla_2012} in the solar neighbourhood, with the spread being significantly smaller than one would expect if the abundance ratios were altered substantially by the effect of stellar rotation.

No previous abundance determinations are available for any of the binary systems analysed in this work. However, we can check our results against published abundances for the open clusters our sample are members of. Photospheric abundances for B-type stars in the open cluster NGC\,6231 were determined by \citet{Kilian_1994} and \citet{Mathys_2002}. Results from these studies are compared to our results for V1034\,Sco in Table~\ref{tab:abucomp}. The parent clusters for the other three systems have not (yet) been subject to a chemical composition study, but the open cluster NGC\,3293 (which is part of young association Car OB1 \citealt{Turner_1980}) is well studied. \citet{Hunter_2009} determined abundances from 50 B-type stars, while in a recent publication \citet{Morel_2022} examined a large sample of about 150 B-type stars in the framework of the Gaia-ESO Survey. Since the dEBs V573\,Car and GL\,Car belongs to the Car OB1 association, we use NGC\,3293 as a proxy for the abundance pattern in Car OB1. It is interesting that the massive sample of B-type stars analysed in \citet{Morel_2022}, with a spread in \vsini\ values,  show a pattern of under-abundances compared to the standard solar abundances \citep{Asplund_2009}, in accordance with our general abundance pattern.

\section{Light curve analysis}\label{sec:lc}

We assembled the available light curves of the four targets in this work and modelled them using a consistent approach in order to determine their physical properties. The light curves were fitted using the Wilson-Devinney (WD) code \citep{WilsonDevinney71apj,Wilson79apj}, which implements Roche geometry to determine the shapes of the stars and thus the brightness of binary systems as a function of orbital phase. We used the 2004 version of the WD code, driven with the {\sc jktwd} wrapper \citep{Me+11mn}.

For each system we performed a series of tests to determine the best approach to modelling it with {\sc jktwd}. Once we had arrived at the preferred solution, we performed further tests to determine the range of plausible solutions and thus the uncertainties in the fitted parameters. This step was taken because we have consistently found that the formal errorbars calculated by the WD code underestimate the true uncertainty of the fitted parameters \citep{Pavlovski_Southworth_2009,Pavlovski_2009,Pavlovski_2018,Me+20mn}, as indicated in the user guide to the code \citep{WilsonVanhamme04}.

Unless otherwise specified we used Mode 0 in the WD code, which is for detached binary systems where the light contributions for each star are fitted individually, simple reflection, and the logarithmic limb darkening (LD) law. We fitted for the potentials and light contributions of the two stars, the orbital inclination and a phase shift with respect to the adopted orbital ephemeris. The mass ratio was fixed at the spectroscopic value, bolometric albedos were set to 1.0, synchronous rotation was assumed, and the gravity brightening exponents were set to 1.0. A circular orbit was assumed for V573\,Car but the possibility of an eccentric orbit was checked. The input LD coefficients were obtained by bilinear interpolation in the tables of \citet{Vanhamme93aj}.

For the purposes of determining the uncertainties in the fitted parameters, we ran a series of alternative solutions for differing choice of WD code mode of operation, choice of numerical resolution, treatment of reflection, choice of LD law, whether the LD coefficients were fixed or fitted, treatment of third light, variation of the mass ratio within the uncertainties, and the possibility of orbital eccentricity (for V573\,Car). We also considered the effects of albedo, rotational velocity and gravity brightening, by fixing them at different values and also attempting to fit for them directly.

The net result of this process was a default solution for each system, accompanied by a measurement of how much each fitted parameter changed between this default solution and each of the alternative solutions. These changes were then added in quadrature to arrive at a final robust uncertainty value for each fitted parameter. The results for all four systems are summarised in Table~\ref{tab:wd}. The fractional radii are volume-equivalent values obtained from the {\sc lc} flavour of the WD code.

\begin{table*} \centering
\caption{\label{tab:wd} Summary of the parameters for the {\sc wd2004} solutions of the light curves of
the systems. }
\begin{tabular}{llcccc} \hline\hline
Parameter                      & {\sc wd2004} name & V1034\,Sco           & GL\,Car              & V573\,Car           & V346\,Cen            \\
\hline
{\it Control and fixed parameters:} \\
{\sc wd2004} operation mode    & {\sc mode}        & 0                   & 0                   & 0                   & 0                   \\
Treatment of reflection        & {\sc mref}        & 1                   & 1                   & 1                   & 1                   \\
Number of reflections          & {\sc nref}        & 1                   & 1                   & 1                   & 1                   \\
Limb darkening law             & {\sc ld}          & 2 (logarithmic)     & 2 (logarithmic)     & 1 (linear)          & 2 (logarithmic)     \\
Numerical grid size (normal)   & {\sc n1, n2}      & 60                  & 50                  & 50                  & 50                  \\
Numerical grid size (coarse)   & {\sc n1l, n2l}    & 50                  & 40                  & 40                  & 40                  \\[3pt]
{\it Fixed parameters:} \\
Orbital period (d)             & {\sc period}      & 2.440646            & 2.4222681           & 1.4693316           & 6.3220088           \\
Primary eclipse time (HJD)     & {\sc hjd0}        & 2451931.2652        & 2459321.2994        & 2450456.8164        & 2459335.5607        \\
Mass ratio                     & {\sc rm}          & 0.563               & 0.943               & 0.818               & 0.712               \\
$T_{\rm eff}$ star\,A (K)      & {\sc tavh}        & 32\,200             & 30\,960             & 31\,900             & 26\,100             \\
$T_{\rm eff}$ star\,B (K)      & {\sc tavh}        & 25\,800             & 30\,390             & 28\,700             & 22\,500             \\
Rotation rates                 & {\sc f1, f2}      & 1.0, 1.0            & 1.67, 1.29          & 1.0, 1.0            & 2.49, 2.70          \\
Gravity darkening              & {\sc gr1, gr2}    & 1.0, 1.0            & 1.0, 1.0            & 1.0, 1.0            & 1.0, 1.0            \\
Bolometric albedos             & {\sc alb1, alb2}  & 1.0, 1.0            & 1.0, 1.0            & 1.0, 1.0            & 1.0, 1.0            \\[3pt]
{\it Fitted parameters:} \\
Phase shift                    & {\sc pshift}      & $-$0.0035           & 0.0413              & 0.0008              & 0.0702              \\
Star\,A potential              & {\sc phsv}        & $3.670 \pm 0.044$   & $5.736 \pm 0.033$   & $3.913 \pm 0.024$   & $5.933 \pm 0.015$   \\
Star\,B potential              & {\sc phsv}        & $4.228 \pm 0.031$   & $5.766 \pm 0.064$   & $4.106 \pm 0.027$   & $8.012 \pm 0.032$   \\
Orbital inclination (\degr)    & {\sc xincl}       & $81.80 \pm 0.32$    & $86.57 \pm 0.17$    & $80.52 \pm 0.14$    & $84.97 \pm 0.12$    \\
Orbital eccentricity           & {\sc e}           & $0.027  \pm 0.013$  & $0.1465 \pm 0.0004$ & 0.0 (fixed)         & $0.2750 \pm 0.0006$ \\
Argument of periastron (\degr) & {\sc perr0}       & $57 \pm 18$         & $22.47 \pm 0.36$    &                     & $27.53 \pm 0.28$    \\
Light from star\,A ($u$ band)  & {\sc hlum}        &                     &                     & 8.006               &                     \\
Light from star\,B ($u$ band)  & {\sc clum}        &                     &                     & 4.269               &                     \\
Light from star\,A ($v$ band)  & {\sc hlum}        &                     &                     & 7.861               &                     \\
Light from star\,B ($v$ band)  & {\sc clum}        &                     &                     & 4.449               &                     \\
Light from star\,A ($b$ band)  & {\sc hlum}        &                     &                     & 7.893               &                     \\
Light from star\,B($b$ band)   & {\sc clum}        &                     &                     & 4.531               &                     \\
Light from star\,A ($y$ band)  & {\sc hlum}        &                     &                     & 7.888               &                     \\
Light from star\,B ($y$ band)  & {\sc clum}        &                     &                     & 4.551               &                     \\
Light from star\,A (TESS band) & {\sc hlum}        & $8.37 \pm 0.10$     & $6.096 \pm 0.081$   &                     & $8.374 \pm 0.086$   \\
Light from star\,B (TESS band) & {\sc clum}        & $1.73 \pm 0.08$     & $5.204 \pm 0.102$   &                     & $1.967 \pm 0.006$   \\
Third light (TESS band)        & {\sc el3}         & $0.239 \pm 0.012$   & $0.166 \pm 0.006$   &                     & $0.214 \pm 0.007$   \\
Fractional radius of star\,A   &                   & $0.3300 \pm 0.0032$ & $0.2203 \pm 0.0015$ & $0.3308 \pm 0.0025$ & $0.2088 \pm 0.0014$ \\
Fractional radius of star\,B   &                   & $0.1901 \pm 0.0022$ & $0.2088 \pm 0.0017$ & $0.2759 \pm 0.0029$ & $0.1106 \pm 0.0008$ \\[3pt]
\hline
\end{tabular} 
\tablefoot{
Detailed descriptions of the control parameters can be found in the WD code user guide
\citep{WilsonVanhamme04}. A and B refer to the primary and secondary stars, respectively. Uncertainties are
only quoted when they have been robustly assessed by comparison with a full set of alternative solutions.
}
\end{table*}

\subsection{V1034\,Sco}

\begin{figure}
\includegraphics[width=\columnwidth]{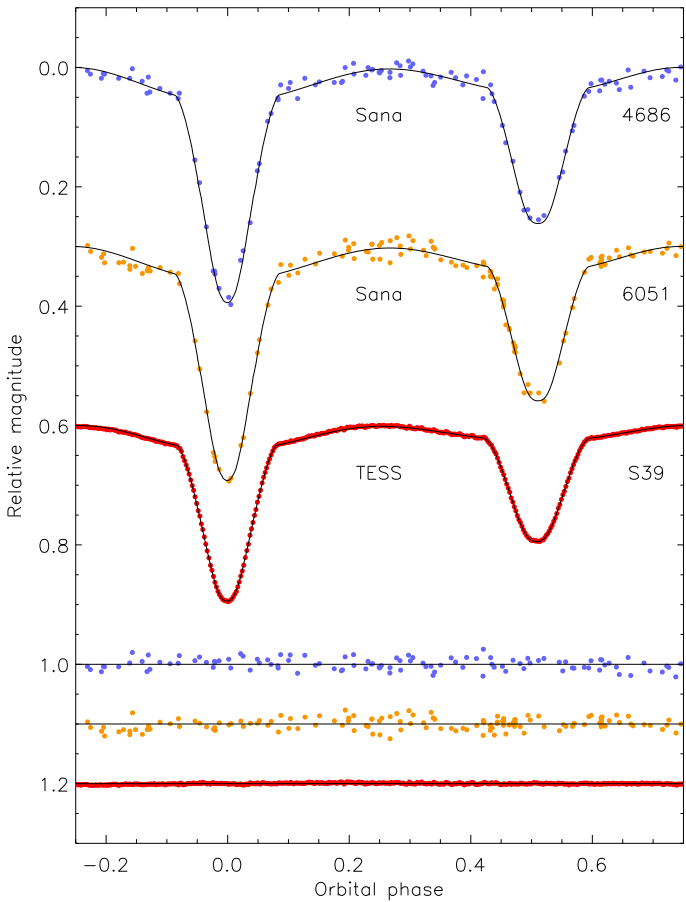} \\
\caption{\label{fig:lc:v1034} The light curves and best WD models for V1034\,Sco.
The differential magnitudes are plotted versus orbital phase and are colour-coded
according to the central wavelengths of the passbands. The source and passband of
each light curve is labelled. The residuals of the fit are shown at the base of the
figure. Offsets have been applied between the light curves and residuals for clarity.}
\end{figure}

Two photometric studies of V1034\,Sco have been published. \citet{Bouzid_2005} presented light curves taken in the Str\"omgren $uvby$ filters, with 409, 645, 1058 and 1036 datapoints, respectively. \citet{Sana_2005} obtained light curves in two narrow-band filters, designated $\lambda$4685 and $\lambda$6051, containing 112 and 138 datapoints, respectively. For our exploratory solutions we used the \citet{Sana_2005} data as the \citet{Bouzid_2005} data are not available.

In the course of this work a new light curve became available from sector 39 of the TESS satellite (see Section~\ref{sec:data}). As the TESS data are of much higher quality than the other photometry, we have based our final results for V1034\,Sco on these data. Before doing so, we performed a preliminary fit with {\sc jktebop} \citep{Me13aa} to obtain an orbital ephemeris then phase-binned these data down to 500 bins to decrease the computation time. Our final solution is for an eccentric orbit, including third light, and the logarithmic LD law and the linear LD coefficient fitted for each star and passband. The main contributors to the uncertainty in the fractional radii are the treatment of albedo and gravity darkening. Uncertainties arising from the choice of numerical resolution, WD program mode, rotation rate and LD were all significantly smaller and therefore contributed negligibly when all uncertainties for each parameter were added in quadrature. The parameters and their uncertainties are given in Table~\ref{tab:wd}, and the best fits are shown in Fig.~\ref{fig:lc:v1034}.

\subsection{GL\,Car}

\begin{figure}
\includegraphics[width=\columnwidth]{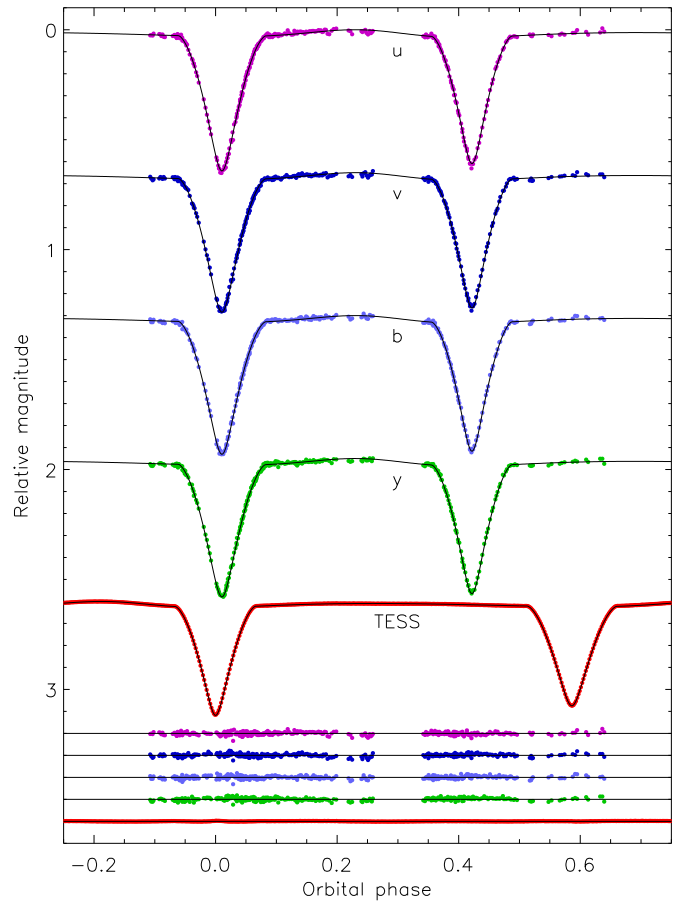} \\
\caption{\label{fig:lc:gl} The light curves and best WD models for
GL\,Car. Other comments are the same as for Fig.~\ref{fig:lc:v1034}.}
\end{figure}

Light curves of GL\,Car in the Str\"omgren $uvby$ system were obtained by \citet{Gimenez_1985} using the 0.5\,m Str\"omgren Automated Telescope at ESO La Silla. They comprise 526 observations through each filter, 234 in the 1982 observing season and 308 in the 1983 season. These data were analysed by \citet{Gimenez_Clausen_1986} using the {\sc wink} code \citep{Wood71aj}. The observations were obtained in electronic form from the archive of J.\ V.\ Clausen and used in the current work to obtain a preliminary solution. We found values and uncertainties for the fitted parameters in good agreement with those from \citet{Gimenez_Clausen_1986}.

Subsequent to our analysis of the $uvby$ data a new light curve of GL\,Car became available from TESS. We phase-binned this and modelled it using {\sc wd}, fitting for an eccentric orbit and third light. The rotation rates (F1 and F2) were set to the ratios of the measured rotational velocities (Table~\ref{tab:atmospar}) and the synchronous values, determined iteratively. Unlike the $uvby$ data, the TESS light curve shows a strong correlation between the light ratio and the amount of third light. We therefore applied the light ratio from our spectroscopic analysis. Because there is no mechanism to explicitly apply a spectroscopic light ratio in {\sc wd2004} we propagated the light ratio from our spectral interval (which corresponds closely to the Johnson $B$ band) to the Str\"omgren $uvby$ bands \citep[see][]{Me10mn} using {\sc atlas9} theoretical spectra \citep{Castelli++97aa} and passband response functions from \citet{Maizapellaniz06aj}. We forced {\sc wd2004} to match them by fixing the {\sc hlum} parameters at the appropriate values. Including this constraint greatly improved the reliability of the results.

We find precise fractional radii for GL\,Car once our spectroscopic light ratio is included (Table~\ref{tab:wd}). The uncertainties are dominated by those from this light ratio, but are still below 1\% and a factor of three smaller than those from the $uvby$ data alone. They also agree well with the less precise results from \citet{Gimenez_Clausen_1986}. The fitted orbital eccentricity is in excellent agreement with that from its apsidal motion ($e = 0.1459 \pm 0.0015$ from \citealt{Wolf_2008}). The orbital phase of secondary eclipse has changed a lot between the $uvby$ and TESS datasets, and the different morphology of the light curve is obvious (see Fig.~\ref{fig:lc:gl}).

\subsection{V573\,Car}

\begin{figure}
\includegraphics[width=\columnwidth]{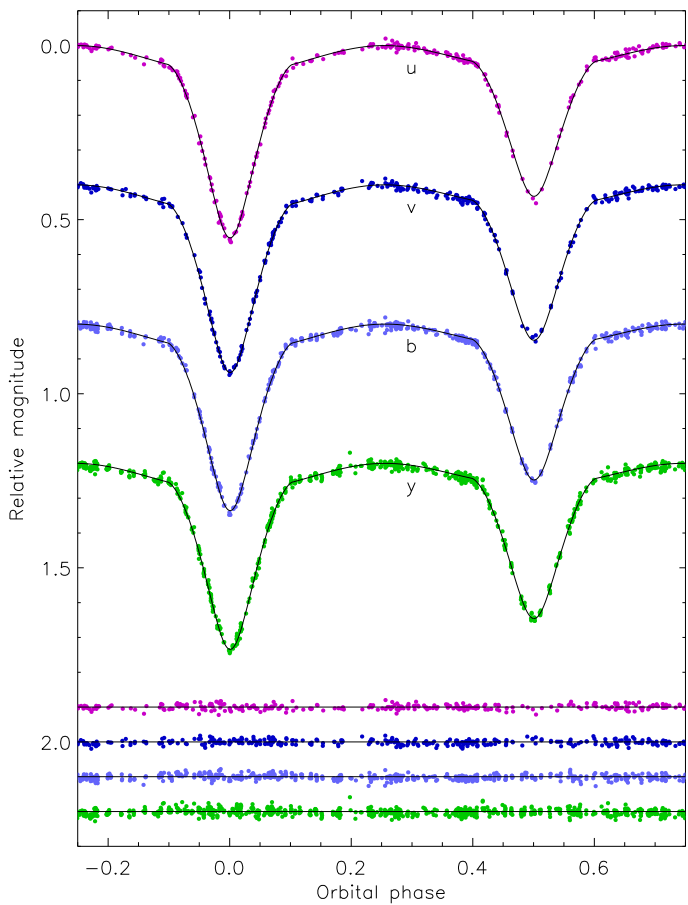} \\
\caption{\label{fig:lc:v573} The light curves and best WD models for V573\,Car.
Other comments are the same as for Fig.~\ref{fig:lc:v1034}.} \end{figure}

V573\,Car was studied by \citet{Freyhammer_2001} using the Dutch 0.9\,m telescope at ESO La Silla. A total of 1910 observations were obtained through the Str\"omgren filters: 763 in $y$, 513 in $b$, 350 in $v$ and 284 in $u$. We fitted all four light curves simultaneously, using the ephemeris from \citet{Freyhammer_2001}. We assumed a circular orbit in most cases, but did run a fit with $e$ and $\omega$ free to check if this led to a better fit to the data (it didn't). We also assumed no third light, after attempts to fit for it had a negligible effect on the results and also led to a slightly negative value for this parameter. The best fit is shown in Fig.~\ref{fig:lc:v573}.

The fitted parameters were the potentials of the two stars, the orbital inclination, a phase shift, and the light contributions of the two stars in each passband. To avoid very small values for the light contributions we renormalised each light curve to be at approximately zero relative magnitude at quadrature. We adopted the linear LD law as it gave results very similar to those for the logarithmic and square-root laws; attempts to fit for the LD coefficients led to unphysical solutions. The rotational velocities of the stars were held to the synchronous values.

We found that the solution of the light curves is degenerate in that significantly different values of the ratio of the radii or the light contributions of the stars led to almost indistinguishable fits. This was also found by \citet{Freyhammer_2001}, who constrained their solution using a light ratio measured from their spectra. We took the same approach.

For our final result (Table~\ref{tab:wd}) we give the solution for fitting all four light curves simultanously, constrained by the spectroscopic light ratio. The uncertainties in the parameters include contributions from the uncertainty in the spectroscopic light ratio, the effect of a change of 5\% in the rotation velocities of the stars, and the treatment of albedo and gravity darkening. Other sources of uncertainty (see above) were checked and found to be negligible. We were able to measure the fractional radii of the stars to precisions of 0.8\% (star A) and 1.1\% (star B); the main contribution to these uncertainties is the spectroscopic light ratio (for star~A) and the treatment of gravity darkening (for star~B). The values we find are in reasonable agreement with those from \citet{Freyhammer_2001}, but our uncertainties are slightly larger.

\begin{table*} \centering
\caption{\label{tab:absdim}   Physical properties measured for the four systems analysed in this work. }
\begin{tabular}{l r@{\,$\pm$\,}l r@{\,$\pm$\,}l r@{\,$\pm$\,}l r@{\,$\pm$\,}l} \hline\hline
Parameter                                       &    \mc{V1034\,Sco}   &    \mc{GL\,Car}      &    \mc{V573\,Car}     &    \mc{V346\,Cen}     \\
\hline
Mass ratio                                      &    0.5628 & 0.0021  &    0.943 & 0.012    &    0.8182 & 0.0037   &    0.7119 & 0.0038   \\
Mass of star A (\Msunnom)                       &     17.01 & 0.14    &    15.86 & 0.31     &     15.11 & 0.13     &     11.74 & 0.12     \\
Mass of star B (\Msunnom)                       &     9.573 & 0.053   &    14.95 & 0.30     &    12.365 & 0.096    &     8.359 & 0.089    \\
Semimajor axis (\Rsunnom)                       &    22.767 & 0.053   &    23.79 & 0.15     &    16.412 & 0.044    &     39.12 & 0.13     \\
Radius of star A (\Rsunnom)                     &     7.513 & 0.075   &    5.242 & 0.048    &     5.429 & 0.043    &     8.278 & 0.079    \\
Radius of star B (\Rsunnom)                     &     4.328 & 0.051   &    4.968 & 0.051    &     4.528 & 0.049    &     4.123 & 0.072    \\
Surface gravity of star A ($\log$[cgs])         &     3.917 & 0.009   &    4.199 & 0.007    &     4.148 & 0.007    &     3.672 & 0.008    \\
Surface gravity of star B ($\log$[cgs])         &     4.147 & 0.010   &    4.220 & 0.008    &     4.218 & 0.009    &     4.130 & 0.015    \\
Synch.\ rotational velocity of star\,A (\kms)   &     155.7 & 1.6     &    109.5 & 1.0      &     186.9 & 1.5      &     66.25 & 0.64     \\
Synch.\ rotational velocity of star\,B (\kms)   &      89.7 & 1.1     &    103.8 & 1.1      &     155.9 & 1.7      &     33.00 & 0.57     \\
\Teff\ of star\,A (K)                           &     32200 & 500     &    30960 & 500      &     31900 & 400      &     26100 & 300      \\
\Teff\ of star\,B (K)                           &     25800 & 300     &    30390 & 500      &     28700 & 350      &     22500 & 300      \\
Luminosity of star\,A $\log(L/\Lsunnom)$        &     4.738 & 0.028   &    4.357 & 0.029    &     4.439 & 0.023    &     4.457 & 0.022    \\
Luminosity of star\,B $\log(L/\Lsunnom)$        &     3.874 & 0.028   &    4.278 & 0.030    &     4.098 & 0.023    &     3.594 & 0.028    \\
Absolute bolometric magnitude of star\,A        &  $-$7.104 & 0.071   &  $-$6.15 & 0.073    &  $-$6.331 & 0.057    &  $-$6.403 & 0.054    \\
Absolute bolometric magnitude of star\,B        &  $-$4.944 & 0.057   &  $-$5.96 & 0.075    &  $-$5.505 & 0.058    &  $-$4.245 & 0.069    \\
Interstellar extinction $E(B-V)$ (mag)          &      0.75 & 0.05    &     0.55 & 0.05     &      0.40 & 0.05     &      0.56 & 0.03     \\
Distance (pc)                                   &      1460 & 50      &     2278 & 63       &      2466 & 78       &      2290 & 60       \\
\textit{Gaia} DR3 parallax (mas)                &    0.6452 & 0.0231  &   0.4232 & 0.0130   &    0.4428 & 0.0200   &    0.4380 & 0.0261   \\
\textit{Gaia} DR3 distance (pc)                 &      1550 & 56      &     2363 & 73       &      2260 & 100      &      2280 & 140      \\
% Bailer-Jones geometric                          & \ermc{1476}{56}{45} & \ermc{2299}{78}{73} & \ermc{2201}{85}{96}  & \ermc{2221}{137}{26} \\           % I suggest not to include these as they are not very informative
% Bailer-Jones photogeometric                     & \ermc{1489}{64}{61} & \ermc{2344}{86}{103}& \ermc{2262}{114}{117}& \ermc{2235}{134}{122}\\           % I suggest not to include these as they are not very informative
\hline
\end{tabular} 
\tablefoot{ The units labelled with a superscripted `N' are
given in terms of the nominal solar quantities defined in IAU 2015 Resolution B3 \citep{Prsa+16aj}.}
\end{table*}

% IDL> print, [0.14,0.053,0.12,0.089,0.13,0.096,0.31,0.30]/[17.01,9.573,11.74,8.359,15.11,12.365,15.86,14.95]*100
%      0.823045     0.553640      1.02215      1.06472     0.860357     0.776385      1.95460      2.00669
% IDL> print, [0.075,0.051,0.079,0.072,0.043,0.049,0.048,0.051]/[7.513,4.328,8.278,4.123,5.429,4.528,5.242,4.968]*100
%      0.998270      1.17837     0.954337      1.74630     0.792043      1.08216     0.915681      1.02657
% so masses to 0.6-2.0% and radii to 0.8-1.7%

\subsection{V346\,Cen}

\begin{figure}
\includegraphics[width=\columnwidth]{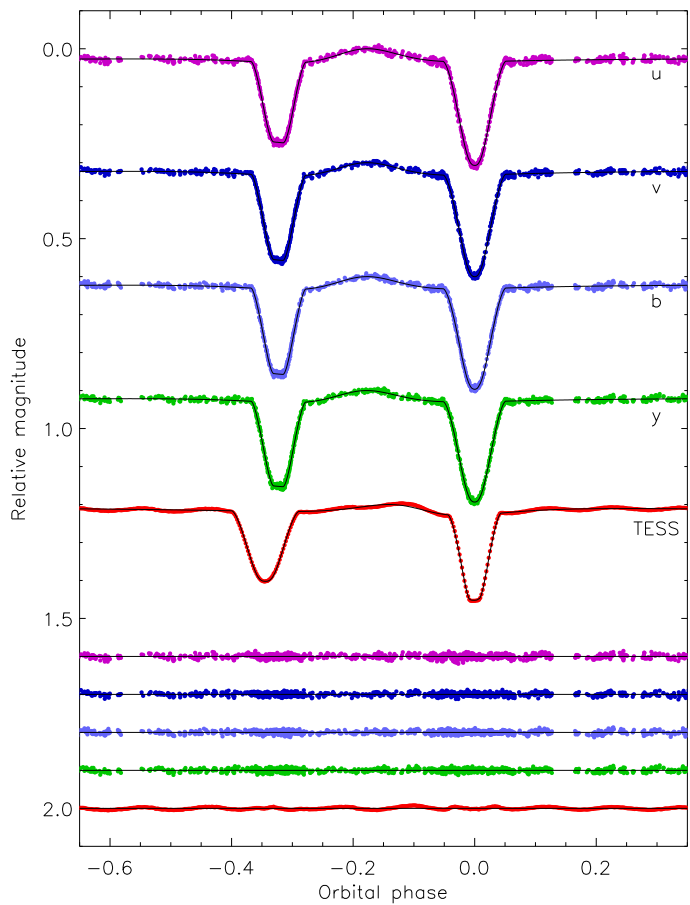} \\
\caption{\label{fig:lc:v346} The light curves and best WD models for V346\,Cen.
Other comments are the same as for Fig.~\ref{fig:lc:v1034}.} \end{figure}

Extensive photometry in the Str\"omgren $uvby$ system was obtained by \citet{Gimenez_1986b}, comprising 1056 observations made simultaneously through all four filters using the Str\"omgren Automated Telescope \citep{Gronbech++76aas}. These data have been analysed by \citet{Gimenez_1986a} using the {\sc wink} model, and by \citet{Mayer_2016} using the {\sc phoebe} code. The two studies agree on the values of the fractional radii to within the uncertainties quoted by \citet{Gimenez_1986a} but not the uncertainties quoted by \citet{Mayer_2016}. We therefore performed our own analysis of these data in order to assess robust errorbars and check the level of agreement with the previous studies.

Our WD code model for the $uvby$ data provided a good fit to the observations (Fig.~\ref{fig:lc:v346}) but required $\omega$ to be fixed at a suitable value to avoid the fit diverging to unphysical solutions. We set the rotation rates to 2.49 and 2.70 based on the rotational velocities of the stars measured from the disentangled spectra. The logarithmic LD law was adopted, although the other two laws gave almost identical results. Third light was fixed at zero because attempts to fit for it returned a small negative value that was consistent with zero. Our results were in excellent agreement with those of \citet{Gimenez_1986b}.

After this work had been performed, light curves from sectors 37 and 38 of the TESS satellite became available. These are of much higher quality so we used them for our final analysis. We performed a preliminary fit with {\sc jktebop} to obtain an orbital ephemeris then phase-binned them into 500 bins to make the computations faster. Our approach was the same as for the $uvby$ data except that we were able to fit for $\omega$ and also needed to fit for third light due to significant contamination of the TESS light curve. We found the best fit to the TESS data to be highly stable against changes in mass ratio, rotation rate, treatment of LD, albedo, gravity darkening and numerical grid size. We had to fix the LD coefficients as they diverged to unphysical values when we attempted to fit for them.

The final parameters and uncertainties of the fit are given in Table~\ref{tab:wd}. The fits are shown in Fig.~\ref{fig:lc:v346}, and two things are worth highlighting. First, the morphology of the light curve has changed between the $uvby$ and TESS epochs due to apsidal motion. The phase of secondary eclipse has changed and it is no longer annular -- the primary eclipse has become a transit instead. Second, the TESS data show a clear pulsation signature. This affected the quality of our solution and was probably why we were unable to fit for LD coefficients. The pulsation almost certainly arises from the EB itself and not from the contaminating light, because they are commensurate with the orbital period (see Fig.~\ref{fig:v346phased}). V346\,Cen is therefore another high-mass EB showing pulsations \citep{MeBowman22mn}. Because our light curve solution did not account for pulsations, we have conservatively doubled the uncertainties in the measured fractional radii.

\subsubsection{Pulsations}

\begin{figure}
\centering
\includegraphics[width=\columnwidth]{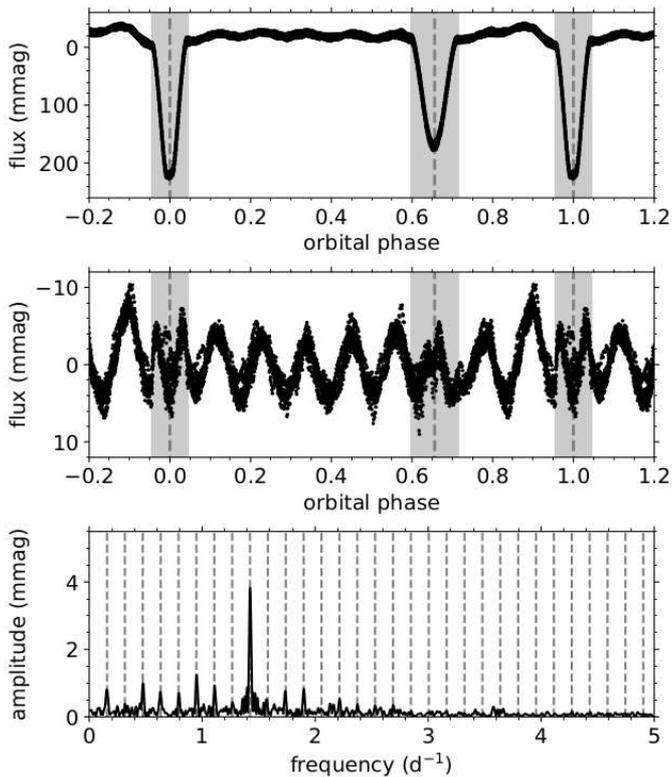}
\caption{\label{fig:phasefold-lc} Tidally excited pulsations of V346\,Cen. {\em Top:} observed light curve of V346\,Cen for sectors 37 and 38, phase-folded with the binary orbital period. The eclipses are indicated by the grey bands. {\em Middle:} Pulsation light curve of V346\,Cen for sectors 37 and 38, phase-folded with the binary orbital period. Data points taken during the eclipses again lie within the grey bands. {\em Bottom:} Lomb-Scargle periodogram, calculated for the out-of-eclipse data points of the pulsation light curve for sectors 37 and 38. The dashed vertical lines indicate harmonics of the orbital frequency.}
\end{figure}

Following the binary analysis, an analysis of the residual light curve (hereafter called the pulsation light curve) revealed the presence of tidally excited pulsations, as illustrated in Fig.\,\ref{fig:phasefold-lc}. To measure this tidally induced variability, we fitted sine waves, corresponding to the 20 lowest-order orbital harmonic frequencies, to the out-of-eclipse part of the pulsation light curve. Fitted orbital harmonics were accepted when the signal-to-noise ratio $S/N \geq 4.0$, where $S/N$ was calculated as the ratio of the amplitude of the fitted sine wave, and the average signal amplitude of the Lomb-Scargle periodogram \citep{Scargle1982} in a $1\,\rm d^{-1}$ window around the considered frequency. Finally, the measured amplitudes, phases and orbital harmonic frequencies were optimised simultaneously by nonlinearly fitting them to the pulsation light curve. Their values are listed in Table \ref{tab:puls_freq}.

From these results, we determined that the tidally induced pulsation corresponds to the $9^{\rm th}$ orbital harmonic, in agreement with what is shown in Fig.\,\ref{fig:phasefold-lc}. The physical origin of the other measured orbital harmonics is less clear. While they may also partially correspond to tidally induced pulsations, this could not be confirmed. At least part of it is likely caused by the non-sinusoidal nature and orbital-phase dependent amplitude modulations of the $9^{\rm th}$ orbital harmonic signal. Moreover, as shown in the middle panel of Fig.\,\ref{fig:phasefold-lc}, this pulsation has a minimum during the primary eclipse and a maximum during the secondary eclipse, which indicates that it belongs to the primary component.

\begin{figure}
\centering
\includegraphics[width=\columnwidth]{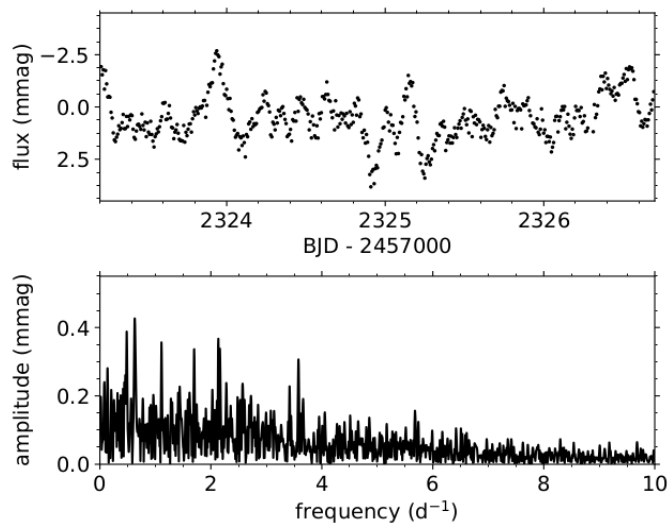}
\caption{\label{fig:slf} Stochastic low-frequency variability of V346\,Cen. {\em Top:} part of the (out-of-eclipse) residual light curve of V346\,Cen, after fitting the orbital harmonics. {\em Bottom:} Lomb-Scargle periodogram, calculated for the out-of-eclipse data points of the residual light curve for sectors 37 and 38.}
\end{figure}

Finally, after the significant tidally excited variability was removed from the pulsation light curve, we evaluated the residuals. As illustrated in Fig.\,\ref{fig:slf}, the remaining data exhibit signatures of stochastic low-frequency variability, as has been reported in the literature for other high-mass stars \citep[e.g.,][]{Bowman2019,Bowman2020}.

\begin{table}
\caption{\label{tab:puls_freq} Values of the amplitudes $A$, frequencies $\nu$, phases $\phi$ and signal-to-noise ratios $S/N$ of the orbital harmonics, calculated for the out-of-eclipse data points in the pulsation light curve of V346~Cen. }
\centering
\setlength{\tabcolsep}{4pt}
\begin{tabular}{lcccc}
\hline\hline
$n_{\rm orb}$ & $A$ (mmag) & $\nu$ ($\rm d^{-1}$) & $\phi$ ($2\pi$ rad) & $S/N$\\
\hline
1  &  $0.813 \pm 0.023$  &  0.15818130  &  $ 0.0811 \pm 0.0005$  &   6.3  \\
5  &  $0.812 \pm 0.026$  &  0.79090648  &  $ 0.378  \pm 0.005 $  &   6.7  \\
6  &  $0.771 \pm 0.025$  &  0.94908778  &  $-0.240  \pm 0.005 $  &   6.5  \\
8  &  $0.483 \pm 0.025$  &  1.26545037  &  $ 0.237  \pm 0.008 $  &   4.0  \\
9  &  $3.938 \pm 0.025$  &  1.42363166  &  $-0.2993 \pm 0.0010$  &  33.4  \\
\hline
\end{tabular}
\tablefoot { 
The frequency values were fixed at the indicated integer multiples of the measured orbital frequency $\nu_{\rm orb}$.
}
\end{table}

\subsection{Physical properties}

We have determined the physical properties of the systems using the results from the spectroscopic and photometric analyses outlined above. For this we used the velocity amplitudes, \Teff\ values, $e$ and $\omega$ from the spectroscopic analysis, and the fractional radii and orbital inclination from the photometric analysis. To perform the calculations we used the {\sc jktabsdim} code \citep{Me++05aa}, which propagates the errorbar from each input parameter using a perturbation analysis. We used a version of {\sc jktabsdim} modified to use the IAU system of nominal solar values \citep{Prsa+16aj} plus the NIST 2018 values for the Newtonian gravitational constant and the Stefan-Boltzmann constant. The results of this analysis are given in Table~\ref{tab:absdim}.

Distances have been derived using the measured radii and \Teff s of the stars, apparent magnitudes of the system in the Johnson-Cousins $UBVRI$ and 2MASS $JHK_s$ bands, and the theoretical bolometric corrections tabulated by \citet{Girardi+02aa}. We adjusted the interstellar extinction $E(B-V)$ to obtain consistent distances in the optical and infrared passbands. These results are given in Table~\ref{tab:absdim} alongside the \textit{Gaia} EDR3 parallaxes \citep{Gaia16aa,Gaia21aa} and the distance from simple inversion of the parallax. We see agreement within the errorbars, the most discrepant (1.6$\sigma$) being for V573\,Car. Very similar conclusions are drawn if we use the geometric or photogeometric distances from \citet{Bailerjones+21aj}. We conclude that our results for all four targets are independently verified by the \textit{Gaia} parallaxes.

\section{The parent clusters}\label{sec:parents}

Knowledge of the properties of stars in dEBs allows the determination of their distance. Moreover, a comparison of the properties of dEBs to stellar evolutionary models constrains their age. The age of stars in our sample, except for GL\,Car, were determined from isochrone fitting in \citet{Tkachenko_2020} for two cases: (i) as a single star; (ii) as a binary where the two components have the same age. Three different interior structures were assumed in these calculations, hence in Table~\ref{tab:clusters} we give lower and upper limits for the age. The distances to the binary systems in our sample are given in Table~\ref{tab:absdim}.

\begin{table*} \centering
\caption{\label{tab:clusters} Distances and ages for the binary systems in the present sample compared to the parent clusters. }
\begin{tabular}{lcclcccc} \hline\hline
Binary     & Distance (pc)  &           Age (Myr)         &   Cluster    &  Distance (pc)  & Ref.& Age (Myr)  & Ref.\\
\hline
V1034\,Sco  & $1460 \pm 50$  &   5.0--7.7 \ | \ 5.3--7.0   & NGC\,6321    & $1538 \pm 20 $  &  1  & 6.3        & 2   \\
GL\,Car     & $2278 \pm 63$  &       $2.0 \pm 0.5 $     & NGC\,3572a   & $2444 \pm 33 $  &  3  & 1--4       & 4   \\
V573\,Car   & $2466 \pm 78$  &   1.5--3.1 \ | \ 2.2--2.7   & Trumpler\,16 & $2360 \pm 505$  &  5  & $2 \pm 1$  & 6   \\
V346\,Cen   & $2290 \pm 60$  & 10.5--16.0 \ | \ 10.7-16.0  & Stock\,14    & $2439 \pm 326$  &  7  & $10 \pm 2$ & 7   \\
\hline
\end{tabular} 
\tablefoot{
The distances to the binary systems are from the present work (Table~\ref{tab:absdim}). The ages were calculated by \citet{Tkachenko_2020} except GL~Car for which the age is adopted from \citet{Gimenez_Clausen_1986}. \citet{Tkachenko_2020} estimated the age for two options: assuming the components are individual stars, and constraining the age to be the same for both components.
Both measurements are given, separated by a vertical line. }
\\
\tablebib{
(1) \citet{Banyard_2022}; (2) \citet{Kuhn_2017}; (3) \citet{Claria_1976}; (4) \citet{Garcia_1994};
(5) \citet{Goppl_Preibisch_2022}; (6) \citet{Hur_2012}; (7) \citet{Paunzen_Netopil_2006}.
} 
\end{table*}

\subsection{V1034\,Sco and NGC\,6231}

The distance to  V1034\,Sco was evaluated by \citet{Sana_2005} who found $d = \er{1528}{117}{109}$\,pc, which is within 1$\sigma$ of the distance we calculated. The light curve solution in \citet{Bouzid_2005} suffers from an ambiguity in setting the primary's \Teff\ so the authors calculated the distance for both cases. The larger one is exactly the same as those reported by \citet{Sana_2005}, while the shorter one is $d = \er{1399}{20}{20}$\,pc. \citet{Mayer_2008} determined the distance to another dEB in this cluster, V1007~Sco, as 1622~pc (no uncertainty given) which is somewhat larger than the other distance estimates mentioned here.

The open cluster NGC\,6231 belongs to the star-formation complex Sco~OB1 \citep{Perry_1991}. The cluster is the oldest and most massive in Sco~OB1 \citep{Damiani_2016}. The ages of the cluster members have been estimated to be between 2 and 8~Myr \citep{Sung_2013, Damiani_2016, Kuhn_2017}, with OB stars  being an older population in the cluster. The cluster is rich in spectroscopic binaries: \citet{Garcia_Mermilliod_2001} listed about 30 systems of which 16 are certain. \citet{Mayer_2008} did an exhaustive search of the cluster members, and listed ten EBs. The most recent distance determinations to NGC\,6231 are based on \textit{Gaia} parallaxes. \citet{Kuhn_2019} quoted $d = \er{1710}{13}{100}$\,pc using \textit{Gaia} DR2, while \citet{Banyard_2022} found the median geometric and photogeometric distances for their sample of about 60 stars in the cluster using \textit{Gaia} EDR3 parallaxes to be 1579 and 1576 pc, respectively.

\citet{Rosu_2022b} determined the age of V1034\,Sco to be $\tau = 6.8 \pm 1.4$ Myr, in perfect agreement with the result of \citet{Tkachenko_2020}. Three other binary systems that are members of this cluster were studied: HD~152248 \citep{Rosu_2020}, HD~152219 \citep{Rosu_2022b} and HD~152218 \citep{Rosu_2022a}. Their ages were determined from the apsidal motion rate and range from 5 to 9.5 Myr.

\subsection{GL\,Car and NGC\,3572/Collinder~240}

\citet{Gimenez_Clausen_1986} found a distance to GL\,Car of $d = 2100$\,pc. They did not give an uncertainty but quoted an error of 100~pc due to bolometric corrections and interstellar reddening. This distance is smaller than our result and that from the \textit{Gaia} DR3 parallax. \citet{Gimenez_Clausen_1986} extensively discussed possible physical relationships to the open clusters in the vicinity of GL\,Car, which is in a region crowded with young open clusters and in the direction of the Sagittarius-Carina spiral arm.

Membership of GL\,Car in NGC\,3572 was proposed by \citet{Sahade_Beron_1963}. \citet{Gimenez_Clausen_1986} rejected this association due to the shorter distance to the dEB than the cluster, and because NGC\,3572 is a compact cluster with a radius of 5$^\prime$ and GL\,Car is at an angular distance of 40$^\prime$. It was recognised that the open cluster NGC\,3572 consists of two overlapping clusters, one at 2.3 kpc and one at 3.0 kpc \citep{Claria_1976}. The nearer cluster is also considered by \citet{Claria_1976} to be the probable nucleus of a scattered group of OB stars located in the vicinity, identified as Collinder 240 and an extension of Car~OB2. This is a region in which the line of sight is tangential to the molecular cloud ridge in the Carina Arm, and is projected on a rather small area in the sky. It shows as a region with a higher concentration of OB stars, but with a radial extension of several kpc.

The age of GL\,Car was determined to be $\tau = 2.0 \pm 0.5$\,Myr \citep{Gimenez_Clausen_1986}. This is compatible with age determinations for Collinder~240, $\tau \sim 1$\,Myr, and Car OB2, $\tau = 4$ Myr \citep{Garcia_1994}.

\subsection{V573\,Car and Trumpler\,16}

\citet{Freyhammer_2001} determined a distance to V573\,Car of $d = 2600\pm120$\,pc, and an age of $\tau = 1.5\pm1.0$\,Myr. Their distance determination is within 1$\sigma$ of ours. Also, the very young age is confirmed with extensive isochrone fitting to different stellar interior structure models in \citet{Tkachenko_2020}, as summarised in Table~\ref{tab:clusters}.

V573\,Car is situated near the centre of the open cluster Trumpler\,16, close to $\eta$~Carinae, the brightest star in the cluster, and one of the most intriguing objects in the Galaxy. The cluster itself, with its neighbouring clusters, Trumpler\,14, and Trumpler\,15, forms a chain of rich clusters in the prominent Carina star-forming complex, a conspicous part of the Carina-Vela spiral arm. The whole region is recognised as the young association Carina OB1, which also includes NGC\,3293 and several small open clusters, the \ion{H}{ii} region and prominent nebula NGC\,3372 powered by $\eta$~Car \citep{Smith_2006, Wright_2020}. Pre-\textit{Gaia} distance estimates relied mostly on multicolour photometry, and gave distances in the range 2.2--2.9~kpc and young ages in the range 1--3~Myr \citep{Hur_2012}. Using \textit{Gaia} EDR3 \citet{Shull_2021}, \citet{Maiz_Apellaniz_2022} and \citet{Goppl_Preibisch_2022} found distances to the cluster Trumpler\,16 at the lower end of the range: $2.32 \pm 0.12$, $2.38 \pm 0.20$ and $2.36 \pm 0.05$~kpc, respectively, all within 1$\sigma$ of our determination.

\subsection{V346\,Cen and Stock~14}

The distance and age of V346\,Cen were also determined by \citet{Gimenez_1986b}, which allows a direct comparison with our results. \citet{Gimenez_1986b} determined the distance $d = 2.38 \pm 0.18$ kpc, which is within 1$\sigma$ of our determination. The age of the binary system they found, $\tau = \er{10.0}{5.8}{3.6}$\, Myr, also agrees well with our result, $\tau = 10.7$--16.0\,Myr (Table~\ref{tab:clusters}).

Stock~14, the parent cluster of V346\,Cen, is described as a loose but clearly defined open cluster \citep{Moffat_Vogt_1975,Eichendorf_Reipurth_1979}. The most recent deep $UBV$ photometry of Stock~14 was obtained by \citet{Drobek_2013} primarily in a search for new variable stars. Their photometry allowed determination of the distance and an estimate of the age for Stock~14, $d = \er{2399}{56}{55}$\,pc, and $\tau = 20\pm10$\,Myr. The authors confirmed the cluster membership of V346\,Cen. Re-evaluation of the photometric distance and age of Stock~14 by \citet{Paunzen_Netopil_2006} also favoured a shorter distance than previous determinations. They obtained a distance of $d = 2439 \pm 326$\,pc, and an age of $\tau = 10\pm2$\,Myr, in fine agreement with the extensive photometric study by \citet{Drobek_2013} as well as our results for V346\,Cen.

%%%%%%%%%%%%%%%%%%%%%%%%%%%%%%%%%%%%%%%%%%%%%%%%%%%%%%%%%%%%%%%%%%%%%%%%%%%%%%%%%%%%%%%%%%%%%%%%%%%%%%%%%%%%%%%%%%%%%%%%%%%%%%%%%%%%%%%%%%%%%%%%%%%%%%%%%%%%%%%%%%%%%%%%%%%%%%%%%%%%%%%%%%%%%%%%%%%%%%%

\section{Discussion}\label{sec:discussion}

The results of the analyses above are summarised in Table~\ref{tab:cno} for elemental abundances and Table~\ref{tab:absdim} for fundamental stellar quantities. The stars in the present work cover a range of mass (8.4--17.1\Msun), radius (4.1--8.3\Rsun), \Teff\ (22\,500 to 32\,200~K), surface gravity (3.7--4.2\,dex) and \vsini\ (90--185\kms) and are all unevolved main sequence stars from late-O to early-B spectral types. 
%%(see Fig.~\ref{fig:sample}). 
We have achieved a high accuracy in the fundamental stellar properties, with uncertainties in mass of 0.6--2.0\%, radius of 0.8--1.7\%, and \logg\ of 0.009--0.021\,dex. Having a precise \logg\ allows us to avoid its degeneracy with \Teff\ in spectral analysis, resulting in uncertainties of 1.7--2.5\% in \Teff. Since \Teff\ and \logg\ are the principal quantities for specifying a model atmosphere, precise values are a prerequisite in measuring chemical abundances to a high precision. We now discuss the implications of our results for two subjects: evolutionary models for high-mass stars, and chemical evolution in high-mass binaries.

% \subsection{Abundances: binary components versus single stars}

Studies of chemical abundances in high-mass stars mostly concentrate on more advanced evolutionary stages, so it is difficult to perform a quantitative comparison between our results and those published elsewhere. \citet{Martins_2017} presented a study of six short-period binary systems. Of them, two are contact or overcontact systems so will have abundances altered by mass transfer, one (DH~Cep) has component stars considerably more massive than our sample (38\Msun\ and 33\Msun), while the remaining three (Y~Cyg, AH~Cep and V478~Cyg) are suitable for the comparison. Of these, AH~Cep and V478~Cyg were analysed in our previous work \citep{Pavlovski_2018} so a direct comparison is possible.
Two studies agreed to within 2$\sigma$ uncertainties in the [N/C] and [N/O], but only because of the large uncertainties quoted by \citet{Martins_2017}. It is hard to trace the reason for this, but it may be related to the large uncertainties in the atmospheric parameters in their study.

Results of a comprehensive analysis of a large sample of binary and/or multiple stars in the Tarantula Nebula have recently been published \citep{Almeida_2017, Mahy_2020a, Mahy_2020b} based on medium-resolution ($R = 6400$) spectra from VLT/FLAMES/GIRAFFE covering 3964--4567\,\AA. A total of 51 SB1 and SB2 systems were studied, of which 13 are eclipsing. The atmospheric parameters were determined using NLTE methods, and He, C and N abundances derived. The objects studied fall into five different groups: (1) long-period systems ($P > 20$~d) with well-detached components; (2) eccentric short-period ($P < 10$~d) detached binaries; (3) circular-orbit short-period ($P < 10$~d) binaries with strong tidal effects; (4) semi-detached systems; and (5) contact systems. No N enrichment was found for binaries in the first two groups, despite the components having \vsini\ values of 50--250\kms. This finding is in disagreement with evolutionary models with rotationally induced mixing \citep{Maeder_Meynet_2000, Heger_Langer_2000, Heger_2000}. Furthermore, a large N abundance was found for apparently slowly rotating stars in binaries. This agrees with initial findings by \citet{Hunter_2007, Hunter_2008, Hunter_2009} who detected three distinctive groups in a diagram of [N/H] versus \vsini\ for single OB stars (sometimes dubbed the ``Hunter diagram''): (1) stars showing N enrichment with \vsini; (2) rapidly rotating stars with no sign of N enrichment; and (3) stars with low \vsini\ and excessive N abundance.

In the third group of binaries from \citet{Mahy_2020a}, N enrichment was found for the fast rotators. This is a group of stars in which the strongest influence of tidal forces on rotationally induced mixing is expected, following theoretical calculations by \citet{deMink_2009}. By far the largest N enhancement was found for stars with almost the lowest \vsini\ in this group ($\sim$50\kms), just as in the case of findings for stars in the first two groups. \citet{Mahy_2020a} concluded that stars in detached binaries (groups 1 to 3) are evolving as single stars. A lack of a clear relationship between N abundance and \vsini\ is in conflict with theoretical models and makes it hard to understand the effect of rotationally induced turbulent mixing in stellar interiors.

A very recent comprehensive spectroscopic analysis of a large set of B-type stars in the young open cluster NGC\,3293 \citep{Morel_2022} also corroborates these results: in the sample of almost 150 B-type stars of which the majority have high \vsini, apparently no star with excess N abundance was detected. Only two stars are found with mild N enhancement, and these stars have a low \vsini. A lack of N enhancement in fast-rotating B stars, and conversely, further evidence for N enhancement in low-\vsini\ B stars, is in clear contradiction with theoretical evolutionary models which incorporate rotationally-induced mixing.

A state-of-the-art statistical analysis was carried out by \citet{Aerts_2014} to identify possible mechanism(s) that could explain the distribution of stars in the Hunter diagram. The authors collected a statistically significant sample of well-studied Galactic single B stars for which seven observables were available (surface N abundance, rotational frequency, magnetic field strength, and the amplitude and frequency of their dominant acoustic and gravity modes of oscillation). A multivariate analysis indicated that the \Teff\ and the frequency of the dominant acoustic oscillation mode have the most predictive power of the surface N abundance, whereas the rotational frequency of the star does not have any predictive power at all. Up to now, no follow-up studies have been undertaken to investigate these unexpected results.

Strong support for rotationally induced mixing has come from detailed abundance study for early B-type stars by \citet{Przybilla_2008} and \citet{Nieva_Przybilla_2012}. The authors selected 20 early-B stars with a low \vsini\ to allow a high precision in determination of the atmospheric parameters and chemical abundances. \citet{Przybilla_2010} confirmed an observationally tight correlation in the plot of abundance {ratios N/C versus N/O, with a slope predicted via nuclear reactions in the CNO process. The targets had a broad evolutionary range, from dwarfs to supergiants, and their CNO abundances followed predictions of the nuclear reaction theory.

\begin{figure*} \centering
\includegraphics[width=8.4cm]{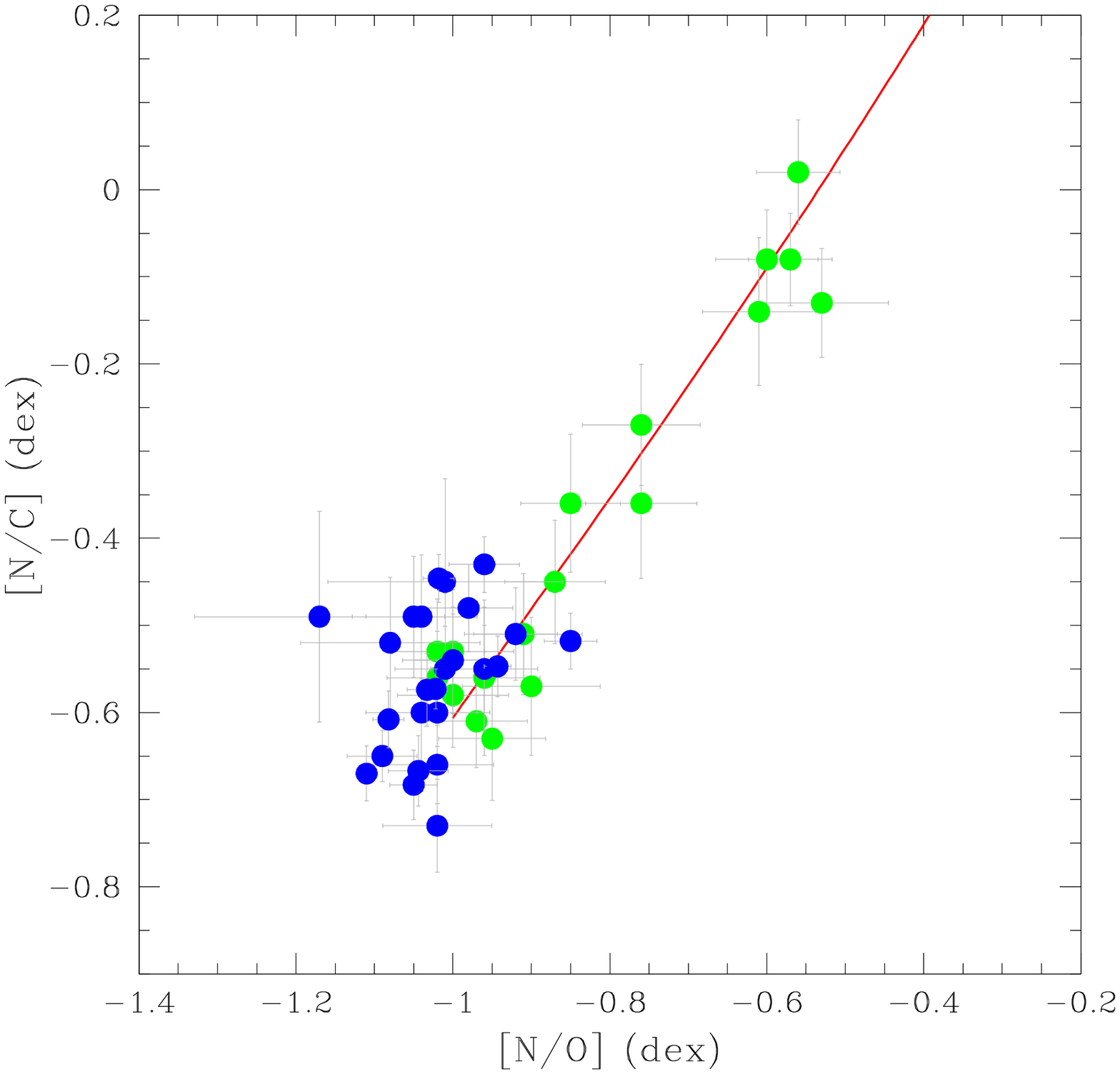} 
\includegraphics[width=8.4cm]{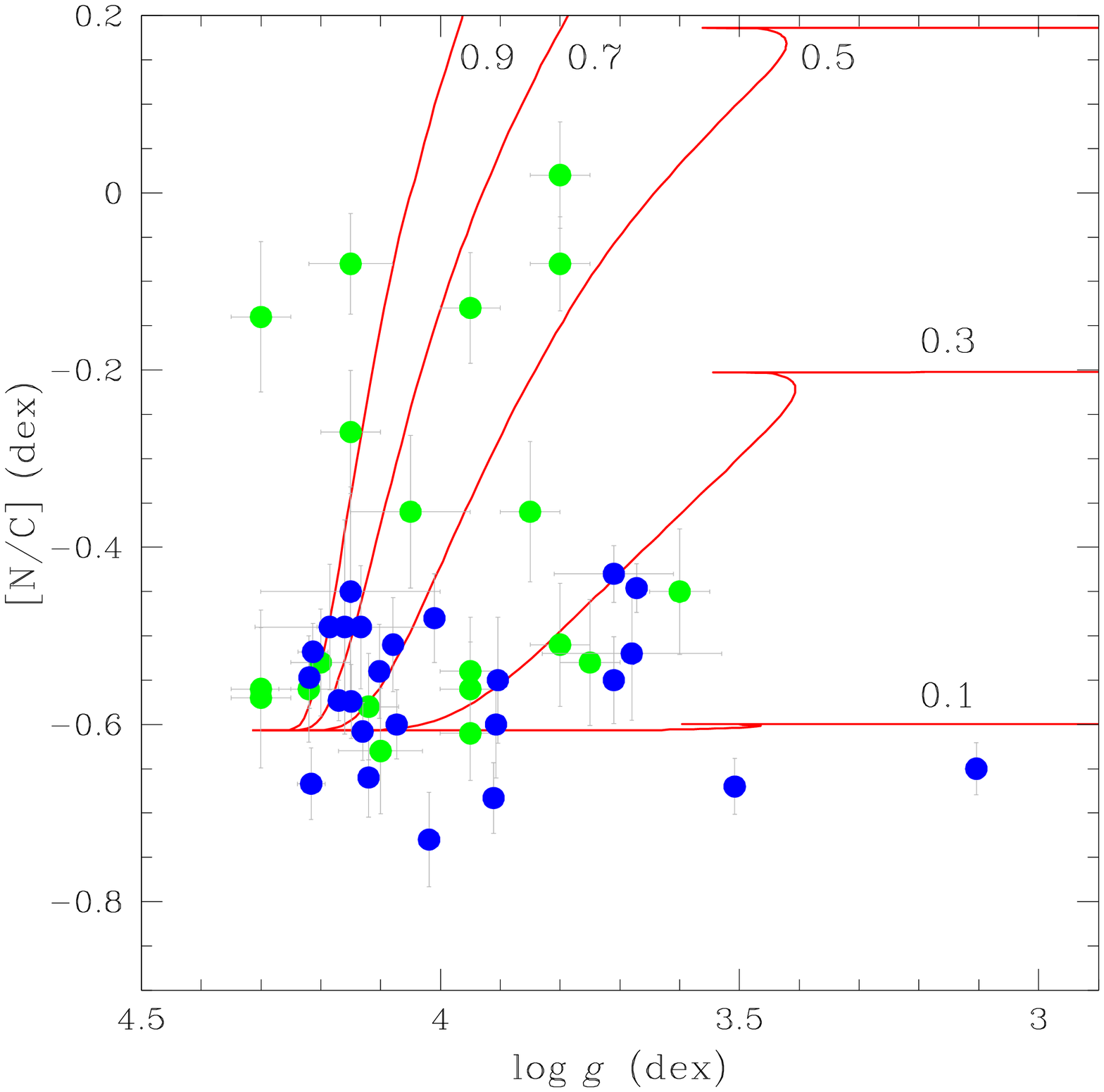}
\caption{\label{fig:cno.final.plot}  Left panel: Abundances of the CNO elements for high-mass stars in a diagram of [N/C] index versus [N/O] index. Stars in binary systems (Section~\ref{sec:abund}) are represented by solid blue circles. For comparison, abundance
determinations for single early-B type stars \citep{Nieva_Przybilla_2012} are represented by green circles.
Single stars obey a trend indicated by an analytical approximation to the nuclear reactions path for the CNO cycle derived in \citet{Przybilla_2010} and \citet{Maeder_2014}.  The slope in this diagram indicates a gradual enhancement of N at the expense of C.
A slight decrease in O abundance is also predicted. Whilst the single and binary stars span almost identical mass and \Teff\
ranges, and are all main-sequence stars, it is clear they do not share the same distribution (see Section~\ref{sec:discussion}). 
 Right panel: The observed [N/C] index for 23 high-mass stars in binaries (solid blue circles), compared to single B-type stars showing in solid green circles \citep{Nieva_Przybilla_2012}, as a function of surface gravity.  Solid red lines in the right panel show theoretical evolutionary tracks for a 15\Msun\ star and five values of the initial rotational velocity $\Omega/\Omega_{\rm crit}$ = 0.1, 0.3, 0.5, 0.7, and 0.9 \citep{Georgy_2013}.  Striking differences between single stars, and stars in binary systems are discussed in Section~\ref{sec:discussion}.} 
\end{figure*}

From our current and previous (see Section~\ref{sec:abund}) studies, we have a sample of 13 dEBs of which 25 components have measured CNO abundances. We compared these to a sample of high-mass stars published in \citet{Przybilla_2010} and \citet{Nieva_Przybilla_2012} in the logarithmic N/C versus N/O diagram (left panel in Fig.\,\ref{fig:cno.final.plot}). This is a powerful diagnostic tool in which the slope between [N/C] and [N/O] represents changes in CNO abundances due to nuclear reactions as derived in \citet{Przybilla_2010}. It is striking that the two samples cover the same mass range (8--20\Msun) but do not fully overlap in the diagram. For binary components there is a cut-off at [N/C]~$\sim -0.4$~dex and [N/O]~$\sim -0.8$~dex and they cluster around values close to solar ([N/C]$_\odot = -0.52$~dex and [N/O]$_\odot = -1.00$~dex), but a slope can be seen. The targets in the current work strengthen our previous conclusion that properties of interior mixing in binary stars are different from and might be less efficient than in single high-mass stars \citep{Pavlovski_2018}. 

 This striking effect is also clearly seen in the diagram of [N/C] versus  $\log g$ (Fig.\,\ref{fig:cno.final.plot}). Theoretical evolutionary tracks are presented for a 15\Msun\ star and five values of the initial rotational velocity $\Omega/\Omega_{\rm crit} = 0.1$, 0.3, 0.5, 0.7, and 0.9 \citep{Georgy_2013}. The overall spread in [N/C] could be interpreted as due to evolutionary changes or (very) high initial rotational velocities. However, only single stars from the sample of \citep{Nieva_Przybilla_2012} tend to be consistent with the large [N/C] ratio predicted by the models for large initial rotational velocity values of $\Omega/\Omega_{\rm crit}\gtrapprox 0.5$ and [N/C] $\gtrapprox -0.4$~dex. The main issue with the interpretation of the observed distribution in the context of the rotationally induced mixing alone is the generally low projected rotational velocity values ($v\,\sin\,i < 30$~\kms) found by \citet{Nieva_Przybilla_2012} for about half of their sample stars. For the effect of rotational mixing being alone responsible for the observed [N/C] and [N/O] abundance ratios, one would require the majority of apparently slow rotators in the sample of \citet{Nieva_Przybilla_2012} to be stars that are seen pole-on. This is a highly improbable scenario, so we conclude that the CNO abundances and their ratios observed in single high-mass stars are altered by multiple processes rather than just a single mechanism of rotational mixing. For example, high-mass stars are know to possess magnetic fields, stellar winds, and pulsations. To this (strong) tidal effects in close high-mass binary systems should be added. All these mechanisms, in one way or another, are expected to impact the efficiency of internal mixing, and hence the surface chemical composition.

However, the comparison between these two sets of empirical data, one with single high-mass stars and the other with high-mass stars in binary systems, is not straightforward. First and foremost, the sets differ in their distributions of $v \sin i$. The set of single stars were deliberately selected to be sharp-lined stars, so contains a mix of intrinsically slowly-rotating stars and ones with small inclinations and thus small $\sin i$ terms. The set of binary stars, on the other hand, contains objects whose equatorial rotational velocities are accurately known, assuming their rotational and orbital axes have been aligned during formation or by tidal effects. Furthermore, even though the $v \sin i$ distribution could be statistically corrected to intrinsic rotational velocities for single stars, there is a substantial difference in the rotational history between single and binary stars that one cannot easily account for. Evolution of stellar rotational velocity from its initial value at the zero-age main sequence, and its subsequent changes in the course of stellar evolution due primarily to changes in radius, is substantially different due to tidal effects. This is particularly important for short-period systems whose rotation is synchronised with and thus governed by their orbital period. Nevertheless, the non-detection of substantial changes in the CNO abundances of stars in binaries contradicts the predictions of single-star rotational evolutionary models. For the sample of \citet{Przybilla_2010} and \citet{Nieva_Przybilla_2012}, i.e.\ single high-mass stars with low observed $v \sin i$, the possibility remains that they agree with the theoretical predictions.

Tidal forces in binary and/or multiple systems affect the geometry of the orbits and the shape and spin of the components \citep{Mazeh_2008}. In order of increasing timescale, the stellar spin axes are aligned first, then their rotation is synchronised, and finally the orbit is circularised. Later evolution is dominated by mass transfer due to the increase in the sizes of the component stars. Our hypothesis that tidal effects suppress the efficiency of rotational mixing is not supported by theoretical calculations \citep{deMink_2013}, which predict precisely the opposite: that short-period circularised binary systems should experience rotationally induced turbulent mixing in stellar interiors.

In looking for possible mechanisms which diminish turbulent mixing in the components of binary systems, \citet{Koenigsberger_2021} examined differential rotation in asynchronous binary systems. If the components in a binary system are not yet in synchronous rotation, tidally-induced and time-variable differential rotation could develop. The calculations by \citet{Koenigsberger_2021} clearly show the role of asynchronism: the induced rotation structure and its temporal variability depend on the degree of departure from synchronism. The authors further speculated that, in this context, slowly-rotating asynchronous binaries could have more efficient mixing than the more rapidly-rotating but tidally locked systems. This shows that processes triggered by asynchronous rotation in binary systems cannot be ignored, while a comparison between samples of single and binary stars should be done with particular care, even when the latter are in a detached configuration. We note that 12 of the 14 binaries in our sample have eccentric orbits but that most of the component stars rotate synchronously.

%%%%%%%%%%%%%%%%%%%%%%%%%%%%%%%%%%%%%%%%%%%%%%%%%%%%%%%%%%%%%%%%%%%%%%%%%%%%%%%%%%%%%%%%%%%%%%%%%%%%%%%%%%%%%%%%%%%%%%%%%%%%%%%%%%%%%%%%%%%%%%%%%%%%%%%%%%%%%%%%%%%%%%%%%%%%%%%%%%%%%%%%%%%%%%%%%%%%%%%

\section{Conclusion}\label{sec:conclusions}

Despite their astrophysical importance, high-precision fundamental stellar quantities (mass, radius, \Teff) have been determined for only a few high-mass stars in binary systems in our galaxy \citep{Southworth_2015}. Even fewer have measurements of their surface chemical composition \citep{Serenelli_2021}. In the present work we have added four more binary systems to this list: V1034\,Sco, V346\,Cen, GL\,Car and V573\,Car, containing stars of masses from 8.4 to 17.1\Msun. Most of these stars are young, with only two in the second half of their MS evolution.

We have presented high-quality HARPS spectra and analysed them using spectral disentangling to determine their spectroscopic orbits and the individual spectra of the component stars. These were analysed using an NLTE approach. We have modelled the available light curves for our systems, comprising $uvby$ photometry in all cases and TESS photometry in three cases, to determine their photometric parameters. Combining these analyses, we have determined high-precision masses, radii, surface gravities, \Teff\ values, \vsini\ values and C, N, O, Mg and Si abundances for all eight stars in the four binary systems. Of particular interest are the CNO abundances since these elements are involved in core hydrogen burning through the CNO cycle. During a star's evolution its N abundance increases and its C abundance decreases. Rotationally induced mixing of stellar material, or some other mixing processes, could bring nuclear-processed material from the stellar core to the surface. Therefore, the [N/C] ratio is a sensitive probe of interior mixing processes during the MS evolutionary stage.

The CNO abundances determined in this work corroborate our previous findings \citep{Pavlovski_2018} that interior mixing is different in binary stars to single stars. A tight correlation of [N/C] with [N/O] versus the predicted evolutionary changes has been found for single early B-type stars \citep{Przybilla_2010, Nieva_Przybilla_2012}, whereas binary systems in our sample show much less variation in both [N/C] and [N/O]. However, care is needed when comparing them with single stars due to the differences in rotational velocity between these types of object. It remains true that the binary sample does not reproduce the results found for a sample of single low-\vsini\ B-type stars.}

On other hand, recent spectroscopic analysis of large samples of high-mass stars in binaries \citep{Mahy_2020b}, and single B-type stars in the young open cluster NGC\,3293 \citep{Morel_2022} apparently confirmed the lack of substantial changes in CNO abundances for high-\vsini\ stars, i.e.\ for intrinsically fast-rotating stars.

We speculate %\textbf{AT: speculate is a better word?}
that proximity effect in binary systems somehow suppress mixing and/or transport of chemical elements from the interior to the surface. However, firmer conclusions will need a substantial expansion of the binary stars sample and an extension to more massive and hotter stars, and/or wider long-period binary systems.

%%%%%%%%%%%%%%%%%%%%%%%%%%%%%%%%%%%%%%%%%%%%%%%%%%%%%%%%%%%%%%%%%%%%%%%%%%%%%%%%%%%%%%%%%%%%%%%%%%%%%%%%%%%%%%%%%%%%%%%%%%%%%%%%%%%%%%%%%%%%%%%%%%%%%%%%%%%%%%%%%%%%%%%%%%%%%%%%%%%%%%%%%%%%%%%%%%%%%%%

\section*{Acknowledgements}

 Careful reading of the manuscript and useful suggestions provided by the referee are acknowledged. We are indebted to Keith Butler and Norbert Przybilla for kindly sharing their codes, and the model atoms used in the present work. 
KP and ET were initially supported by the Croatian Science Foundation through research grant IP-2014-09-8656. 
The research leading to these results has (partially) received funding from the KU~Leuven Research Council (grant C16/18/005: PARADISE) and from the BELgian federal Science Policy Office (BELSPO) through PRODEX grant PLATO.
TVR gratefully acknowledges support from the Research Foundation Flanders (FWO) under grant agreement number 12ZB620N.

\bibliographystyle{aa} % style aa.bst
\bibliography{harps_high-mass_binaries}

\newpage

\appendix

\section{Additional plots}

\begin{figure*} \centering
\includegraphics[width=\textwidth]{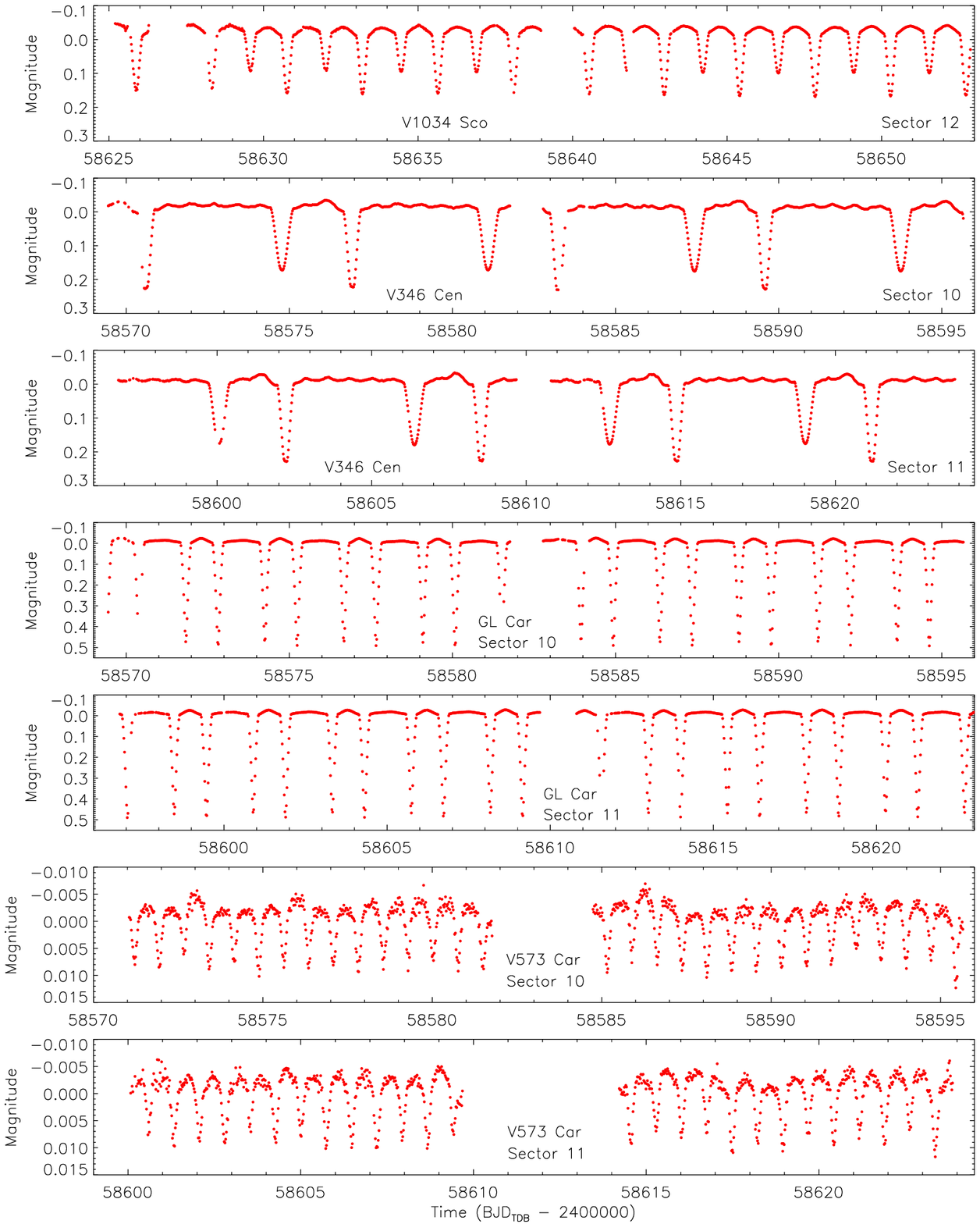}
\caption{\label{fig:tessnotused} Light curves of our target stars, from our own reduction of data from the TESS satellite, that were not included in the work in this paper, but could useful for studies of the period changes, and apsidal motion. The reduced photometric data are given in Table~A.1 only available in electronic form at the CDS (see article front page).}
\end{figure*}

\begin{figure*} \centering
\includegraphics[width=\textwidth]{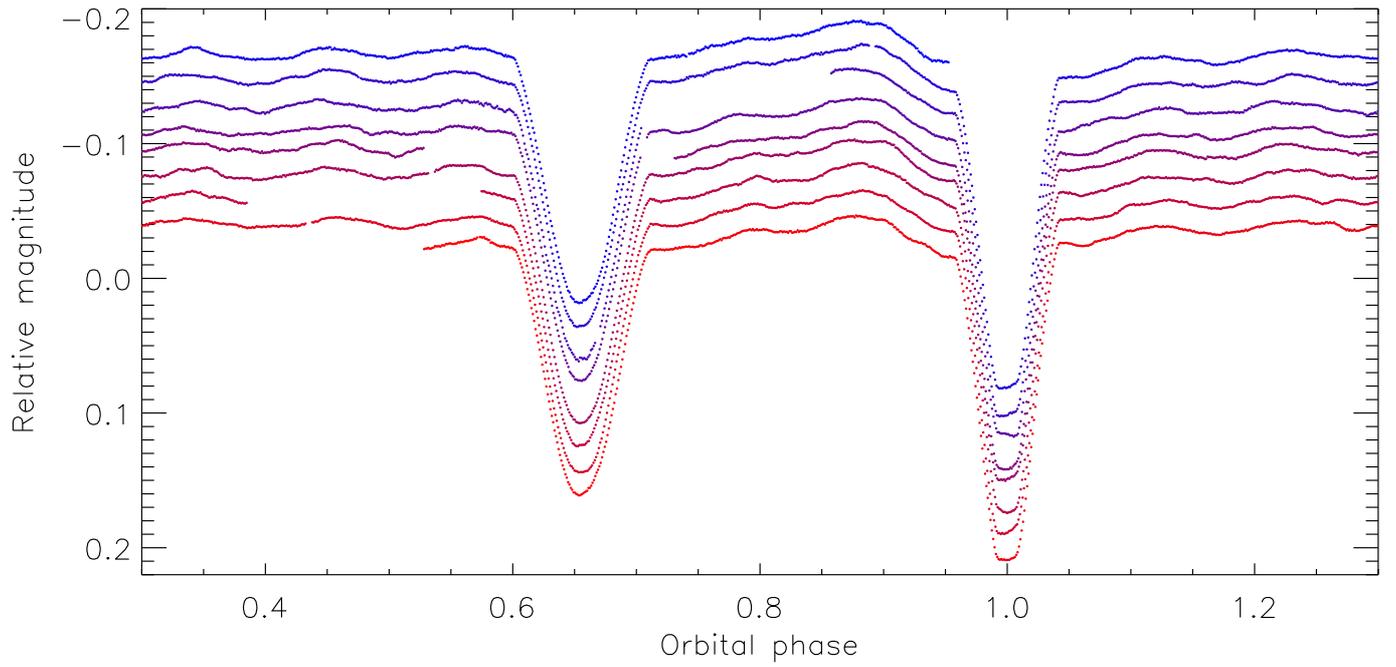}
\caption{\label{fig:v346phased} Light curve of V346\,Cen from the TESS satellite, plotted
versus orbital phase but with a small magnitude offset linearly dependent on time to shift
successive cycles upward in the figure. The earliest points are coloured red and the latest
points are coloured blue. The repetition of the pulsation signature with orbital phase is
easy to see.}
\end{figure*}

\begin{figure*}
\centering
\includegraphics[width=84mm]{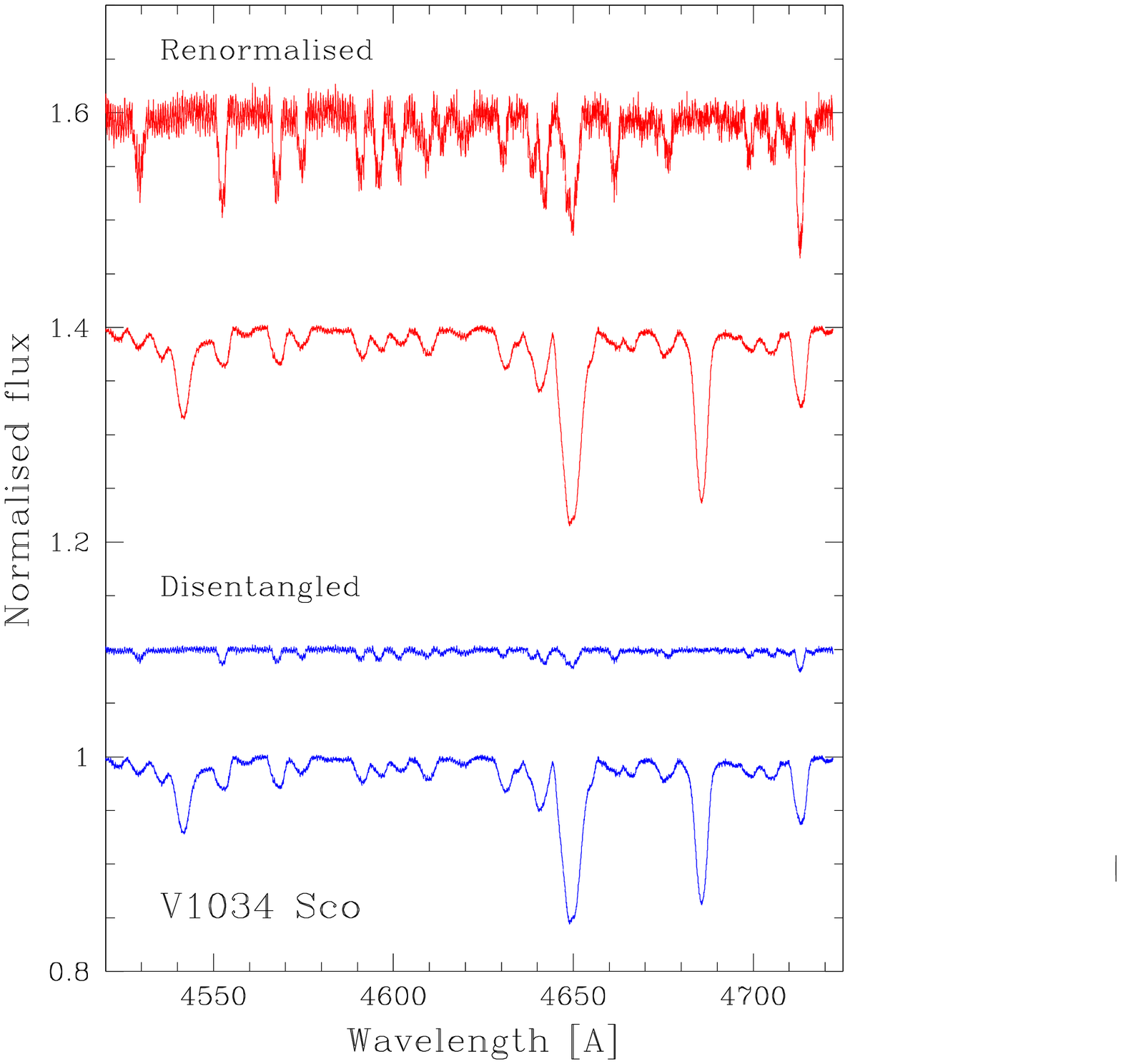}
\includegraphics[width=84mm]{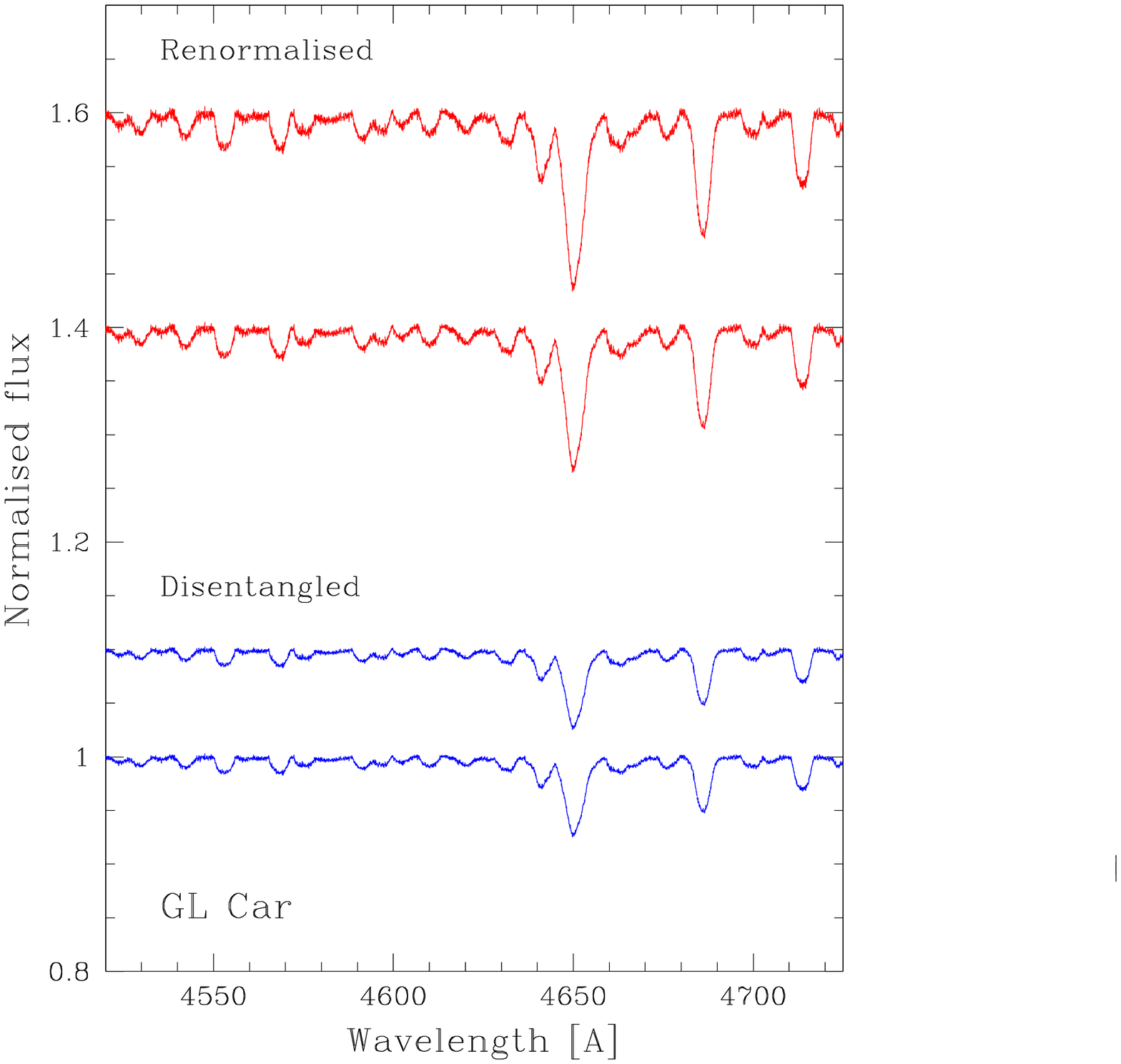} \\
\includegraphics[width=84mm]{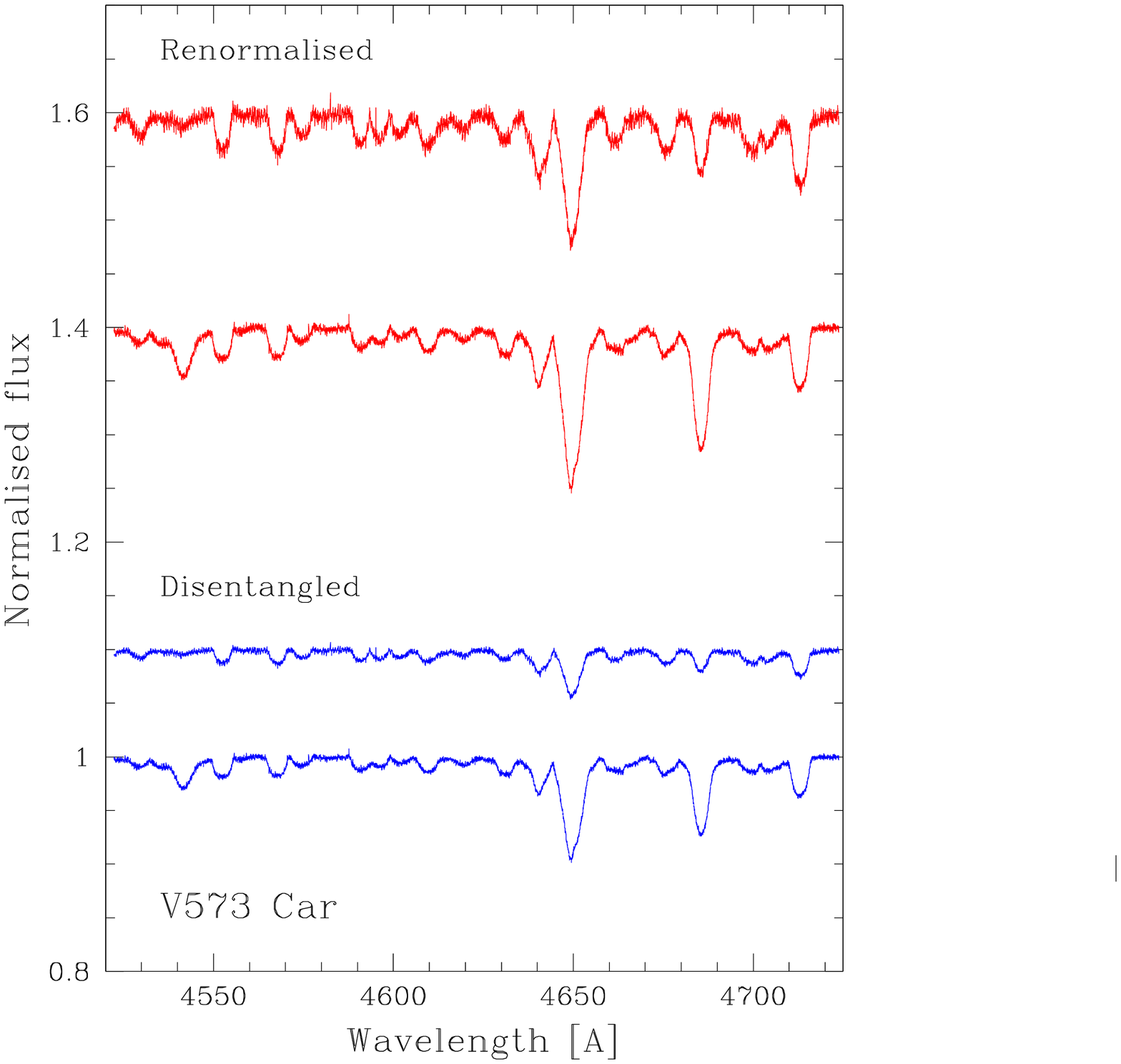}
\includegraphics[width=84mm]{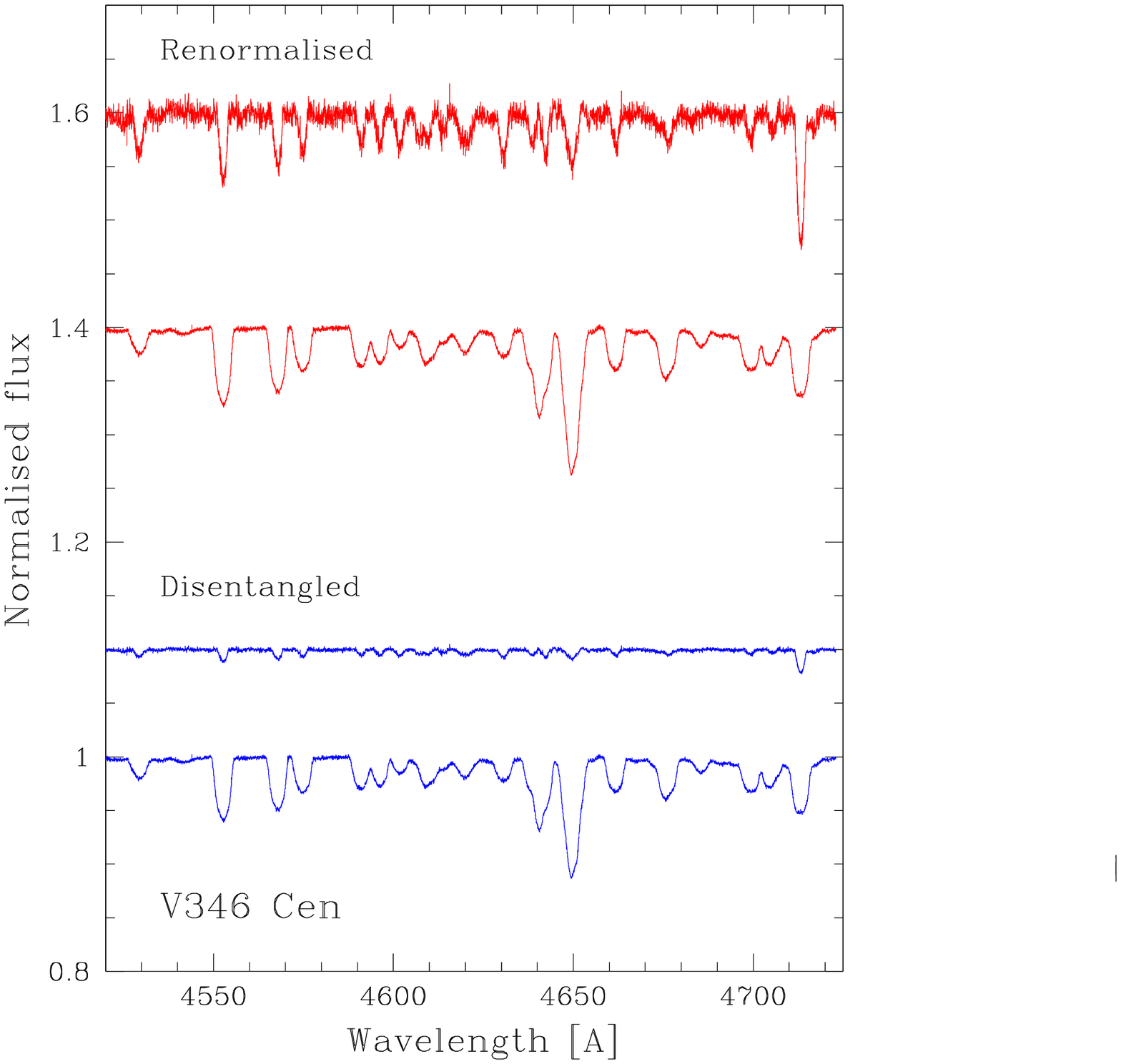} \\
\caption{Portions of the disentangled spectra of the stars (labelled) studied in this work.}
\label{fig:plotspe}
\end{figure*}

\label{lastpage}
\end{document}